\documentclass[12pt]{article}

\usepackage{psfrag}
\usepackage[small]{caption}
\usepackage{color}
\definecolor{darkblue}{rgb}{0.1,0.1,.7}
\usepackage[dvips, colorlinks, linkcolor=darkblue, citecolor=darkblue, urlcolor=black, linktocpage]{hyperref}
\usepackage{epsfig,amsmath,amssymb,verbatim,mathrsfs}
\usepackage[square, comma, sort&compress,numbers]{natbib}
\usepackage{dsdshorthand}
\usepackage{feynmp}
\usepackage{subfigure}
\usepackage{simplewick}

\numberwithin{equation}{section}
\newcommand\cN{\mathcal{N}}
\newcommand\cS{\mathcal{S}}
\newcommand\cW{\mathcal{W}}
\newcommand\bc{\mathbf{c}}
\newcommand\nn{\nonumber}

\newcommand\biblink[2]{{\tt\href{#1}{#2}}}

\renewcommand\be{\begin{eqnarray}}
\renewcommand\ee{\end{eqnarray}}
\newcommand\coeffs{\mathbf{coeffs}}

\newcommand\cV{\mathcal{V}}
\newcommand\vecsix[6]{\p{\begin{array}{c}\!\!#1\!\!\\\!\!#2\!\!\\\!\!#3\!\!\\\!\!#4\!\!\\\!\!#5\!\!\\\!\!#6\!\!\end{array}}}
\newcommand\Ad{\mathrm{Ad}}
\newcommand\EW{\mathrm{EW}}
\newcommand\UV{\mathrm{UV}}
\newcommand\TeV{\mathrm{TeV}}
\newcommand\AdS{\mathrm{AdS}}
\newcommand\cK{\mathcal{K}}
\newcommand\eff{\mathrm{eff}}
\newcommand\lrptl{\raise .8ex\hbox{$^\leftrightarrow$} \hspace{-9pt} \partial}
\newcommand\lptl{\raise .8ex\hbox{$^\leftarrow$} \hspace{-9pt} \partial}
\newcommand\rptl{\raise .8ex\hbox{$^\rightarrow$} \hspace{-9pt} \partial}

\title{{\Large {\bf Carving Out the Space of 4D CFTs}}}
\author{David Poland$^{1,2}$, David Simmons-Duffin$^1$, and Alessandro Vichi$^{3,4}$ \\ \\
{\it \normalsize ${}^1$Jefferson Physical Laboratory, Harvard University, Cambridge, MA 02138, USA}\\
{\it \normalsize ${}^2$School of Natural Sciences, Institute for Advanced Study, Princeton, NJ 08540, USA}\\
{\it \normalsize ${}^3$ Institut de Th\'eorie des Ph\'enom\`enes Physiques, EPFL, Lausanne, Switzerland}\\
{\it \normalsize ${}^4$ Ernest Orlando Lawrence Berkeley National Laboratory,}\\
{\it \normalsize University of California, Berkeley, CA 94720, USA}
}

\begin{document}

\begin{titlepage}

\noindent

\vspace{1cm}

\maketitle
\thispagestyle{empty}

\begin{abstract}
We introduce a new numerical algorithm based on semidefinite programming to efficiently compute bounds on operator dimensions, central charges, and OPE coefficients in 4D conformal and $\cN=1$ superconformal field theories.  Using our algorithm, we dramatically improve previous bounds on a number of CFT quantities, particularly for theories with global symmetries. In the case of $\SO(4)$ or $\SU(2)$ symmetry, our bounds severely constrain models of conformal technicolor.  In $\cN=1$ superconformal theories, we place strong bounds on $\dim(\Phi^\dagger\Phi)$, where $\Phi$ is a chiral operator.  These bounds  asymptote to the line $\dim(\Phi^\dagger\Phi) \leq 2 \dim(\Phi)$ near $\dim(\Phi) \simeq 1$, forbidding positive anomalous dimensions in this region.  We also place novel upper and lower bounds on OPE coefficients of protected operators in the $\Phi \x\Phi$ OPE.  Finally, we find examples of lower bounds on central charges and flavor current two-point functions that scale with the size of global symmetry representations.  In the case of $\cN=1$ theories with an $\SU(N)$ flavor symmetry, our bounds on current two-point functions lie within an $O(1)$ factor of the values realized in supersymmetric QCD in the conformal window.
\end{abstract}

\end{titlepage}

\setcounter{page}{1}

\setcounter{tocdepth}{3}
\tableofcontents

\vfill\eject


\newpage

\section{Introduction}
\label{sec:intro}

Conformal phases in four dimensions are ubiquitous and may play a crucial role in beyond the Standard Model physics.  Some simple examples include walking~\cite{Holdom:1984sk,Akiba:1985rr,Appelquist:1986an,Yamawaki:1985zg,Appelquist:1986tr,Appelquist:1987fc} or conformal~\cite{Luty:2004ye,Luty:2008vs,Galloway:2010bp,Evans:2010ed,Azatov:2011ht,Azatov:2011ps} technicolor, explanations of the flavor hierarchies~\cite{Georgi:1983mq, Nelson:2000sn, Poland:2009yb, Craig:2010ip}, and solutions to the supersymmetric flavor problem~\cite{Kobayashi:2001kz, Nelson:2001mq, Luty:2001jh, Luty:2001zv, Kobayashi:2002iz, Dine:2004dv, Sundrum:2004un, Ibe:2005pj, Ibe:2005qv, Schmaltz:2006qs, Kachru:2007xp, Aharony:2010ch, Kobayashi:2010ye, Dudas:2010yh}, the $\mu/B\mu$ problem in gauge mediation~\cite{Roy:2007nz, Murayama:2007ge, Perez:2008ng,Kim:2009sy,Craig:2009rk,Hanaki:2010xf}, or the $\eta$ problem in inflation~\cite{Baumann:2010ys}.  Moreover, studying conformal field theories (CFTs) can also give us important insights into quantum gravity and string theory via the AdS/CFT correspondence~\cite{Maldacena:1997re,Gubser:1998bc,Witten:1998qj}, which in turn provides a simple framework for describing many new physics scenarios via effective field theories in AdS~\cite{Randall:1999ee,ArkaniHamed:2000ds,Rattazzi:2000hs} (dual to `effective CFTs'~\cite{Fitzpatrick:2010zm}).

However, in recent years it has been realized that the restrictions imposed by conformal symmetry are not very well understood.  While constraints on the form of simple correlation functions (e.g.,~\cite{3-pt,Osborn:1993cr}) and unitarity restrictions on operator dimensions~\cite{unitarity4D,Mack:1975je} were worked out long ago, it was pointed out in~\cite{Rattazzi:2008pe} that crossing symmetry of four-point functions combined with the constraints of unitarity imply additional bounds on operator dimensions that must be satisfied in any consistent CFT.  These bounds were soon strengthened~\cite{Rychkov:2009ij} and extended to bounds on scalar operator product expansion (OPE) coefficients~\cite{Caracciolo:2009bx}.  In~\cite{Poland:2010wg} the bounds were also extended to $\cN=1$ superconformal field theories (SCFTs); bounds on central charges in general CFTs and SCFTs were also explored in~\cite{Poland:2010wg} and~\cite{Rattazzi:2010gj}.  In addition, progress on incorporating global symmetries into the program (important for both phenomenological applications and to have a more direct comparison with known theories) was made in~\cite{Rattazzi:2010yc}, and improved bounds (both for general CFTs with global symmetries and for SCFTs) were presented in~\cite{Vichi:2011ux}. 

The methods used in~\cite{Rattazzi:2008pe,Rychkov:2009ij,Caracciolo:2009bx,Poland:2010wg,Rattazzi:2010gj,Rattazzi:2010yc,Vichi:2011ux} to obtain bounds involve applying linear functionals to CFT crossing relations, which in practice means taking linear combinations of derivatives of the crossing relations evaluated at a particular point.  By searching for linear functionals that are positive when acting on the contributions of all possible primary operators in the spectrum other than the unit operator, one can obtain bounds on OPE coefficients (and sometimes operator dimensions).  However, to implement this positivity condition, the authors of \cite{Rattazzi:2008pe,Rychkov:2009ij,Caracciolo:2009bx,Poland:2010wg,Rattazzi:2010gj,Rattazzi:2010yc,Vichi:2011ux} introduced a finely-spaced discretization of the set of possible operator dimensions, making the resulting {\it linear programming} problem numerically difficult and limiting how far the idea could be pushed.  This numerical limitation was particularly apparent when considering systems of crossing relations that occur in theories with global symmetries, where the bounds obtained so far still seem to be quite far from their optimal values.

In the present paper we will present an alternate approach that completely avoids this discretization of dimensions.  We will use the fact that linear combinations of derivatives of conformal blocks can be arbitrarily-well approximated by ratios of polynomials in the operator dimensions, which allows us to convert the problem of obtaining bounds into a {\it semidefinite programming} problem that is numerically much more efficient.  This then allows us to obtain much stronger bounds on CFTs and SCFTs, particularly in the presence of global symmetries.  

More concretely, for general CFTs we will consider four-point functions of scalar operators $\phi$, as well as collections of operators $\phi_i$ transforming as fundamentals under $\SO(N)$ or $\SU(N)$ global symmetries.  For theories with $\cN=1$ supersymmetry we will focus on the case of chiral superconformal primary operators $\Phi$, as well as on collections of chiral operators $\Phi_i$ transforming as $\SU(N)$ fundamentals.  We start by reviewing the relevant crossing relations and representation theory in section~\ref{sec:review}.  There we will also introduce our new method to obtain bounds on operator dimensions and OPE coefficients based on semidefinite programming.  

In section~\ref{sec:dimbounds} we use this method to derive general bounds on operator dimensions.  In the case of general CFTs with $\SO(N)$ global symmetries, we will place upper bounds on the dimension of the lowest-dimension $\SO(N)$-singlet operator appearing in the $\phi_i \times \phi_j$ OPE.  This greatly improves upon the bounds in the presence of global symmetries previously presented in~\cite{Rattazzi:2010yc,Vichi:2011ux}.  We also place similar bounds on the lowest-dimension $\SO(N)$ symmetric tensor $\f_{(i}\f_{j)}$.  In the case of $\SU(N)$ global symmetries we can additionally place bounds on $\SU(N)$-singlet or $\SU(N)$-adjoint operators appearing in the $\phi_{i} \times \phi^{\bar{\jmath} \dagger}$ OPE.  Somewhat surprisingly, we find that $\SU(N)$-singlet bounds turn out to be identical to $\SO(2N)$-singlet bounds using the present method.

The special case of an $\SO(4)$ or $\SU(2)$ global symmetry is relevant for the scenario of conformal technicolor~\cite{Luty:2004ye}, with or without custodial symmetry.  In this scenario one would like the dimension of the Higgs operator $H$ to be somewhat close to 1, while the dimension of $H^{\dagger} H$ should be close to or greater than 4.  On the other hand, the bounds in this paper show that requiring $\textrm{dim}(H^{\dagger} H) \geq 4$ forces one to have at least $\textrm{dim}(H) \gtrsim 1.52$, excluding flavor-generic versions of this scenario and placing significant constraints on models where Yukawa-like suppressions are generated in four-fermion operators.

In $\cN=1$ superconformal theories we also place bounds on the lowest-dimension scalar superconformal primary appearing in the $\Phi \times \Phi^{\dagger}$ OPE, where $\Phi$ is a chiral operator.  This greatly strengthens the bounds presented in~\cite{Poland:2010wg,Vichi:2011ux}.  In fact, we will see that the bound appears to asymptote to the line $\dim(\Phi^{\dagger}\Phi) \leq 2\dim(\Phi)$ near $\dim(\Phi) \sim 1$, essentially excluding the possibility of `positive anomalous dimensions' (as recently discussed in~\cite{Fitzpatrick:2011hh}) in this region.  This also implies that the solution to the $\mu/B\mu$-problem proposed in~\cite{Roy:2007nz,Murayama:2007ge} cannot easily work near $\dim(\Phi) \sim 1$.  

In section~\ref{sec:opebounds} we explore bounds on OPE coefficients.  First we strengthen the upper bounds presented in~\cite{Caracciolo:2009bx} on the sizes of OPE coefficients of scalars $\cO$ appearing in the $\phi \times \phi$ OPE in non-supersymmetric theories.  Then, as a new application of these methods in superconformal theories, we place both upper and {\it lower} bounds the OPE coefficient of the chiral $\Phi^2$ operator which always appears in the $\Phi \times \Phi$ OPE.  In this case, lower bounds are possible because unitarity requires that there is a gap in the spectrum of dimensions, so no other nearby operators can mimic the effects of the $\Phi^2$ operator in the conformal block decomposition.  We similarly place upper and lower bounds on the OPE coefficients of the other higher-spin protected operators that can appear in the $\Phi \times \Phi$ OPE.  These bounds have interesting implications for Banks-Zaks theories or CFTs with weakly-coupled AdS$_5$ duals, where they can be checked in perturbation theory.

Next, in section~\ref{sec:ccbounds} we place lower bounds on the central charge $c$, which appears as the coefficient in the two-point function of the stress tensor: $\<T T\> \propto c$.  These bounds strengthen and expand upon those previously explored in~\cite{Poland:2010wg,Rattazzi:2010gj,Vichi:2011ux}.  In theories with operators of dimension $d$ transforming as fundamentals under $\SO(N)$ or $\SU(N)$ global symmetries, we find that the bounds scale linearly with $N$ near $d \sim 1$, consistent with our intuition from free CFTs.  We explore these bounds on $c$ in both general CFTs and $\cN=1$ SCFTs.  In the latter case one can calculate $c$ using 't Hooft anomaly matching in many known SCFTs, and our bounds are satisfied in all such examples that we have checked.

In section~\ref{sec:currentbounds} we place similar bounds on the coefficient $\ka$ appearing in the two-point function of a global symmetry current: $\<J^A J^B \> \propto \kappa \Tr(T^A T^B)$.  Here we extend the previous results of~\cite{Poland:2010wg} to include the full information about global symmetries.  In the case of scalar operators transforming as fundamentals of $\SO(N)$, we place lower bounds on $\kappa_{\SO(N)}$.  In the case of $\SU(N)$ global symmetries, one can either bound the OPE coefficient appearing in front of the $\SU(N)$ (adjoint) current or the coefficient in front of an $\SU(N)$-singlet current corresponding to a different global symmetry.  In the latter case, the bounds again scale linearly with $N$ near $d \sim 1$ in accordance with our intuition from free CFTs.  We also compute similar bounds in $\cN=1$ SCFTs where $\kappa$ can be computed using 't Hooft anomaly matching, and present a comparison of our results with supersymmetric QCD in the conformal window~\cite{Seiberg:1994pq}.  We conclude in section~\ref{sec:conclusions}.

\section{Bounds from Crossing Relations}
\label{sec:review}

\subsection{CFT Review}

Let us begin by reviewing some basic aspects of conformal field theories that will be important for our discussion.  The conformal algebra contains, in addition to Poincar\'e generators, a dilatation generator $D$ and special conformal generators $K_\mu$.  Operators in a CFT can be classified into primaries $\cO^I$ satisfying $K_\mu \cO^I(0)=0$, and their descendants $P^{\mu}\cdots P^\nu \cO^I(0)$.\footnote{We leave the adjoint action of charges on operators implicit, i.e. $K_\mu \cO \equiv [K_\mu,\cO]$.}  Here, $I$ denotes possible Lorentz indices.  We will be primarily concerned with spin-$\ell$ operators which transform as traceless symmetric tensors of the Lorentz group, $\cO^I=\cO^{\mu_1\dots\mu_\ell}$.

Correlation functions of a conformal field theory on $\R^n$ are completely determined by some simple discrete data: the spectrum of operator dimensions and spins, and the coefficients appearing in the operator product expansion (OPE).  Knowledge of the spectrum is sufficient to determine all two-point functions. For primary operators $\cO_i^I$ and $\cO_j^J$ with equal dimensions and spins $\{\De,\ell\}$, we have
\be
\<\cO_i^I(x_1)\cO_j^J(x_2)\> &\propto&
\frac{w^{IJ}(x_{12})}{x_{12}^{2\De}},
\label{eq:conformal2ptfunction}
\ee
where $w^{IJ}(x)$ is a tensor whose form is fixed by conformal symmetry (e.g., for spin-$1$ operators $w^{\mu\nu}(x)=\eta^{\mu\nu}-\frac{2x^\mu x^\nu}{x^2}$).  When the dimensions and spins are not equal, the two-point function must vanish.  In addition, unitarity constrains  $\De$ to satisfy~\cite{unitarity4D,Mack:1975je}
\be
\De &\geq& 1\qquad\ \  (\ell=0),\nn\\
\De &\geq& \ell+2 \quad(\ell \geq 1).
\ee
These bounds can sometimes be strengthened if the conformal algebra is enhanced, as in superconformal theories.  We will see some examples of this shortly.

Let us choose an orthonormal basis of primaries $\cO_i$, so that the constant of proportionality in Eq.~(\ref{eq:conformal2ptfunction}) is $\de_{ij}$. Having done so, the remaining $n$-point functions of the theory are determined by coefficients in the operator product expansion.  For real scalars $\phi_1$ and $\phi_2$, this takes the form~\cite{Ferrara:1971zy}
\be
\phi_1(x) \phi_2(0) &=& \sum_{\cO\in\f_1\x\f_2} \l_{\f_1\f_2\cO} C_I(x,P)\cO^I(0),
\label{eq:scalarOPE}
\ee
where $\l_{\f_1\f_2\cO}$ are constants that must be real in a unitary theory.  The notation $\cO\in \f_1\x\f_2$ indicates that $\cO$ is a primary operator in the OPE of $\f_1$ and $\f_2$.  We have grouped together each primary $\cO$ and its descendants $P\cO,P^2\cO,\dots$ into a single term using the operator $C_I(x,P)$ (which depends on the dimensions and spins of $\f_1,\f_2,$ and $\cO$, though we are suppressing that dependence for brevity).  One can show that the form of $C_I(x,P)$ is completely fixed by conformal symmetry.  For instance, applying special conformal generators $K_\mu$ to both sides of Eq.~(\ref{eq:scalarOPE}) gives a recursion relation for the terms in $C_I$ which can be solved order-by-order.  When $\f_1=\f_2=\f$, Bose symmetry dictates that only even-spin operators may enter the OPE~(\ref{eq:scalarOPE}).

In general field theories, the OPE is an asymptotic expansion, valid only at short distances.  However in a CFT, because of the absence of scales, the OPE is an exact equality that can be used to simplify products of operators with arbitrary separation inside correlation functions.\footnote{This is true provided there are no other operators `nearby' in a sense that can be made precise.}  
A key example for us is a four-point function of a scalar operator $\f$ of dimension $d$, which can be evaluated as follows:
\be
\<
\contraction{}{\f(x_1)}{}{\f(x_2)}
\f(x_1)\f(x_2)
\contraction{}{\f(x_3)}{}{\f(x_4)}
\f(x_3)\f(x_4)
\>
&=& \mathop{\sum_{\cO\in \f\x\f}}_{\cO'\in \f\x\f} \l_{\cO}\l_{\cO'} C_I(x_{12},\ptl_2)C_J(x_{34},\ptl_4)\<\cO^I(x_2)\cO'^J(x_4)\>\nn
\\
&=& \sum_{\cO\in\f\x\f} \l_{\cO}^2\frac{1}{x_{12}^{2d}x_{34}^{2d}} g_{\De,\ell}(u,v),\\
g_{\De,\ell}(u,v) &\equiv& x_{12}^{2d}x_{34}^{2d} C_I(x_{12},\ptl_2)C_J(x_{34},\ptl_4)\frac{w^{IJ}(x_{24})}{x_{24}^\De},
\label{eq:realscalarconformalblockexpansion}
\ee
where we have inserted Eq.~(\ref{eq:scalarOPE}) twice and used Eq.~(\ref{eq:conformal2ptfunction}) together with orthonormality of the $\cO$'s.  Here, $\De$ and $\ell$ are the dimension and spin of $\cO$, $u\equiv\frac{x_{12}^2x_{34}^2}{x_{13}^2 x_{24}^2}$ and $v\equiv\frac{x_{14}^2 x_{23}^2}{x_{13}^2 x_{24}^2}$ are conformal cross-ratios, and the functions $g_{\De,\ell}(u,v)$ are called {\it conformal blocks}.  Since conformal symmetry completely fixes $C_I$ and $w^{IJ}$, it also determines $g_{\De,\ell}$.  An exact expression in four dimensions, computed by Dolan and Osborn~\cite{Dolan:2000ut,Dolan:2003hv}, is given by
\be
\label{eq:conformalblocks}
g_{\De,\ell}(u,v) &=& \frac{z\bar z}{z-\bar z}(k_{\De+\ell}(z)k_{\De-\ell-2}(\bar z)-(z\leftrightarrow \bar z)),\\
k_\b(x) &\equiv& x^{\b/2}{}_2F_1(\b/2,\b/2,\b,x),
\ee
where $u=z\bar z$ and $v=(1-z)(1-\bar z)$.\footnote{Our convention for conformal blocks here differs by a factor of $(-2)^\ell$ from the one used in~\cite{Dolan:2000ut,Dolan:2003hv}.}  The unit operator is an important special case, with $g_{0,0}(u,v)=1$.

\subsection{Crossing Relations for Singlets, $\SO(N)$, and $\SU(N)$}

While a set of dimensions, spins, and OPE coefficients is enough to compute any correlation function, this data must satisfy additional consistency relations in a sensible CFT.  To simplify $\<\f\f\f\f\>$ using the OPE, we had to choose some way of pairing up the operators, and this choice necessarily broke manifest permutation symmetry among the $\f(x_i)$'s.  Nevertheless it should be the case that the end result remains permutation-symmetric, a requirement known as crossing symmetry.  As an example, switching $x_1\leftrightarrow x_3$ in the conformal block expansion Eq.~(\ref{eq:realscalarconformalblockexpansion}) leads to the crossing relation
\be
\sum_{\cO\in \f\x\f} \l_\cO^2 g_{\De,\ell}(u,v) &=& \p{\frac{u}{v}}^d \sum_{\cO\in \f\x\f} \l_\cO^2 g_{\De,\ell}(v,u).
\label{eq:crossingrelation}
\ee
Meanwhile, switching $x_1\leftrightarrow x_2$ reproduces the statement that only even-spin primaries appear in $\f\x\f$. Other permutations give no new information in this case.\footnote{Note that crossing symmetry of all four-point functions is equivalent to associativity of the OPE, which is enough to guarantee that higher $n$-point functions are crossing-symmetric as well.}

Recall that the $\l_\cO$ are real by unitarity, which means that the coefficients $\l_\cO^2$ are nonnegative.  This is a source of tension in Eq.~(\ref{eq:crossingrelation}), which can be expressed most clearly by rewriting our crossing relation as a `sum rule' with positive coefficients,
\be
\label{eq:sumrule}
F_{0,0}(u,v)+\sum_{\cO\in \f\x\f} \l_\cO^2 F_{\De,\ell}(u,v) &=& 0,
\ee
where
\be
F_{\De,\ell}(u,v) &\equiv & \frac{v^d g_{\De,\ell}(u,v)-u^d g_{\De,\ell}(v,u)}{u^d-v^d},\\
F_\mathrm{0,0}(u,v) &=& -1,
\ee
and we are suppressing the $d$-dependence of $F_{\De,\ell}$ for brevity.  Note that we have isolated the term  corresponding to the unit operator, whose OPE coefficient is fixed by the fact that $\f$ has a canonically normalized two-point function.  The unit operator contributes to Eq.~(\ref{eq:sumrule}) in its own particular way. Requiring that this contribution be cancelled by $F_{\De,\ell}$'s with positive coefficients leads to nontrivial constraints on the allowed $\De,\ell$ appearing in $\f\x\f$.  For some explicit examples and many details about the structure of the sum rule, see~\cite{Rattazzi:2008pe,Rychkov:2009ij}.  In sections \ref{sec:boundsfromcrossing} and \ref{sec:semidefiniteprogramming}, we will explain our improved method for extracting bounds on CFT data from Eq.~(\ref{eq:sumrule}).  For now, let us present some generalizations of the sum rule for other kinds of operators.

\subsubsection{$\SO(N)$ Crossing Relations}

An analysis of crossing relations in theories with $\SO(N)$ and $\SU(N)$ global symmetries was performed in \cite{Rattazzi:2010yc}, and improved bounds for $\SO(N)$ were presented in \cite{Vichi:2011ux}.  We will make extensive use of these results, so let us review them here.  

Consider a real scalar primary $\f_i$ transforming in the fundamental representation of an $\SO(N)$ global symmetry group.  A complex scalar is a special case with symmetry group $\SO(2)\cong U(1)$.  Operators in $\f_i\x\f_j$ can be organized into singlets $S$, symmetric tensors $T$, and antisymmetric tensors $A$ of $\SO(N)$.  Schematically,
\be
\f_i\x\f_j &\sim& \sum_{S^+}\de_{ij}\cO+\sum_{T^+} \cO_{(ij)} + \sum_{A^-} \cO_{[ij]}.
\ee
The notation $S^\pm,T^\pm,A^\pm$ indicates that the sum is restricted to even-spin ($+$) or odd-spin ($-$) primaries in $\f_i \x \f_j$ with the given representation, as dictated by Bose symmetry.  Keeping track of the $\SO(N)$ indices, each representation contributes differently to the conformal block decomposition of a four-point function,
\be
\label{eq:SONconformalblock}
x_{12}^{2d} x_{34}^{2d}\<
\contraction{}{\f_i(x_1)}{}{\f_j(x_2)}
\f_i(x_1)\f_j(x_2)
\contraction{}{\f_k(x_3)}{}{\f_l(x_4)}
\f_k(x_3)\f_l(x_4)
\> &=& \sum_{S^+} \l_\cO^2 (\de_{ij}\de_{kl}) g_{\De,\ell}(u,v)\nn\\
&&+\sum_{T^+}\l_\cO^2 \p{\de_{il}\de_{jk}+\de_{ik}\de_{jl}-\frac 2 N \de_{ij}\de_{kl}} g_{\De,\ell}(u,v)\nn\\
&&+\sum_{A^-}\l_\cO^2 \p{\de_{il}\de_{jk}-\de_{ik}\de_{jl}}g_{\De,\ell}(u,v) .
\ee
If we recompute this four-point function using a different operator pairing, each primary contributes again, but with the conformal cross-ratios $u$ and $v$ switched, and the tensor structures $\de_{ij}\de_{ki}$, $\de_{ik}\de_{jl},$ $\de_{il}\de_{jk}$ permuted.  Picking out the coefficients of each tensor structure then leads to three sum rules, which we can write in vectorial form
\be
\label{eq:SONvectorialsumrule}
\sum_{S^+}\l_\cO^2 \p{
\begin{array}{c}
0 \\
F_{\De,\ell}\\
H_{\De,\ell}
\end{array}
}
+
\sum_{T^+}\l_\cO^2 \p{
\begin{array}{c}
F_{\De,\ell}\\
(1-\frac 2 N)F_{\De,\ell}\\
-(1+\frac 2 N)H_{\De,\ell}
\end{array}
}
+
\sum_{A^-}\l_\cO^2 \p{
\begin{array}{c}
-F_{\De,\ell}\\
F_{\De,\ell}\\
-H_{\De,\ell}
\end{array}
}
&=& 0.
\ee
Here $H_{\De,\ell}(u,v)$ is a symmetrized version of $F_{\De,\ell}(u,v)$,
\be
H_{\De,\ell}(u,v) &\equiv& \frac{v^d g_{\De,\ell}(u,v)+u^d g_{\De,\ell}(v,u)}{u^d+v^d}, \\
H_{0,0}(u,v) &=& 1.
\ee
For brevity, we have not isolated the unit operator in Eq.~(\ref{eq:SONvectorialsumrule}); it is included with the even-spin singlets $S^+$.

The case of $\SO(4)$ is special, since one can additionally decompose antisymmetric tensors into self-dual and anti-self-dual parts $A_\pm$.  Let us quickly summarize the consequences, though they will turn out to be irrelevant for this work.  A new tensor structure can now appear in $\<\f_i\f_j\f_k\f_l\>$, namely $\e_{ijkl}$.  In Eq.~(\ref{eq:SONconformalblock}) we must replace 
\be
\sum_{A^-}\l_\cO^2 \p{\de_{il}\de_{jk}-\de_{ik}\de_{jl}}g_{\De,\ell}(u,v)
&\rightarrow&
\sum_{A^-_\pm}\l_\cO^2 \p{\de_{il}\de_{jk}-\de_{ik}\de_{jl}\pm \e_{ijkl}}g_{\De,\ell}(u,v).
\ee
Since $\e_{ijkl}$ maps to itself under permutations, the sum rule  Eq.~(\ref{eq:SONvectorialsumrule}) is unaffected.  We must simply supplement it with
\be
\label{eq:SO4sumrule}
\sum_{A_+^-}\l_\cO^2 F_{\De,\ell} - \sum_{A_-^-}\l_\cO^2 F_{\De,\ell} &=& 0.
\ee

Before we proceed, it is worth mentioning that all three of the sum rules given in Eq.~(\ref{eq:SONvectorialsumrule}) can be derived from a single `master' crossing relation
\be\label{eq:master}
\sum_{S^+} g_{\De,\ell}(u,v)  -\frac{2}{N} \sum_{T^+} g_{\De,\ell}(u,v) = \left(\frac{u}{v}\right)^d \left( \sum_{T^+} g_{\De,\ell}(v,u) + \sum_{A^-} g_{\De,\ell}(v,u)\right) .
\ee
Adding Eq.~(\ref{eq:master}) to itself with $u \leftrightarrow v$ gives the second row of Eq.~(\ref{eq:SONvectorialsumrule}) and subtracting it from itself with $u \leftrightarrow v$ gives the third row.  To obtain the first row, we must make repeated use of the identity $g(u,v) = (-1)^l g(u/v,1/v)$:
\be\label{eq:ident}
\sum_{T^+} g_{\De,\ell}(u,v) - \sum_{A^-} g_{\De,\ell}(u,v) &=& \sum_{T^+} g_{\De,\ell}(u/v,1/v)+\sum_{A^-} g_{\De,\ell}(u/v,1/v) \nonumber\\
&=& u^d \left(\sum_{S^+} g_{\De,\ell}(1/v,u/v) - \frac{2}{N} \sum_{T^+} g_{\De,\ell}(1/v,u/v)\right) \nonumber\\
&=& u^d\left(\sum_{S^+} g_{\De,\ell}(1/u,v/u) - \frac{2}{N} \sum_{T^+} g_{\De,\ell}(1/u,v/u)\right) \nonumber\\
&=& \left(\frac{u}{v}\right)^d \left(\sum_{T^+} g_{\De,\ell}(v/u,1/u)+\sum_{A^-} g_{\De,\ell}(v/u,1/u)\right) \nonumber\\
&=& \left(\frac{u}{v}\right)^d \left(\sum_{T^+} g_{\De,\ell}(v,u) - \sum_{A^-} g_{\De,\ell}(v,u)\right) .
\ee
This implies in particular that the first sum rule is not independent from the other two.  However, in practice we find it useful to retain all three sum rules, since we will keep only a finite number of terms in their Taylor expansions around a single point in $(u,v)$-space.  Since the derivation Eq.~(\ref{eq:ident}) requires transformation between different $(u,v)$ points, the exact equivalence between the third sum rule and the other two is only visible with an infinite number of terms in the Taylor expansion.  However, it will be important to clarify the meaning of this `master' sum rule (and its generalization to other symmetries) in future studies.

\subsubsection{$\SU(N)$ Crossing Relations}

Let us now consider a complex scalar $\f_i$ transforming in the fundamental representation of an $\SU(N)$ global symmetry.  For this paper, we will only analyze four-point functions $\<\f_i\f^{\bar\jmath\dag}\f_k\f^{\bar l\dag}\>$ that would be invariant under an additional $U(1)$ acting on $\f$.  Note that this is not tantamount to assuming such a $U(1)$ exists --- rather, we are restricting our attention to a subset of CFT correlators.  The various channels for decomposing our four-point function now involve two different kinds of OPEs.  Firstly,
\be
\label{eq:SUNOPE1}
\f_i \x \f^{\bar\jmath\dag} &\sim& \sum_{S^\pm} \de_i^{\bar\jmath}\cO + \sum_{\Ad^{\pm}} \cO_i^{\bar\jmath}
\ee
which can contain $\SU(N)$ singlets and adjoints of any spin.  We also have
\be
\label{eq:SUNOPE2}
\f_i \x \f_j &\sim& \sum_{T^+}\cO_{(ij)}+\sum_{A^-}\cO_{[ij]}
\ee
containing symmetric and antisymmetric tensors with even and odd spins, respectively, and its complex conjugate $\f^{\bar\imath\dag}\x\f^{\bar\jmath\dag}$ containing the conjugate operators in dual representations.  Extracting the coefficients of different tensor structures in all possible ways of evaluating $\<\f_i\f^{\bar\jmath\dag}\f_k\f^{\bar l\dag}\>$ leads to the six-fold sum rule
\be
\label{eq:SUNsumrule}
\sum_{S^\pm}\l_\cO^2 V^{S^\pm}_{\De,\ell}+\sum_{\Ad^\pm}\l_\cO^2 V^{\Ad^\pm}_{\De,\ell}+\sum_{T^+}\l_\cO^2 V^{T^+}_{\De,\ell}+\sum_{A^-}\l_\cO^2 V^{A^-}_{\De,\ell} &=& 0,
\ee
where
\be
V^{S^\pm} = \vecsix{
F}{
H}{
(-)^\ell F}{
(-)^\ell H}{
0}{
0}
,\ \ 
V^{\Ad^\pm}= \vecsix{
(1-\frac 1 N)F}{
-(1+\frac 1 N)H}{
-(-)^{\ell}\frac 1 N F}{
-(-)^{\ell}\frac 1 N H}{
(-)^{\ell}F}{
(-)^{\ell}H}
,\ \ 
V^{T^+}=\vecsix{
0}{
0}{
F}{
-H}{
F}{
-H}
,\ \ 
V^{A^-}=\vecsix{
0}{
0}{
F}{
-H}{
-F}{
H} .
\ee
Once again, the unit operator is included among even-spin singlets $S^+$.

\subsection{Crossing Relations in Superconformal Theories}
\label{sec:crossrelSCFTs}

The 4D $\cN=1$ superconformal algebra extends the conformal algebra to include supersymmetry generators $Q_\a,\bar Q_{\dot\a}$, superconformal generators $S_\a,\bar S_{\dot\a}$, and a $U(1)$ $R$-charge generator.
SCFT operators admit a more refined classification into {\it superconformal primaries} satisfying $S\cO(0)=\bar S\cO(0)=0$, with their superconformal descendants obtained by acting with any combination of $Q,\bar Q,$ and $P$.  It's easy to see using $\{S,\bar S\}\sim K$ that a superconformal primary is also a conformal primary.  But the converse is not necessarily true.  A multiplet built from a single superconformal primary generally contains several (though finitely many) conformal primaries whose dimensions, spins, and OPE coefficients are related by supersymmetry.

A principle example for this work is a chiral superconformal primary scalar $\Phi$ of dimension $d$, which satisfies $S\Phi(0)=\bar S\Phi(0)=\bar Q\Phi(0)$.  Unitarity implies that its dimension is proportional to its $R$-charge, $d=\frac 3 2 R_\Phi$.  Below we will review the structure of the OPEs needed to decompose four-point functions of $\Phi$ and $\Phi^\dag$ into conformal blocks.  We also refer the reader to~\cite{Poland:2010wg,Vichi:2011ux,Fortin:2011nq} for additional discussions of these OPEs.

First, superconformal primaries $\cO$ appearing in $\Phi\x\Phi^\dag$ are restricted to have vanishing $R$-charge and a dimension satisfying the unitarity bound $\De\geq 2+\ell$, where $\ell\geq 0$ is the spin of $\cO$.  Each superconformal primary generically comes with three superconformal descendants of definite spin which are also primaries under the conformal subalgebra.  (When the unitarity bound is saturated, $\De=2+\ell$, two of these descendants vanish and the multiplet is shortened.)  Schematically, the OPE takes the form
\be
\label{eq:phiphidaggersuperOPE}
\Phi\x\Phi^\dag &\sim& \sum_{\cO\in \Phi\x\Phi^\dag}\left[ \cO+(Q\bar Q\cO)_{\ell-1} + (Q\bar Q\cO)_{\ell+1} + Q^2 \bar Q^2 \cO\right],
\ee
where $\cO\in \Phi\x\Phi^\dag$ denotes that the sum is over {\it super}conformal primaries in $\Phi\x\Phi^\dag$, and the subscript on $Q\bar Q\cO$ indicates the spin.  We are being somewhat sketchy in our notation; the exact form of these conformal primaries depends on $\De$ and $\ell$ and is given in~\cite{Poland:2010wg}.  Superconformal symmetry imposes the following relations between their OPE coefficients,\footnote{The difference in normalization from the formulae in~\cite{Poland:2010wg} is due to our different convention for conformal blocks Eq.~(\ref{eq:conformalblocks}).}
\be
\label{eq:superconformalrelations}
\l_{(Q\bar Q\cO)_{\ell+1}}^2 &=& \frac{(\De+\ell)}{4(\De+\ell+1)}\l_\cO^2,\\
\l_{(Q\bar Q\cO)_{\ell-1}}^2 &=& \frac{(\De-\ell-2)}{4(\De-\ell-1)}\l_\cO^2,\\
\l_{Q^2 \bar Q^2 \cO}^2 &=& \frac{(\De+\ell)(\De-\ell-2)}{16(\De+\ell+1)(\De-\ell-1)}\l_\cO^2.
\ee
Note that $\l_{(Q\bar Q\cO)_{\ell-1}}^2$ and $\l_{Q^2 \bar Q^2 \cO}^2$ vanish when $\De=\ell+2$, consistent with shortening of the superconformal multiplet.

Meanwhile, the $\Phi\x\Phi$ OPE can only contain operators which are killed by $\bar Q$.  First and foremost, we have the chiral primary $\Phi^2$, whose dimension is exactly $2d$, by virtue of the relation between dimension and $R$-charge for chiral operators.  All other operators are $\bar Q$-descendants.  Schematically,
\be
\label{eq:phiphisuperOPE}
\Phi\x\Phi &\sim& \Phi^2 + \sum_{\ell=2,4,\dots}\bar Q \cO_\ell + \sum_{\cO} \bar Q^2 \cO.
\ee
The operators $\cO_\ell$ transform in $(\frac \ell 2,\frac {\ell-1} 2)$ representations of the Lorentz group $\SO(4)\cong \SU(2)\x\SU(2)$, and satisfy the BPS shortening condition $\bar Q_{\dot\a}\cO^{\dot\a\dot\a_3\dots\dot\a_\ell,\a_1\dots\a_\ell}_\ell = 0$.  The product $\bar Q\cO_\ell$ is then a spin-$\ell$ operator, which is required by the superconformal algebra to have dimension $2d+\ell$.  Finally, the remaining operators $\bar Q^2\cO$ are not protected by a BPS condition, and can have any dimension satisfying $\De\geq |2d-3|+3+\ell$.  Note that when $d<3/2$, a gap in dimensions exists between the protected operators $\Phi^2,\bar Q\cO_\ell$ and the non-protected operators $\bar Q^2\cO$.  In contrast to the situation for $\Phi\x\Phi^\dag$, each conformal primary in Eq.~(\ref{eq:phiphisuperOPE}) appears with an independent coefficient --- there are no additional relations imposed by supersymmetry among operators in $\Phi\x\Phi$.

Because of the $U(1)_R$ symmetry, crossing symmetry of the four-point function $\<\Phi\Phi^\dag\Phi\Phi^\dag\>$ is a special case of crossing symmetry for $\SO(2)$.  Note that the antisymmetric tensor representation of $\SO(2)$ is the trivial representation, so that we may equivalently write $S^-$ (odd-spin singlets) for $A^-$ (odd-spin antisymmetric tensors).  We are also free to multiply the sum rule by any invertible matrix without changing its content.\footnote{Specifically, we will replace the middle row with itself plus twice the top row, and the top row with the middle row.}  Consequently, we can rewrite Eq.~(\ref{eq:SONvectorialsumrule}) for $\SO(2)$ as
\be
\label{eq:SO2vectorialsumrule}
\sum_{S^\pm}\l_\cO^2 \p{
\begin{array}{c}
F_{\De,\ell}\\
(-)^\ell F_{\De,\ell} \\
(-)^\ell H_{\De,\ell}
\end{array}
}
+
\sum_{T^+} 2\l_\cO^2 \p{
\begin{array}{c}
0\\
F_{\De,\ell}\\
-H_{\De,\ell}
\end{array}
}
&=& 0.
\ee
In our superconformal four-point function $\<\Phi\Phi^\dag\Phi\Phi^\dag\>$, the $S^\pm$ terms will come from the OPE (\ref{eq:phiphidaggersuperOPE}), while the $T^+$ terms come from (\ref{eq:phiphisuperOPE}).  Making use of the relations between OPE coefficients Eq.~(\ref{eq:superconformalrelations}), the specialization of Eq.~(\ref{eq:SO2vectorialsumrule}) to the superconformal case is
\be
\label{eq:SCFTcrossing}
\sum_{S^\pm} \l_\cO^2 \p{
\begin{array}{c}
\cF_{\De,\ell}\\
\tl\cF_{\De,\ell} \\
\tl \cH_{\De,\ell}
\end{array}
}
+
\sum_{T^+_\textrm{BPS}} \l_{\cO}^2 \p{
\begin{array}{c}
0\\
F_{2d+\ell,\ell}\\
-H_{2d+\ell,\ell}
\end{array}
}
+
\sum_{T^+_\textrm{non-BPS}} \l_{\cO}^2 \p{
\begin{array}{c}
0\\
F_{\De,\ell}\\
-H_{\De,\ell}
\end{array}
}
&=& 0,
\ee
where
\be
\cF_{\De,\ell} &\equiv& F_{\De,\ell}+ \frac{(\De+\ell)}{4(\De+\ell+1)} F_{\De+1,\ell+1}
+\frac{(\De-\ell-2)}{4(\De-\ell-1)}F_{\De+1,\ell-1}\nn\\
&&+\frac{(\De+\ell)(\De-\ell-2)}{16(\De+\ell+1)(\De-\ell-1)}F_{\De+2,\ell}.
\ee
In addition, $\tl \cF$ is $\cF$ with odd spins flipped $F_{\De,\ell}\to(-)^\ell F_{\De,\ell}$ throughout, and $\tl \cH$ is $\tl \cF$ with $F_{\De,\ell}\to H_{\De,\ell}$.  The set $T^+_\mathrm{BPS}$ consists of the BPS operators appearing in $\Phi\x\Phi$, namely $\Phi^2$ and $\bar Q \cO_\ell$ for $\ell\in\{2,4,\dots\}$.  $T^+_\mathrm{non-BPS}$ consists of the remaining operators in $\Phi\x\Phi$.  In going from Eq.~(\ref{eq:SO2vectorialsumrule}) to Eq.~(\ref{eq:SCFTcrossing}), we have removed the factors of $2$ in front of the symmetric tensor contributions because their conventional normalization differs between $\SO(2)$ and $U(1)$, $[\l^2_{T^+}]_{U(1)}=2[\l^2_{T^+}]_{\SO(2)}$.  

\subsubsection{Superconformal $\SU(N)$ Crossing Relations}
It is straightforward to generalize this analysis to the case of a scalar superconformal primary $\Phi_i$ transforming as a fundamental under an $\SU(N)$ global symmetry.  The index structure of the OPE is the same as is given in Eqs.~(\ref{eq:SUNOPE1}) and~(\ref{eq:SUNOPE2}), but with the additional constraints imposed by supersymmetry discussed above.  Note that now both BPS and non-BPS odd-spin operators can appear in the $\Phi_i \times \Phi_j$ OPE as $\SU(N)$ antisymmetric tensors. Including these constraints, the six-fold sum rule of Eq.~(\ref{eq:SUNsumrule}) becomes
\be
\label{eq:SUSYSUNsumrule}
\sum_{S^\pm}\l_\cO^2 \cV^{S^\pm}_{\De,\ell}+\sum_{\Ad^\pm}\l_\cO^2 \cV^{\Ad^\pm}_{\De,\ell}+\sum_{T^+_\textrm{BPS}}\l_\cO^2 V^{T^+}_{2d+\ell,\ell}+\sum_{A^-_\textrm{BPS}}\l_\cO^2 V^{A^-}_{2d+\ell,\ell} && \nonumber \\
+\sum_{T^+_\textrm{non-BPS}}\l_\cO^2 V^{T^+}_{\De,\ell}+\sum_{A^-_\textrm{non-BPS}}\l_\cO^2 V^{A^-}_{\De,\ell} &=& 0,
\ee
where
\be
\cV^{S^\pm} = \vecsix{
\cF}{
\cH}{
\tl \cF}{
\tl \cH}{
0}{
0}
,\ \ 
\cV^{\Ad^\pm}= \vecsix{
(1-\frac 1 N) \cF}{
-(1+\frac 1 N) \cH}{
-\frac 1 N \tl \cF}{
-\frac 1 N \tl \cH}{
\tl \cF}{
\tl \cH}
,\ \ 
V^{T^+}=\vecsix{
0}{
0}{
F}{
-H}{
F}{
-H}
,\ \ 
V^{A^-}=\vecsix{
0}{
0}{
F}{
-H}{
-F}{
H} .
\ee

\subsection{Bounds from Crossing Relations}
\label{sec:boundsfromcrossing}

Crossing symmetry of four-point functions encodes an infinite number of relations between OPE coefficients --- one for each value of the conformal cross-ratios $u$ and $v$.  In~\cite{Rattazzi:2008pe} a general method was outlined for extracting bounds on CFT data using these relations, together with the constraints of unitarity.  We will now review this method for the simplest case of a real scalar $\f$ of dimension $d$.  Subsequently, we will discuss how the original method can be improved using semidefinite programming.

Suppose we would like to bound the OPE coefficient of a particular operator $\cO_0$ of dimension $\De_0$ and spin $\ell_0$ appearing in $\f\x\f$.  The first step is to isolate $\l_{\cO_0}^2$ on one side of the sum rule Eq.~(\ref{eq:sumrule}),
\be
\l_{\cO_0}^2 F_{\De_0,\ell_0}(u,v) &=& -F_{0,0}(u,v)-\sum_{\cO\neq\cO_0} \l_\cO^2 F_{\De,\ell}(u,v).
\label{eq:rewrittencrossingrelation}
\ee
We can obtain different expressions for $\l_{\cO_0}^2$ in terms of the other OPE coefficients by evaluating Eq.~(\ref{eq:rewrittencrossingrelation}) at different values of $u$ and $v$.  We could also take some number of $u$- and $v$-derivatives first, and then evaluate. And in general, we can apply any linear functional $\a$ to both sides,
\be
\label{eq:linfunctionalapplied}
\l_{\cO_0}^2\a(F_{\De_0,\ell_0}) &=& -\a(F_{0,0})-\sum_{\cO\neq \cO_0} \l_{\cO}^2 \a(F_{\De,\ell}).
\ee
A key insight of~\cite{Rattazzi:2008pe} is that the functions $F_{\De,\ell}$ share certain positivity properties, so that it's sometimes possible to find a linear functional $\a$ such that
\be
\label{eq:alphaconstraint1}
\a(F_{\De_0,\ell_0}) &=& 1,\qquad\textrm{and}\\
\label{eq:alphaconstraint2}
\a(F_{\De,\ell}) &\geq& 0,\qquad\textrm{for all other (non-unit) operators in the spectrum.}
\ee
Eq.~(\ref{eq:alphaconstraint1}) is simply a normalization condition, but to satisfy Eq.~(\ref{eq:alphaconstraint2}) one must choose $\a$ carefully.  If $\a$ satisfies these constraints, then since the $\l_\cO^2$ are positive by unitarity, Eq.~(\ref{eq:linfunctionalapplied}) becomes an upper bound on $\l_{\cO_0}^2$,
\be
\label{eq:upperboundonlambda}
\l_{\cO_0}^2 =-\a(F_{0,0})-\sum_{\cO\neq \cO_0}\textrm{pos.}\x\textrm{pos.} \leq -\a(F_{0,0}).
\ee
The space of viable $\a$'s depends on precisely what assumptions one makes about the spectrum of the CFT.  If one makes an assumption about the spectrum of operator dimensions that makes it easier to satisfy Eq.~(\ref{eq:alphaconstraint2}) (e.g., all scalars have a dimension greater than some $\De_{\textrm{min}}$) and then finds a linear functional $\alpha$ such that the bound of Eq.~(\ref{eq:upperboundonlambda}) violates the unitarity constraint $\l_{\cO_0}^2 \geq 0$, one can rule out that assumption about the spectrum.  

Now, to make the bound~(\ref{eq:upperboundonlambda}) as strong as possible, we should minimize $-\a(F_{0,0})$ over the set $\cS$ of all $\a$ satisfying the constraints~(\ref{eq:alphaconstraint1}, \ref{eq:alphaconstraint2}).  These constraints carve out a convex subset of the space of linear functionals, so the task of determining the best $\a$ is an infinite-dimensional {\it convex optimization problem}.  It would be extremely interesting to develop analytical techniques for finding solutions.  However, the most successful approaches to date, including the one we present here, involve simplifying the problem to make it tractable on a computer, and then determining solutions numerically. 

Putting our optimization problem on a computer requires surmounting two difficulties:

\begin{enumerate}
\item The search space $\cS$ of $\a$'s satisfying Eqs.~(\ref{eq:alphaconstraint1}, \ref{eq:alphaconstraint2}) is infinite dimensional.
\item The number of constraints $\a(F_{\De,\ell})\geq 0$ is infinite --- there's one for each $\De,\ell$.
\end{enumerate}

The first difficulty is easy enough to address: we can restrict to a finite-dimensional subspace $\cW$ of linear functionals.  Then, minimizing $-\a(F_{0,0})$ over all $\a\in \cW\cap \cS$ will give a possibly sub-optimal, but still valid bound $\l_{\cO_0}^2\leq -\a(F_{0,0})$.  The choice of $\cW$ is somewhat arbitrary, and it would be interesting to explore a wider variety of functionals than we do here.  Following~\cite{Rattazzi:2008pe,Rychkov:2009ij,Caracciolo:2009bx,Poland:2010wg,Rattazzi:2010gj,Rattazzi:2010yc,Vichi:2011ux}, we will simply take linear combinations of derivatives around the symmetric point $z=\bar z=1/2$.  That is, we define $\cW_k$ to be the space of functionals
\ben
\label{eq:Wksuspaces}
\a : F(z,\bar{z}) &\mto& \sum_{m+n\leq 2k} a_{mn}\ptl_z^m\ptl_{\bar z}^nF(1/2,1/2),
\een
with real coefficients $a_{mn}$.  This choice is computationally convenient, and will prove useful in our solution to the second difficulty in a moment. One hopes that as we increase $k$ to include more and more derivatives, our search will cover more and more of $\cS$, and our bound will converge to the optimal one.

\begin{figure}[h!]
\begin{center}
\begin{psfrags}
\psfrag{s}[B][B][1][0]{$\mathcal{S}$}
\psfrag{e}[l][B][1][0]{$\a(F_{\De_0,\ell_0})=1$}
\psfrag{b}[l][B][1][0]{$\a(F_{\De_1,\ell_1})\geq0$}
\psfrag{a}[l][B][1][0]{$\a(F_{\De_2,\ell_2})\geq0$}
\psfrag{d}[l][B][1][0]{$\a(F_{\De,\ell_1})\geq0$}
\psfrag{c}[l][B][1][0]{$\a(F_{\De,\ell_2})\geq0$}
\psfrag{r}[l][B][1][0]{$\dots$}
\psfrag{V}[B][B][1.2][0]{$\cW$}
\psfrag{F}[B][B][1][0]{$\a(F_{\De,\ell})\geq0$}
\includegraphics[width=150mm]{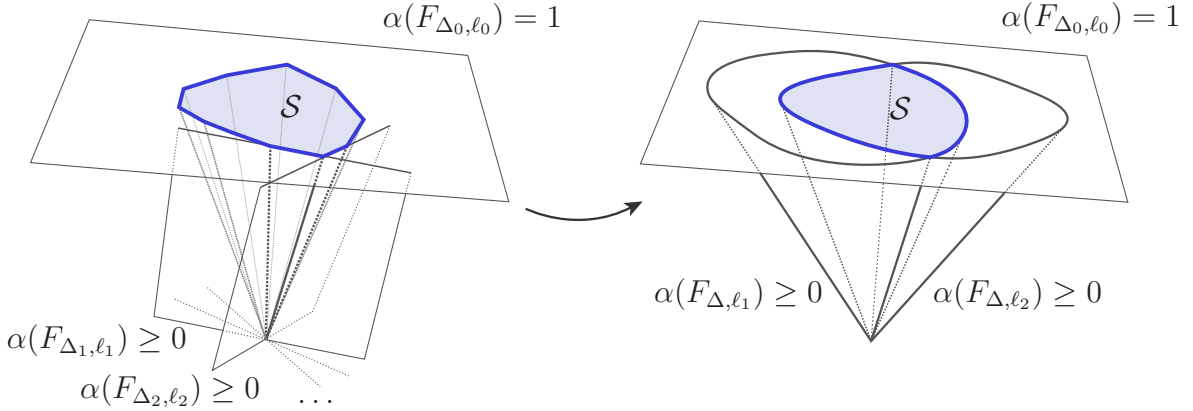}
\end{psfrags}
\end{center}
\caption{The `search space' $\cS$ (shown in blue) is the intersection of the hyperplane $\a(F_{\De_0,\ell_0})=1$ with the convex cone of linear functionals $\a$ satisfying $\a(F_{\De,\ell})\geq 0$ for all $\{\De,\ell\}$ in the spectrum.  Previous methods discretized $\De$ to some finite set $\{\De_i\}$, thus approximating $\cS$ as an intersection of a finite number of hyperplanes and half-spaces (left).  Our approach is to approximate $\cS$ as the intersection of a smaller number of curved spaces --- specifically cones of semidefinite matrices (right).  Such intersections are sometimes called {\it spectrahedra}.}
\label{fig:searchspace}
\end{figure}

The second difficulty is more problematic.  Since angular momentum $\ell$ is discrete, it's reasonable to include constraints with $\ell=0,1,\dots,L$, for some large $L$.  But the dimension $\De$ can vary continuously, and the  constraints $\a(F_{\De,\ell})\geq 0$ carve out a complicated shape $\cS$ inside $\cW$ as $\De$ varies.  The computer has to know about this shape, which means we must encode it with some finite amount of data.  The approach used in~\cite{Rattazzi:2008pe,Rychkov:2009ij,Caracciolo:2009bx,Poland:2010wg,Rattazzi:2010gj,Rattazzi:2010yc,Vichi:2011ux} is to approximate the shape by a convex polytope --- namely discretize $\De$ to lie in some finite set $\{\De_i\}$, so that the constraints $\a(F_{\De_i,\ell})\geq 0$ become a finite number of linear inequalities for $\a$.  Then the problem of minimizing $-\a(F_{0,0})$ becomes a {\it linear programming} problem, which can be solved by jumping from vertex to vertex on the boundary of the polytope, following the direction of steepest descent.  As one makes the set $\{\De_i\}$ larger, the approximation of $\cS$ as a polytope gets more and more refined, and the solution should converge to the correct one.

This method can be quite powerful if one chooses the $\{\De_i\}$ carefully.  However, some basic tensions limit how far it can be pushed.  For example, consider increasing $k$ to obtain a stronger bound.  At higher $k$, the space $\cW_k$ can include wilder linear functionals, and one must include more $\De_i$ to ensure that a constraint $\a(F_{\De,l})\geq 0$ isn't violated.  However, the running time of the usual search algorithm is {\it cubic} in the number of constraints, which means that computations become quickly unwieldy.

Our approach in the present paper is to approximate $\cS$ with a different kind of shape that is more efficient to encode than a polytope, one that naturally respects the properties of conformal blocks (specifically the differential equation that they satisfy), and also admits fast searches.  In the process, we will do away with the discretization $\De\in\{\De_i\}$ entirely.

\subsection{Semidefinite Programming}
\label{sec:semidefiniteprogramming}

Semidefinite programs (SDPs) \cite{Vandenberghe:1996} are linear optimization problems that can contain positive-semidefiniteness constraints for matrices, along with the usual linear inequalities included in linear programs.  As we'll see momentarily, positive-semidefiniteness lets us express the condition that a collection of polynomials be nonnegative for all values of their arguments.  This is useful for us because there is a systematic approximation for the derivatives of $F_{\De,l}$ in terms of polynomials. Specifically, there exist positive functions $\chi_\ell(\De)$ and polynomials $P^{m,n}_\ell(\De)$ such that
\be
\label{eq:polapprox}
\ptl_z^m\ptl_{\bar z}^n F_{\De,\ell}(1/2,1/2) &\aeq& \chi_\ell(\De)P^{mn}_\ell(\De),
\ee
where the approximation can be made arbitrarily good, at the cost of increasing the degree of $P^{mn}_\ell$.  The details of this approximation, which follows from the differential equation for conformal blocks along with some basic facts about hypergeometric functions, are explained in appendix~\ref{app:polynomialapproximations}.

For now, let us assume that such an approximation exists, and understand how to phrase our problem as an SDP.  We will write $F^{mn}_\ell(\De)\equiv \ptl_z^m\ptl_{\bar z}^n F_{\De,\ell}(1/2,1/2)$ for brevity.  Once again, we would like to minimize $-a_{mn} F^{mn}_{0}(0)$ subject to the constraints
\be
a_{mn} F^{mn}_{\ell_0}(\De_0) &=& 1,\\
a_{mn} F^{mn}_{\ell}(\De) &\geq& 0 \qquad \textrm{for $\De\geq \De_\ell$, for all $0\leq \ell\leq L$},
\label{eq:inequalityconstraint}
\ee
where $\De_\ell$ is a lower bound on $\De$ depending on the spin $\ell$.

Using Eq.~(\ref{eq:polapprox}) along with the fact that $\chi_\ell(\De)$ is positive, Eq.~(\ref{eq:inequalityconstraint}) becomes  the statement that each polynomial $a_{mn} P^{mn}_\ell(\De_\ell(1+x))$ is nonnegative on the interval $x\in[0,\oo)$.  Such statements are naturally written in terms of positive-semidefinite matrices, a fact which is well-known in the optimization literature and has been exploited to solve a wide variety of problems (see, e.g.,~\cite{Parillo:2000}).  The rewriting proceeds as follows.  Firstly, a theorem due to Hilbert \cite{Hilbert:1888} states that a polynomial $p(x)$ is nonnegative on $[0,\oo)$ if and only if
\be
p(x) &=& f(x) + x g(x),
\ee
where both $f(x)$ and $g(x)$ are sums of squares of polynomials.  Now suppose $f$ and $g$ have degrees $2d$ and $2d'$ respectively, and let $[x]_d$ denote the vector with entries $(1,x,\dots,x^d)$.  If $f(x)$ is a sum of squares of polynomials with coefficients $\bc_i=(c_{i0},\dots,c_{id})$, then we have
\be
f(x) &=& \sum_i (\bc_i^T[x]_d)^2 \ \ =\ \ [x]_d^T \p{\sum_i \bc_i \bc_i^T} [x]_d\ \ =\ \ [x]_d^T A [x]_d,
\ee
where $A\equiv \sum_i \bc_i \bc_i^T$ is positive-semidefinite.  Conversely, any positive-semidefinite matrix $A$ admits a Cholesky decomposition $A=\sum_i \bc_i \bc_i^T$, so that $[x]_d^T A[x]_d$ is a sum of squares.  Thus, the condition that $p(x)$ be nonnegative on $[0,\oo)$ can be written
\be
p(x) &=& [x]_{d}^TA[x]_{d}+x([x]_{d'}^T B[x]_{d'}),\qquad \textrm{with}\qquad A,B \succeq 0,
\ee
where the $\succeq$ symbol means `positive-semidefinite.' 

Returning to OPE bounds, we now have the following presentation of our convex optimization problem as an SDP: minimize $-a_{mn} F^{mn}_0(0)$, subject to the constraints
\begin{align}
a_{mn} F_{\ell_0}^{mn}(\De_0) &= 1,\\
a_{mn} P^{mn}_\ell(\De_\ell(1+x)) &= [x]_{d_\ell}^TA_\ell[x]_{d_\ell}+x([x]_{d'_\ell}^T B_\ell[x]_{d'_\ell})&\textrm{for $0\leq \ell\leq L$},\\
A_\ell,B_\ell &\succeq 0 & \textrm{for $0\leq \ell\leq L$}.
\end{align}

There are numerous advantages to this formulation.  Firstly, we avoid discretizing the set of operator dimensions $\De$, and thus evade the trade-off between refining $\{\De_i\}$ and improving the running time.  Further, small and large $\De$ are accounted for equally well, so there is no need for separate checks on the asymptotic behavior of $\a(F_{\De,\ell})$ at large dimensions.  Most importantly, there exist efficient algorithms for solving semidefinite programs using interior point methods, with some excellent implementations (see appendix~\ref{app:SDP}).  Their complexity scales much less sharply with the dimension of the search space than the linear programming algorithms used in~\cite{Rattazzi:2008pe,Rychkov:2009ij,Caracciolo:2009bx,Poland:2010wg,Rattazzi:2010gj,Rattazzi:2010yc,Vichi:2011ux}.  Consequently, we have been able to push the previous state-of-the-art searches from 55 dimensions to almost 400 dimensions in some cases.

\subsection{Generalizations for Global Symmetries}
\label{sec:optimizationgeneralization}

While we specialized the above discussion to the case of the singlet sum rule Eq.~(\ref{eq:sumrule}), it is straightforward to modify it for situations with global symmetries.  E.g., if we wish to place a bound on the OPE coefficient of an $S^+$ operator appearing in the $\SO(N)$ sum rule of Eq.~(\ref{eq:SONvectorialsumrule}), we should look for a vectorial linear functional $\alpha$ satisfying
\be
\label{eq:SONalphaconstraint1}
\a \p{
\begin{array}{c}
0\\
 F_{\De_0,\ell_0}\\
H_{\De_0,\ell_0}
\end{array}
} &=& 1,\qquad\textrm{}\\
\label{eq:SONalphaconstraint2}
\a  \p{
\begin{array}{c}
0\\
 F_{\De,\ell}\\
 H_{\De,\ell}
\end{array}
} & \geq & 0,\qquad\textrm{for all other (non-unit) operators in $S^+$,}\\
\label{eq:SONalphaconstraint3}
\a  \p{
\begin{array}{c}
F_{\De,\ell}\\
 \left(1-\frac{2}{N}\right) F_{\De,\ell}\\
- \left(1+ \frac{2}{N}\right) H_{\De,\ell}
\end{array}
}& \geq & 0,\qquad\textrm{for all operators in $T^+$, and}\\
\label{eq:SONalphaconstraint4}
\a  \p{
\begin{array}{c}
-F_{\De,\ell}\\
 F_{\De,\ell}\\
- H_{\De,\ell}
\end{array}
}& \geq & 0,\qquad\textrm{for all operators in $A^-$} .
\ee
Any such linear functional then leads to the upper bound
\be
\l_{\cO_0}^2 &\leq& -\a \p{
\begin{array}{c}
0\\
F_{0,0} \\
H_{0,0}
\end{array}
} .
\ee
The modification for alternatively placing bounds on the OPE coefficients of $T^+$ or $A^-$ operators should be clear.  As in the singlet case, we can also rule out an assumption about the spectrum of operator dimensions by making the assumption and then finding a linear functional that leads to a violation of the unitarity constraint $\l_{\cO_0}^2 \geq 0$.

Similarly, we can bound the OPE coefficient of an $S^\pm$ operator appearing in the $\SU(N)$ sum rule of Eq.~(\ref{eq:SUNsumrule}) by finding an $\alpha$ satisfying
\be
\label{eq:SUNalphaconstraint1}
\a \left( V^{S^\pm}_{\De_0,\ell_0} \right) &=& 1,\qquad\textrm{}\\
\label{eq:SUNalphaconstraint2}
\a \left( V^I_{\De,\ell}  \right) & \geq & 0,\qquad\textrm{for all other (non-unit) operators in the spectrum,}
\ee
where $I=\{S^{\pm},\Ad^\pm,T^+,A^-\}$.  This leads to the upper bound
\be
\l_{\cO_0}^2 \leq - \a \left( V^{S^+}_{0,0} \right) .
\ee
The appropriate generalization of this logic for placing bounds on operators in other $\SU(N)$ representations, and also for obtaining bounds using the superconformal sum rules given in Eqs.~(\ref{eq:SCFTcrossing}) and~(\ref{eq:SUSYSUNsumrule}), should be clear.  

In all of these situations, the task of numerically finding the optimal $\alpha$ can be recast in terms of a semidefinite program.  Similar to what we described in the previous section, to do this we use the fact that derivatives of any of the functions $\{F_{\De,\ell}, H_{\De,\ell}, \cF_{\De,\ell}, \cH_{\De,\ell},\tl\cF_{\De,\ell}, \tl\cH_{\De,\ell}\}$ at $(1/2,1/2)$ can be arbitrarily-well approximated by positive functions times polynomials in $\De$.  The details of these approximations can be found in appendix~\ref{app:polynomialapproximations}.

\subsection{Coincidence Between $\SU(N)$ and $\SO(2N)$ Singlet Bounds}
\label{sec:SONSUNcoincidence}

In the course of running the above algorithm, we found that our bounds on singlet operators appearing in an OPE between $\SU(N)$ fundamentals were numerically identical to bounds on singlets appearing in an OPE between $\SO(2N)$ fundamentals.  This exact coincidence is surprising given the rather different structure of the crossing symmetry constraints.  It hadn't been previously observed because $\SU(N)$ computations were too difficult to perform with previous techniques.  In this section, we'll discuss the relations between those bounds in more detail.

Let us consider more generally a CFT with global symmetry group $\mathcal G$. Suppose we want to obtain a dimension bound on a singlet scalar operator entering a given OPE. The $\mathcal G$ crossing symmetry constraints produce a bound $\Delta_{\mathcal G}(d)$. Now consider a subgroup $\mathcal H\subset \mathcal G$ and repeat the procedure. This time, the $\mathcal H$ crossing symmetry constraints will produce a bound $\Delta_{\mathcal H}(d)$. At this point we must distinguish two cases: 1) all $\mathcal H$-singlets are also $\mathcal G$-singlets, 2) some nontrivial representation of $\mathcal G$, once decomposed with respect to the subgroup, contains $\mathcal H$-singlets. In the first case we can immediately conclude
\be\label{GHrelation}
	\Delta_{\mathcal G}(d) \leq \Delta_{\mathcal H}(d)	\qquad \text{($\mathcal G$-bound stronger)}.
\ee
The above inequality is clear: there are no CFT's with global symmetry $\mathcal H$ where the first scalar singlet operator entering a given OPE has dimension larger than $\Delta_{\mathcal H}(d)$.  Thus in particular there are no CFT's with a larger global symmetry. 

An example of such a group and subgroup is given precisely by $\SU(N)\subset \SO(2N)$. In the decomposition with respect to the subgroup, the only singlets come from $\SO(2N)$-singlets: the symmetric tensor goes to a symmetric tensor and an adjoint while the antisymmetric tensor goes to an antisymmetric tensor and an adjoint. Thus, it is natural to expect the triple sum rule Eq.~(\ref{eq:SONvectorialsumrule}) to give a bound stronger than or equal to the sextuple sum rule Eq.~(\ref{eq:SUNsumrule}).  Indeed, one can verify this explicitly at the level of the optimization problem for $\a$.

To prove the equality of $\SU(N)$ and $\SO(2N)$ bounds one should also show that whenever a linear functional satisfying Eqs.~(\ref{eq:SONalphaconstraint1}-\ref{eq:SONalphaconstraint4}) exists, it is possible to construct a second linear functional satisfying Eq.~(\ref{eq:SUNalphaconstraint1}). Unfortunately, we have not been able to find an analytic proof of this result.  However, we find numerically that it is always possible -- it would be good in future studies to gain a deeper understanding of why this is the case. 

In the case 2) the two bounds are unrelated, since the $\mathcal H$-bound could in principle be determined by representations coming from the decomposition of nontrivial representations of the larger symmetry group.  This is the case for $\SO(N)$ and $\SO(N')$ or $\SU(N)$ and $\SU(N')$, with $N>N'$. In these examples we numerically observe behavior opposite to (\ref{GHrelation}).

\section{Bounds on Operator Dimensions}
\label{sec:dimbounds}

\subsection{General Theories}

\begin{figure}[h!]
\begin{center}
\begin{psfrags}
\def\PFGstripminus-#1{#1}%
\def\PFGshift(#1,#2)#3{\raisebox{#2}[\height][\depth]{\hbox{%
  \ifdim#1<0pt\kern#1 #3\kern\PFGstripminus#1\else\kern#1 #3\kern-#1\fi}}}%
\providecommand{\PFGstyle}{}%
%
\psfrag{d}[cl][cl]{\PFGstyle $d$}%
\psfrag{maxD0}[bc][bc]{\PFGstyle $\De_0$}%
\psfrag{maxdimforr}[bc][bc]{\PFGstyle $\text{Upper bound on $\dim(\f^2)$}$}%
\psfrag{x11}[tc][tc]{\PFGstyle $1$}%
\psfrag{x121}[tc][tc]{\PFGstyle $1.2$}%
\psfrag{x141}[tc][tc]{\PFGstyle $1.4$}%
\psfrag{x161}[tc][tc]{\PFGstyle $1.6$}%
\psfrag{x181}[tc][tc]{\PFGstyle $1.8$}%
\psfrag{y21}[cr][cr]{\PFGstyle $2$}%
\psfrag{y251}[cr][cr]{\PFGstyle $2.5$}%
\psfrag{y31}[cr][cr]{\PFGstyle $3$}%
\psfrag{y351}[cr][cr]{\PFGstyle $3.5$}%
\psfrag{y41}[cr][cr]{\PFGstyle $4$}%
\psfrag{y451}[cr][cr]{\PFGstyle $4.5$}%
\psfrag{y51}[cr][cr]{\PFGstyle $5$}%
\psfrag{y551}[cr][cr]{\PFGstyle $5.5$}%
\includegraphics[width=.9\textwidth]{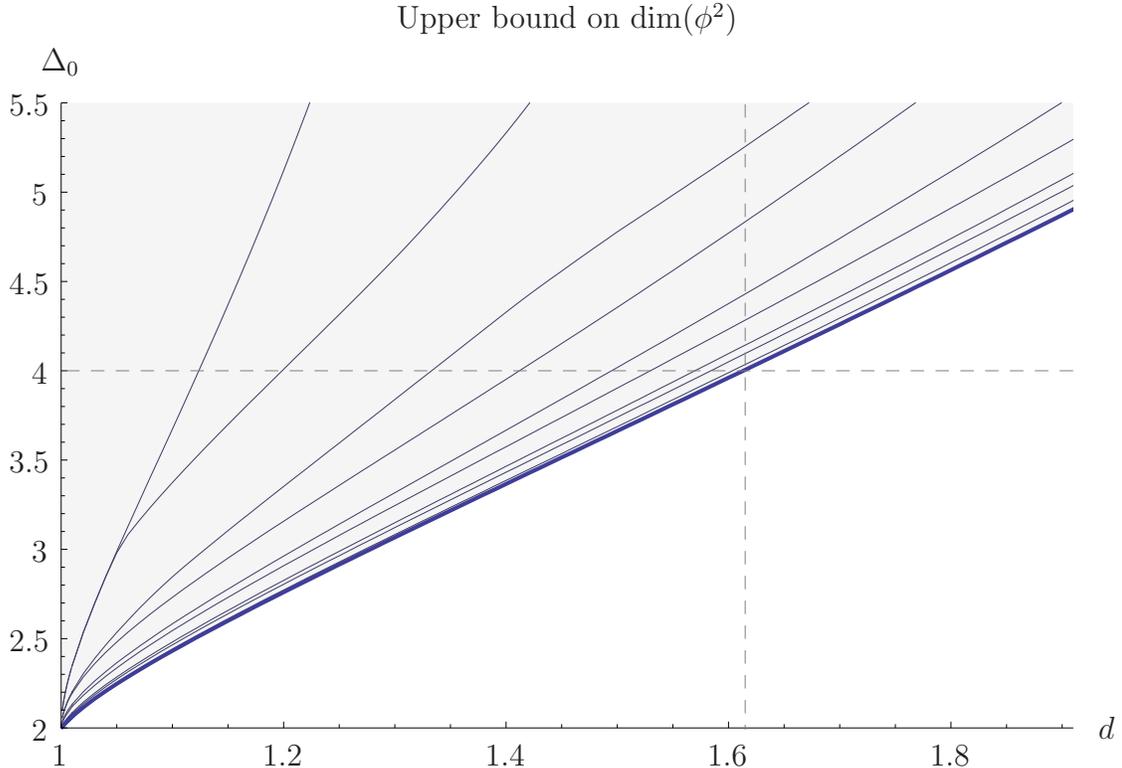}
\end{psfrags}
\end{center}
\caption{An upper bound on the dimension of $\f^2$, the lowest dimension scalar appearing in $\f\x\f$.  Curves for $k=2,\dots,11$ are shown, with the $k=11$ bound being the strongest.}
\label{fig:realscalardim}
\end{figure}

As a first application of our semidefinite programming algorithm, let us reproduce the singlet dimension bound first derived in~\cite{Rattazzi:2008pe}, and later improved in~\cite{Rychkov:2009ij}.  We let $\f$ be a real scalar of dimension $d$ in a general CFT, and seek to place an upper bound on the dimension of $\f^2$, the lowest dimension scalar appearing in $\f\x\f$.  The procedure is precisely as described in section~\ref{sec:boundsfromcrossing}.  In figure~\ref{fig:realscalardim}, we show the resulting bounds for $k=2,\dots,11$, with $k=10$ (a $55$-dimensional search-space) being the previous state-of-the-art.  We find perfect agreement with older linear programming-based calculations for each $k=2,\dots,10$.

The curves appear to converge at large $k$, which is perhaps indicative that they are approaching the best possible bound given our assumptions (referred to as $f_\oo(d)$ in \cite{Rychkov:2009ij}).\footnote{However, since the full optimization problem involves an infinite-dimensional search space, it's always possible a new search direction could open up at higher $k$.  Fully establishing convergence would require more detailed analysis than we do here.}  We will see this kind of convergence in many other plots in this paper.  An approximate fit to the strongest ($k=11$) bound is given by\footnote{While it gives a good description of the shape, we have chosen this functional form somewhat arbitrarily; it is possible that a different basis of functions should be used when describing the optimal bound.}
\be
\dim(\f^2)&\leq& 2+3.006\e+0.160(1-e^{-20\e}),
\ee
where $d=1+\e$, with $\e$ between $0$ and $1$.  Notice that the behavior for both small and large $\e$ is approximately linear. The bound crosses $\dim(\f^2)=4$ around $d\approx 1.61$.

\subsection{Singlet Operators in $\SO(N)$ and $\SU(N)$ Theories}

We can also place bounds on the lowest dimension singlet appearing in $\f_i\x\f_j$, where $\f_i$ transforms as a vector of an $\SO(N)$ global symmetry.  The procedure is as described in section~\ref{sec:optimizationgeneralization}, where we must assume that $\De>\De_\mathrm{min}$ for all scalars in $S^+$, and then scan over $\De_\mathrm{min}$ to obtain a dimension bound.  Recall from section~\ref{sec:SONSUNcoincidence} that our bounds on singlets of $\SU(N)$ turn out to be identical to those for singlets of $\SO(2N)$.  Hence, we will present all $\SU$ and $\SO$ singlet bounds together, with even values of $N$ standing for both $\SO(N)$ and $\SU(N/2)$.

Previous attempts to compute bounds for theories with global symmetries have been somewhat hindered by the need to optimize over very high-dimensional spaces.  Since the vectorial sum rule Eq.~(\ref{eq:SONvectorialsumrule}) has three components, a given $k$ corresponds to
\be
\frac{k(k+1)}{2}\x 3
\ee
different linear functionals.  The linear programming methods implemented so far are essentially limited to a search space dimension that is not much larger than $\sim 50$, or $k\sim 5$ for $\SO(N)$.  Worse, $\SU(N)$ vectorial sum rules have six components, making them even harder to explore.  However, our semidefinite programming algorithm appears to have few problems with large search spaces, and we will present most of our bounds up to $k=11$, regardless of the type of global symmetry group.

\begin{figure}[h!]
\begin{center}
\begin{psfrags}
\def\PFGstripminus-#1{#1}%
\def\PFGshift(#1,#2)#3{\raisebox{#2}[\height][\depth]{\hbox{%
  \ifdim#1<0pt\kern#1 #3\kern\PFGstripminus#1\else\kern#1 #3\kern-#1\fi}}}%
\providecommand{\PFGstyle}{}%
%
\psfrag{d}[cl][cl]{\PFGstyle $d$}%
\psfrag{maxD0}[bc][bc]{\PFGstyle $\De_0$}%
\psfrag{maxdimforS}[bc][bc]{\PFGstyle $\text{Upper bound on $\textrm{dim}(|\f|^2)$ for $\SO(4)$ or $\SU(2)$}$}%
\psfrag{x11}[tc][tc]{\PFGstyle $1$}%
\psfrag{x121}[tc][tc]{\PFGstyle $1.2$}%
\psfrag{x141}[tc][tc]{\PFGstyle $1.4$}%
\psfrag{x161}[tc][tc]{\PFGstyle $1.6$}%
\psfrag{x181}[tc][tc]{\PFGstyle $1.8$}%
\psfrag{y21}[cr][cr]{\PFGstyle $2$}%
\psfrag{y251}[cr][cr]{\PFGstyle $2.5$}%
\psfrag{y31}[cr][cr]{\PFGstyle $3$}%
\psfrag{y351}[cr][cr]{\PFGstyle $3.5$}%
\psfrag{y41}[cr][cr]{\PFGstyle $4$}%
\psfrag{y451}[cr][cr]{\PFGstyle $4.5$}%
\psfrag{y51}[cr][cr]{\PFGstyle $5$}%
\psfrag{y551}[cr][cr]{\PFGstyle $5.5$}%
\includegraphics[width=.9\textwidth]{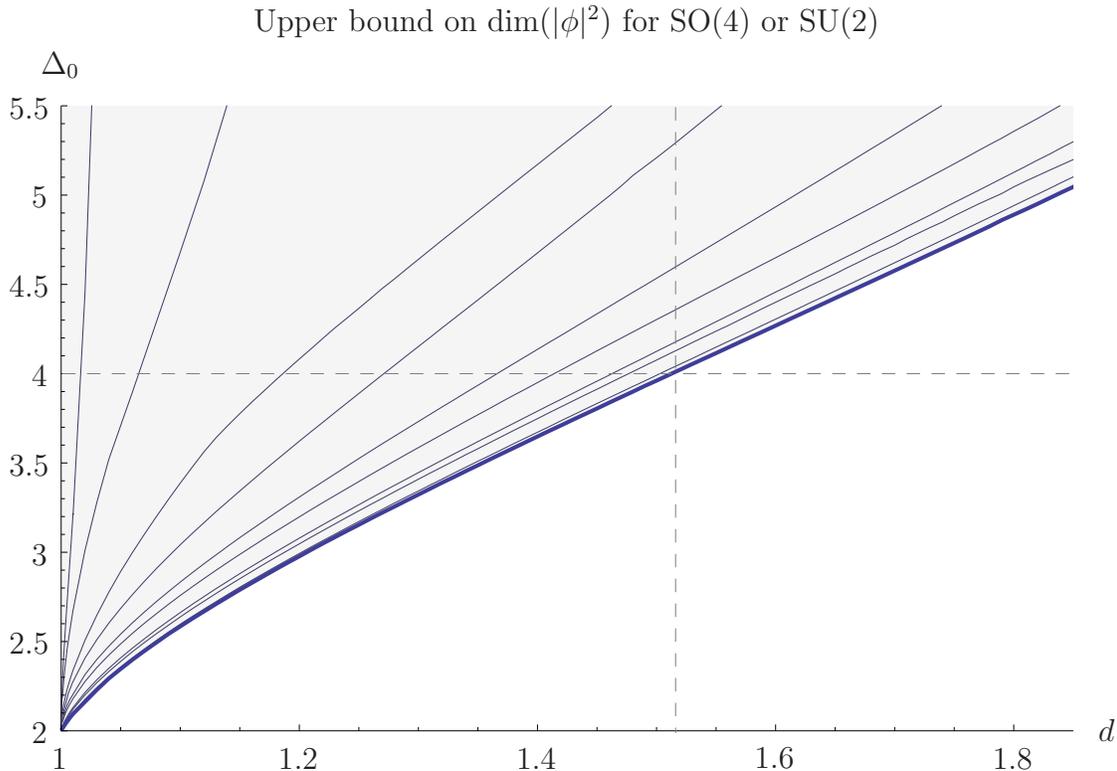}
\end{psfrags}
\end{center}
\caption{An upper bound on the dimension of $\f^\dag\f$, the lowest dimension singlet scalar appearing in $\f^\dag\x\f$, where $\f$ transforms in the fundamental representation of an $\SO(4)$ or an $\SU(2)$ global symmetry.  Curves are shown for $k=2,\dots,11$.  The bounds for $\SO(4)$ and $\SU(2)$ are identical in each case.  The strongest bound crosses $\De_0=4$ around $d=1.52$.}
\label{fig:SO4dim}
\end{figure}

As an example, figure~\ref{fig:SO4dim} shows a bound on the lowest dimension singlet in theories with an $\SU(2)$ or $\SO(4)$ global symmetry.\footnote{Note that to compute the $\SO(4)$ bound, we have only used the triple sum rule of Eq.~(\ref{eq:SONvectorialsumrule}).  It is straightforward to verify that including the fourth sum rule of Eq.~(\ref{eq:SO4sumrule}) leads to a redundant set of constraints, and is therefore unnecessary.} This bound is particularly interesting for conformal technicolor models, as we will discuss in detail in the following section.  Notice again that the curves start to converge at large $k$.  An approximate fit to the strongest ($k=11$) bound is given by
\be
\dim(|\f|^2)&\leq& 2+3.119\e+0.398(1-e^{-12\e}),
\ee
where $d=1+\e$, with $\e$ between $0$ and $1$.  This bound crosses $\De_0=4$ around $d\approx 1.52$.

\begin{figure}[h!]
\begin{center}
\begin{psfrags}
\def\PFGstripminus-#1{#1}%
\def\PFGshift(#1,#2)#3{\raisebox{#2}[\height][\depth]{\hbox{%
  \ifdim#1<0pt\kern#1 #3\kern\PFGstripminus#1\else\kern#1 #3\kern-#1\fi}}}%
\providecommand{\PFGstyle}{}%
%
\psfrag{d}[cl][cl]{\PFGstyle $d$}%
\psfrag{maxD0}[bc][bc]{\PFGstyle $\De_0$}%
\psfrag{maxdimforS}[bc][bc]{\PFGstyle $\text{Upper bound on $\dim(|\f|^2)$ for $\SO(N)$ or $\SU(N/2)$, $N=2,\dots,14$}$}%
\psfrag{x11}[tc][tc]{\PFGstyle $1$}%
\psfrag{x121}[tc][tc]{\PFGstyle $1.2$}%
\psfrag{x141}[tc][tc]{\PFGstyle $1.4$}%
\psfrag{x161}[tc][tc]{\PFGstyle $1.6$}%
\psfrag{x181}[tc][tc]{\PFGstyle $1.8$}%
\psfrag{y21}[cr][cr]{\PFGstyle $2$}%
\psfrag{y251}[cr][cr]{\PFGstyle $2.5$}%
\psfrag{y31}[cr][cr]{\PFGstyle $3$}%
\psfrag{y351}[cr][cr]{\PFGstyle $3.5$}%
\psfrag{y41}[cr][cr]{\PFGstyle $4$}%
\psfrag{y451}[cr][cr]{\PFGstyle $4.5$}%
\psfrag{y51}[cr][cr]{\PFGstyle $5$}%
\psfrag{y551}[cr][cr]{\PFGstyle $5.5$}%
\includegraphics[width=.9\textwidth]{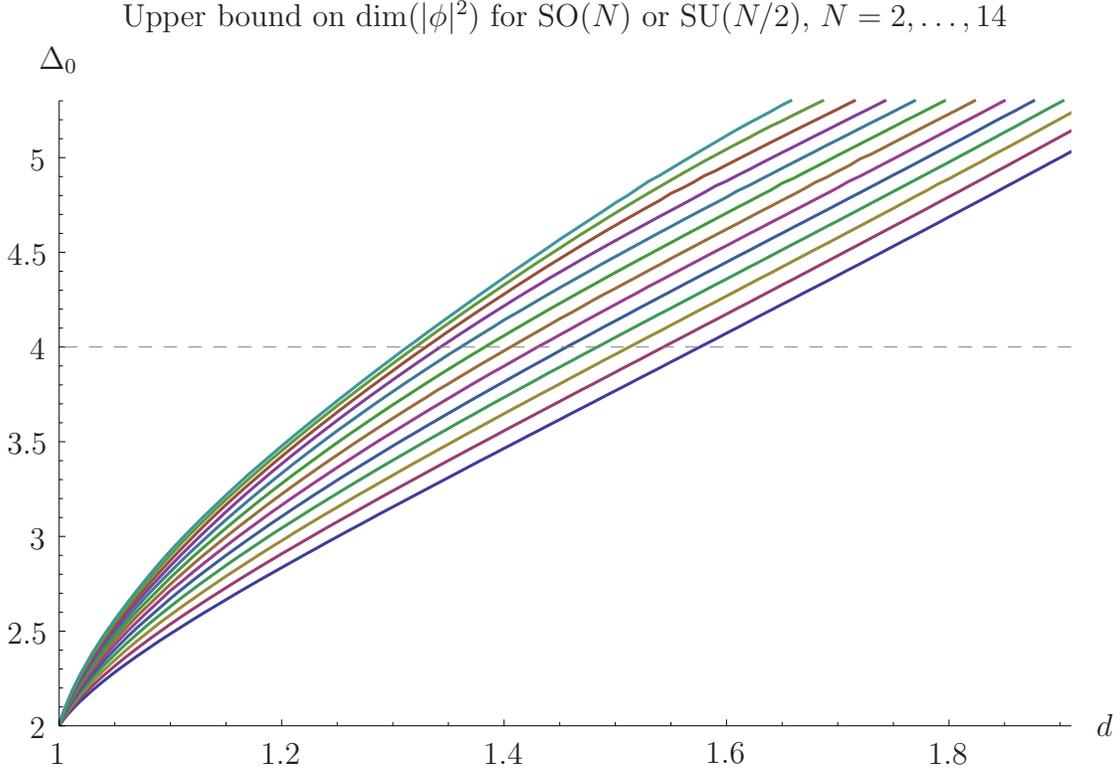}
\end{psfrags}
\end{center}
\caption{An upper bound on the dimension of $|\f|^2$, the lowest dimension singlet scalar appearing in $\f^T\x\f$ (or $\f^\dag\x\f)$, where $\f$ transforms in the fundamental representation of an $\SO(N)$ global symmetry or an $\SU(N/2)$ global symmetry (when $N\geq 4$ is even).  Curves are shown for $N=2,\dots,14$, with $N=2$ being the strongest bound.}
\label{fig:SONdimbounds}
\end{figure}

Figure~\ref{fig:SONdimbounds} shows dimension bounds for $\SO(N)$ with $N=2,\dots,14$ and $\SU(N)$ with $N=2,\dots,7$.  The strongest bound corresponds to the global symmetry group $\SO(2)\cong U(1)$, and the bounds weaken as $N$ increases.  One might na\"ively expect a larger symmetry group to produce a stronger bound.  For instance, a theory with an $\SO(N)$ symmetry certainly also has an $\SO(N-1)$ symmetry, so why shouldn't all bounds from the former apply to the latter?  However, as discussed in section~\ref{sec:SONSUNcoincidence}, the problem we are solving actually changes with $N$, and this turns out to be a more important effect than the enhanced symmetry.  Note that the lowest dimension singlet under an $\SO(N-1)$ subgroup of $\SO(N)$ is not necessarily a singlet at all under the full $\SO(N)$.  Thus, $\SO(N)$ bounds for larger $N$ apply to the operator with lowest dimension among a more restricted class of operators, and consequently can be weaker.

\subsubsection{Implications for Conformal Technicolor}

Let us briefly discuss some phenomenological implications of the bounds presented in figures~\ref{fig:SO4dim} and \ref{fig:SONdimbounds}.  A more detailed discussion of these implications will also appear in~\cite{rychkov:future}, and our analysis draws heavily on the previous discussions of~\cite{Luty:2004ye,Luty:2008vs,Galloway:2010bp,Evans:2010ed,Rattazzi:2008pe,Rattazzi:2010yc,Vichi:2011ux}, as well as the recent talk of~\cite{rychkov:talk}.

Arguably the most interesting operator dimension in the Standard Model is $\dim(H^\dag H)$, the dimension of the Higgs mass operator, where $H$ transforms as a bifundamental under $\SU(2)_L\x U(1)_Y\subset \SU(2)_L\x \SU(2)_R$.  In a weakly-coupled theory with a scalar Higgs, this dimension is approximately $2$, which leads to the hierarchy problem and its associated puzzles.

The idea of increasing $\dim(H^\dag H)$ to ameliorate the hierarchy problem is an old one.  In  traditional Technicolor models, the role of the Higgs is played by a fermion condensate $\bar\psi\psi$ with dimension $3$, so that the `mass' term $(\bar\psi\psi)^2$ is irrelevant.  A basic tension in this setup is that the `Yukawa' terms $(\bar\psi\psi)\bar qu$ which generate fermion masses after EWSB are also irrelevant.  To correctly account for the top-mass, we must imagine that such terms are suppressed by a low scale in the Lagrangian $\cL_\mathrm{Yuk.}\supset \frac 1 {\L_\mathrm{low}^2}(\bar\psi\psi)\bar qu$.  But this same low scale would then generically appear in other four-fermion operators, leading to dangerous flavor-changing neutral currents.

Conformal Technicolor (CTC)~\cite{Luty:2004ye} seeks to avoid this tension by assuming that $H$ participates in strong conformal dynamics above the electroweak scale, which generates a large dimension for $H^\dag H$, while the dimension of $H$ remains near 1.  While this idea is intriguing, we will show that it needs additional assumptions to work in practice.  In particular, our bounds definitively rule out the simplest `flavor-generic' CTC models.

To begin, let us determine the range of $d=\dim(H)$ and $\De=\dim(H^\dag H)$ that is phenomenologically viable in CTC.  Firstly, we must require that $y_t$ remain perturbative throughout the conformal regime, which places an upper bound on the possible running distance.  Indeed, suppose conformal dynamics occurs between $\L_\EW\approx 4\pi v$ and some higher scale $\L_\UV$.  Within this range of energies, Yukawa couplings run according to
\be
\label{eq:yukawaconformalrunning}
y_i(\mu) &=& \p{\frac{\mu}{\L_\EW}}^{d-1}y_i(\L_\EW)
\ee
(ignoring corrections from small perturbations away from exact conformal symmetry, like SM gauge couplings and  other Yukawa couplings).  Requiring $y_t\lesssim 4\pi$ for all $\mu\in[\L_\EW,\L_\UV]$ then gives
\be
\label{eq:topperturbativity}
\frac{\L_\UV}{\L_\EW} &\lesssim& \p{\frac{\L_\EW}{m_t}}^{\frac 1{d-1}}.
\ee

Secondly, we must ensure that small perturbations of the theory by the Higgs mass operator $H^\dag H$ don't destabilize the conformal dynamics.  This is certainly the case if $H^\dag H$ is irrelevant, $\De\geq 4$.  If on the other hand $\De<4$, then we must also impose the lower bound,
\be
\label{eq:CFTstability}
\frac{\L_\UV}{\L_\EW} & \lesssim& \p{\frac 1 {c(\L_\UV)}}^{\frac 1 {4-\De}},
\ee
where $c(\L_\UV)$ is the coefficient of $H^\dag H$ in the perturbation $\de \cL = c(\L_\UV) H^\dag H$ at $\L_\UV$.  The strength of the bound Eq.~(\ref{eq:CFTstability}) varies, depending on the amount of tuning we're willing to tolerate in this coefficient.

Finally, while Eqs.~(\ref{eq:topperturbativity}) and (\ref{eq:CFTstability}) prefer a small running distance, $\L_\UV$ must also be sufficiently large to suppress problematic flavor-changing operators, such as $(d s^c)(\bar s \bar{d^c})$ which contributes to $K$-$\bar K$ mixing.  In a `flavor-generic' model, we should demand
\be
\label{eq:genericflavor}
\frac 1 {\L_\UV^2} &\lesssim& \frac{1}{\L_F^2}\qquad\textrm{(generically)},
\ee
where $\L_F\sim 3.2\x 10^5\,\TeV$ for CP-violating contributions to $K$-$\bar K$ mixing~\cite{Isidori:2010kg}.  More optimistically, we might imagine that $(d s^c)(\bar s \bar{d^c})$ is generated with Yukawa suppression, so that the constraint above gets modified to
\be
\label{eq:optimisticflavor}
\frac {y_d(\L_\UV)y_s(\L_\UV)} {\L_\UV^2} &\lesssim& \frac{1}{\L_F^2}\qquad\textrm{(optimistically)},
\ee
with $y_i(\L_\UV)$ given by Eq.~(\ref{eq:yukawaconformalrunning}).

Together, these requirements restrict viable models to a particular region of the $d$-$\De$ plane, which can then be compared with our bounds.  In models where the conformal dynamics is custodially-symmetric, $H$ transforms as a fundamental of $\SO(4)\cong \SU(2)_L\x\SU(2)_R$ (which is weakly gauged by SM gauge fields).  However, the assumption of custodial symmetry is not actually necessary for us because our bound for $\SU(2)_L$ alone is identical to our bound for $\SO(4)$.  

\begin{figure}[h!]
\begin{center}
\begin{psfrags}
\def\PFGstripminus-#1{#1}%
\def\PFGshift(#1,#2)#3{\raisebox{#2}[\height][\depth]{\hbox{%
  \ifdim#1<0pt\kern#1 #3\kern\PFGstripminus#1\else\kern#1 #3\kern-#1\fi}}}%
\providecommand{\PFGstyle}{}%
%
\psfrag{CDelta}[bc][bc]{\PFGstyle $\Delta $}%
\psfrag{d}[cl][cl]{\PFGstyle $d$}%
\psfrag{ViableRegi}[bc][bc]{\PFGstyle $\text{Viable regions for Conformal Technicolor models}$}%
\psfrag{x11}[tc][tc]{\PFGstyle $1$}%
\psfrag{x121}[tc][tc]{\PFGstyle $1.2$}%
\psfrag{x141}[tc][tc]{\PFGstyle $1.4$}%
\psfrag{x161}[tc][tc]{\PFGstyle $1.6$}%
\psfrag{x181}[tc][tc]{\PFGstyle $1.8$}%
\psfrag{y21}[cr][cr]{\PFGstyle $2$}%
\psfrag{y251}[cr][cr]{\PFGstyle $2.5$}%
\psfrag{y31}[cr][cr]{\PFGstyle $3$}%
\psfrag{y351}[cr][cr]{\PFGstyle $3.5$}%
\psfrag{y41}[cr][cr]{\PFGstyle $4$}%
\psfrag{y451}[cr][cr]{\PFGstyle $4.5$}%
\psfrag{y51}[cr][cr]{\PFGstyle $5$}%
\psfrag{y551}[cr][cr]{\PFGstyle $5.5$}%
\includegraphics[width=0.9\textwidth]{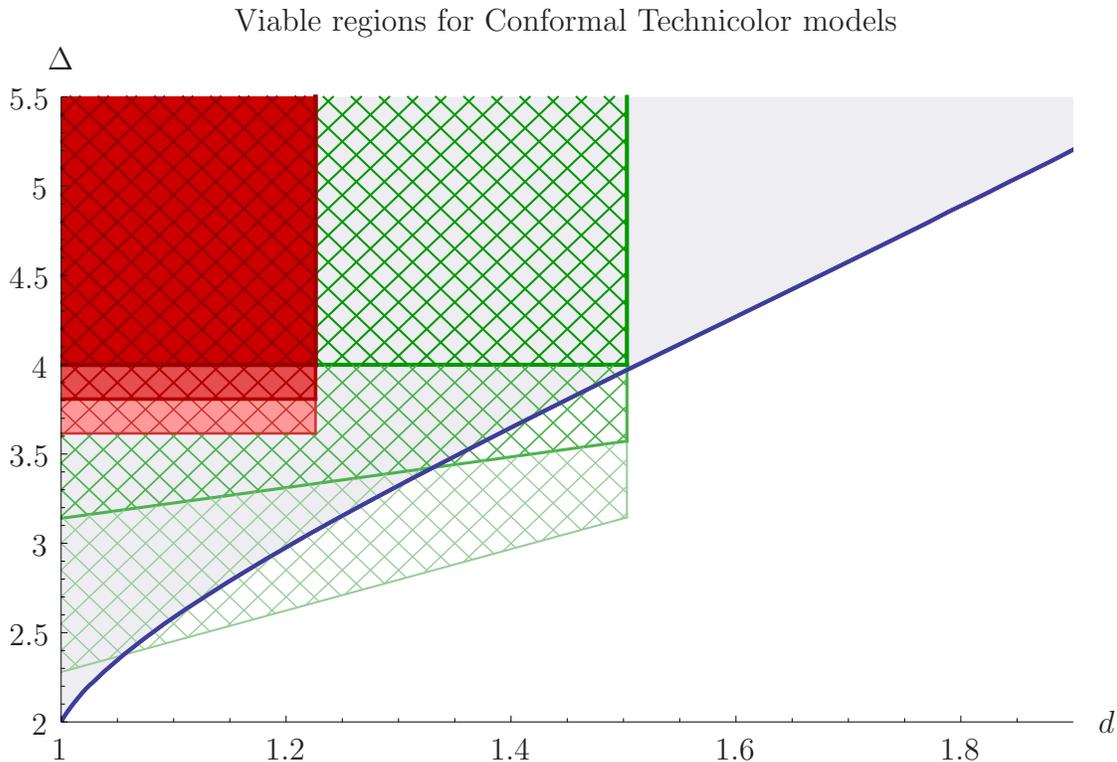}
\end{psfrags}
\end{center}
\caption{Viable regions for conformal technicolor models in the flavor-generic (red) and flavor-optimistic (cross-hatched green) cases are shown superimposed with our bound (blue, excluding the gray-shaded region).  Regions for $c(\L_\UV)=1,\ 0.1,$ and $0.01$ are shown in successively lighter shades of each color, with the largest region corresponding to $c(\L_\UV)=0.01$ in each case. Flavor-generic models are ruled out.}
\label{fig:conformaltechicolor}
\end{figure}

The viable regions for flavor-generic and flavor-optimistic CTC models are shown in figure~\ref{fig:conformaltechicolor}, superimposed with our strongest $\SU(2)$ dimension bound.  The right-hand edge of the viable regions comes from the combination of Eq.~(\ref{eq:genericflavor}) with Eq.~(\ref{eq:topperturbativity}), while the bottom edges come from the combination of Eq.~(\ref{eq:genericflavor}) with Eq.~(\ref{eq:CFTstability}) for different values of $c(\L_\UV)$.  We see that for reasonable assumptions about the coefficient $c(\L_\UV)$, flavor-generic models are ruled out.  This conclusion remains true even if the conformal dynamics respects CP symmetry, in which case the effective flavor scale can be closer to $\L_F\sim 10^4\,\TeV$.

By contrast, flavor-optimistic models with reasonable tunings $c(\L_\UV)\lesssim 0.1$ and somewhat large dimensions $d\sim 1.3$-$1.5$ are not necessarily ruled out.  Our bound does place an upper limit on the scale of new physics $\L_\UV$, but with sufficient Yukawa suppression these upper limits can be phenomenologically acceptable.  For instance, with $c=0.01$, $\L_\UV$ must lie below $6.8 \x 10^3\,\TeV$, while $c=0.1$ gives $\L_\UV\lesssim 1.6\x 10^3\,\TeV$.  At some point however, the predictions for these models become essentially those of minimal flavor violation with a low flavor scale, and strong conformal dynamics seems more and more like a gratuitous assumption.

\subsection{Symmetric Tensors in $\SO(N)$ Theories}

It is straightforward to modify our procedure to obtain bounds on symmetric tensors $\cO_{(ij)}$ appearing in $\f_i\x\f_j$.  To bound a symmetric tensor with dimension $\De_0$ and spin $\ell_0$, we look for a linear functional satisfying the normalization condition
\be
\a  \p{
\begin{array}{c}
F_{\De_0,\ell_0}\\
 \left(1-\frac{2}{N}\right) F_{\De_0,\ell_0}\\
- \left(1+ \frac{2}{N}\right) H_{\De_0,\ell_0}
\end{array}
}& = & 1,
\ee
as well as $\a(V)\geq 0$ for all other vectors $V$ in the $\SO(N)$ sum rule.

\begin{figure}[h!]
\begin{center}
\begin{psfrags}
\def\PFGstripminus-#1{#1}%
\def\PFGshift(#1,#2)#3{\raisebox{#2}[\height][\depth]{\hbox{%
  \ifdim#1<0pt\kern#1 #3\kern\PFGstripminus#1\else\kern#1 #3\kern-#1\fi}}}%
\providecommand{\PFGstyle}{}%
%
\psfrag{d}[cl][cl]{\PFGstyle $d$}%
\psfrag{maxD0}[bc][bc]{\PFGstyle $\De_0$}%
\psfrag{maxdimforS}[bc][bc]{\PFGstyle $\text{Upper bound on $\dim(\phi_{(i}\phi_{j)})$ for $\SO(4)$}$}%
\psfrag{x11}[tc][tc]{\PFGstyle $1$}%
\psfrag{x121}[tc][tc]{\PFGstyle $1.2$}%
\psfrag{x141}[tc][tc]{\PFGstyle $1.4$}%
\psfrag{x161}[tc][tc]{\PFGstyle $1.6$}%
\psfrag{x181}[tc][tc]{\PFGstyle $1.8$}%
\psfrag{y21}[cr][cr]{\PFGstyle $2$}%
\psfrag{y251}[cr][cr]{\PFGstyle $2.5$}%
\psfrag{y31}[cr][cr]{\PFGstyle $3$}%
\psfrag{y351}[cr][cr]{\PFGstyle $3.5$}%
\psfrag{y41}[cr][cr]{\PFGstyle $4$}%
\psfrag{y451}[cr][cr]{\PFGstyle $4.5$}%
\psfrag{y51}[cr][cr]{\PFGstyle $5$}%
\psfrag{y551}[cr][cr]{\PFGstyle $5.5$}%
\includegraphics[width=0.9\textwidth]{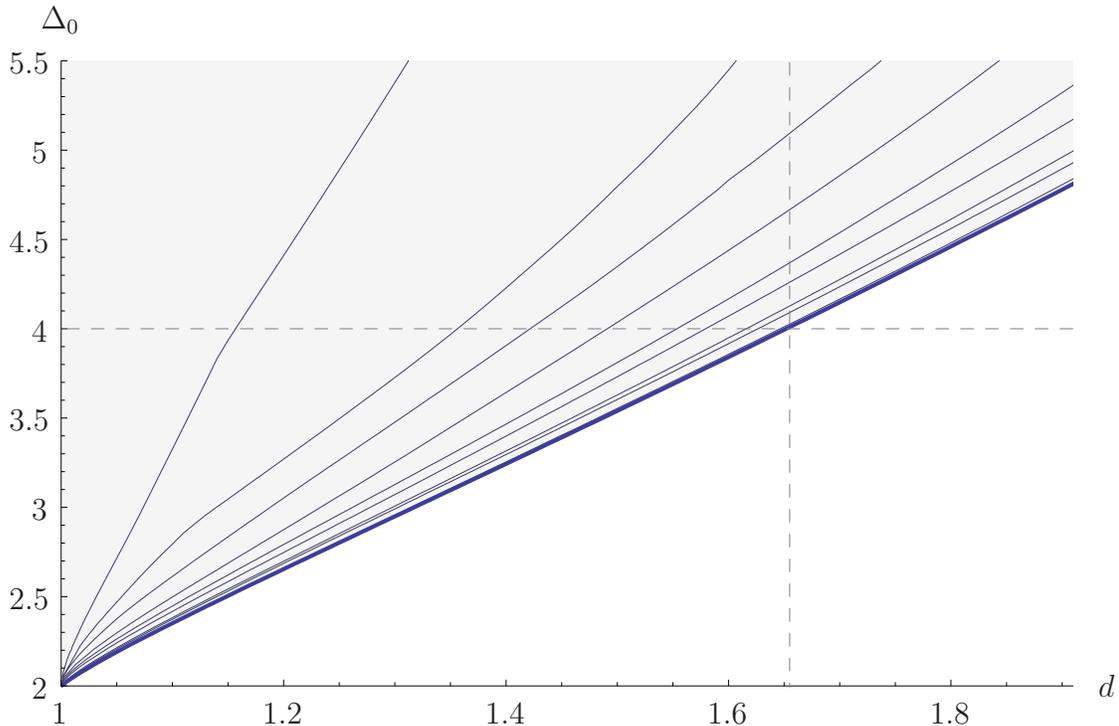}
\end{psfrags}
\end{center}
\caption{An upper bound on the lowest dimension symmetric tensor scalar appearing in $\f\x\f$, where $\f$ transforms in the fundamental of $\SO(4)$.  Here we show $k=2,\dots,11$.}
\label{fig:SO4symtensor}
\end{figure}

Figure~\ref{fig:SO4symtensor} shows the resulting dimension bound on $\f_{(i}\f_{j)}$ (the lowest dimension scalar symmetric tensor appearing in $\f_i\x\f_j$) in the case of $\SO(4)$ symmetry.  Note that this bound does not apply in a simple way to operators in theories with $\SU(2)$ symmetries, because there is no coincidence between $\SU(N)$ and $\SO(2N)$ bounds for non-singlets.

\subsection{Superconformal Theories}

\begin{figure}[h!]
\begin{center}
\begin{psfrags}
\def\PFGstripminus-#1{#1}%
\def\PFGshift(#1,#2)#3{\raisebox{#2}[\height][\depth]{\hbox{%
  \ifdim#1<0pt\kern#1 #3\kern\PFGstripminus#1\else\kern#1 #3\kern-#1\fi}}}%
\providecommand{\PFGstyle}{}%
%
\psfrag{D0min}[bc][bc]{\PFGstyle $\De_0$}%
\psfrag{D2d}[cc][cc]{\PFGstyle $\ \ \ \De_0=2d$}%
\psfrag{d}[cl][cl]{\PFGstyle $d$}%
\psfrag{SUSYU1dime}[bc][bc]{\PFGstyle $\text{Upper bound on $\dim(\Phi^\dag\Phi)$}$}%
\psfrag{x11}[tc][tc]{\PFGstyle $1$}%
\psfrag{x121}[tc][tc]{\PFGstyle $1.2$}%
\psfrag{x141}[tc][tc]{\PFGstyle $1.4$}%
\psfrag{x161}[tc][tc]{\PFGstyle $1.6$}%
\psfrag{x181}[tc][tc]{\PFGstyle $1.8$}%
\psfrag{y21}[cr][cr]{\PFGstyle $2$}%
\psfrag{y251}[cr][cr]{\PFGstyle $2.5$}%
\psfrag{y31}[cr][cr]{\PFGstyle $3$}%
\psfrag{y351}[cr][cr]{\PFGstyle $3.5$}%
\psfrag{y41}[cr][cr]{\PFGstyle $4$}%
\psfrag{y451}[cr][cr]{\PFGstyle $4.5$}%
\psfrag{y51}[cr][cr]{\PFGstyle $5$}%
\psfrag{y551}[cr][cr]{\PFGstyle $5.5$}%
\includegraphics[width=0.9\textwidth]{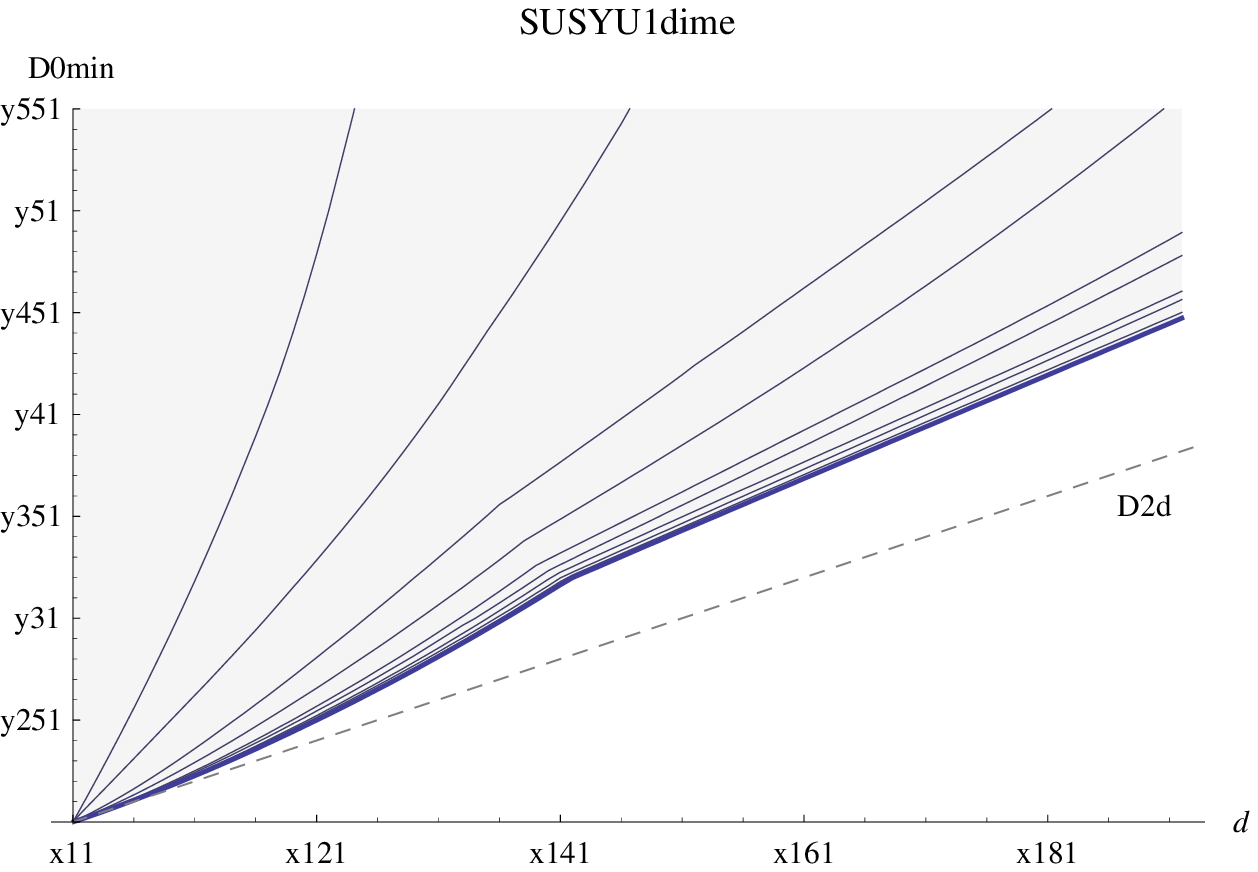}
\end{psfrags}
\end{center}
\caption{An upper bound on the dimension of $\Phi^\dag\Phi$, where $\Phi$ is a chiral primary scalar of dimension $d$ in an SCFT.  The dashed line is the factorization value $\De=2d$.  Here we show $k=2,\dots,11$.}
\label{fig:SUSYU1dim}
\end{figure}

Now let us turn to bounding operator dimensions in superconformal theories, using the sum rule Eq.~(\ref{eq:SCFTcrossing}).  A bound on $\dim(\Phi^\dag \Phi)$ in terms of $\dim(\Phi)$ was first obtained in~\cite{Poland:2010wg} using only the middle row of Eq.~(\ref{eq:SCFTcrossing}).  In~\cite{Vichi:2011ux}, it was shown that the bound could be improved by incorporating the other rows, and linear programming calculations were given up to $k=4$.  In figure~\ref{fig:SUSYU1dim}, we present a new version of these bounds for $k$ up to $11$, corresponding to a 198-dimensional search space.

\begin{figure}[h!]
\begin{center}
\begin{psfrags}
\def\PFGstripminus-#1{#1}%
\def\PFGshift(#1,#2)#3{\raisebox{#2}[\height][\depth]{\hbox{%
  \ifdim#1<0pt\kern#1 #3\kern\PFGstripminus#1\else\kern#1 #3\kern-#1\fi}}}%
\providecommand{\PFGstyle}{}%
%
\psfrag{D0min}[bc][bc]{\PFGstyle $\De_0$}%
\psfrag{D2d}[cc][cc]{\PFGstyle $\ \ \ \De_0=2d$}%
\psfrag{d}[cl][cl]{\PFGstyle $d$}%
\psfrag{SUSYU1dime}[bc][bc]{\PFGstyle $\text{Upper bound on $\dim(\Phi^\dag\Phi)$}$}%
\psfrag{x1021}[tc][tc]{\PFGstyle $1.02$}%
\psfrag{x1041}[tc][tc]{\PFGstyle $1.04$}%
\psfrag{x1061}[tc][tc]{\PFGstyle $1.06$}%
\psfrag{x1081}[tc][tc]{\PFGstyle $1.08$}%
\psfrag{x111}[tc][tc]{\PFGstyle $1.1$}%
\psfrag{x11}[tc][tc]{\PFGstyle $1$}%
\psfrag{y2051}[cr][cr]{\PFGstyle $2.05$}%
\psfrag{y211}[cr][cr]{\PFGstyle $2.1$}%
\psfrag{y2151}[cr][cr]{\PFGstyle $2.15$}%
\psfrag{y21}[cr][cr]{\PFGstyle $2$}%
\psfrag{y221}[cr][cr]{\PFGstyle $2.2$}%
\includegraphics[width=0.9\textwidth]{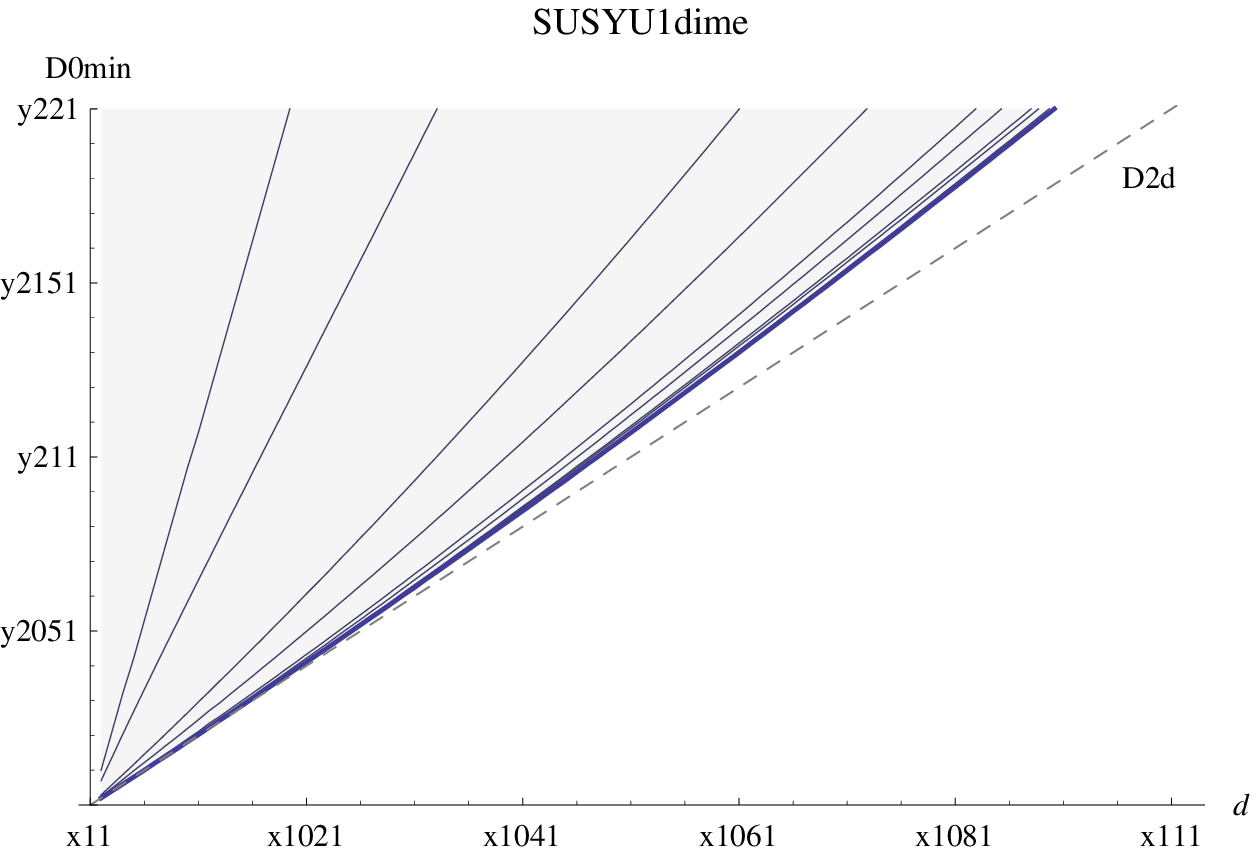}
\end{psfrags}
\end{center}
\caption{A zoom in on the region of figure~\ref{fig:SUSYU1dim} near $\dim(\Phi)=1$.}
\label{fig:SUSYU1dimzoom}
\end{figure}

Several interesting new features emerge at large $k$.  Most strikingly, the bound appears to be tangent to the factorization line $\De_0=2d$ near $d=1$.  Figure~\ref{fig:SUSYU1dimzoom} shows a higher-resolution plot for small values of $d$, which displays this behavior more clearly.  An approximate fit to the $k=11$ curve in figure~\ref{fig:SUSYU1dimzoom} is given by
\be
\label{eq:SUSYU1fit}
\De_0 \leq 2(1+\e) + 2.683\,\e^2+\dots\qquad (\e\ll 1),
\ee
where $d=1+\e$.  Note that known superconformal theories populate the entire factorization line,\footnote{Namely supersymmetric mean field theories, which satisfy the necessary requirements of unitarity and crossing symmetry, and exist for each $d\geq 1$.  They occur in the infinite-$N$ limit of supersymmetric gauge theories.} so it is impossible to have a bound stronger than $\De_0\leq 2d$.  Our bound on $\dim (\Phi^\dag\Phi)$ is one of the few examples computed to date that approaches the provably best possible bound for some nontrivial range of $d$'s.

Eq.~(\ref{eq:SUSYU1fit}) can be directly tested in theories that admit a perturbative Banks-Zaks limit and contain a chiral operator with dimension near $1$.  As far as we are aware, there are no known examples of perturbative theories living above the factorization line.  Here we have shown numerically that this can be understood purely from the constraints of crossing symmetry and unitarity.  It would be very interesting to understand this fact analytically.

It is amusing to speculate on the form of the bound as $k\to \oo$.  A simple and intriguing possibility is that the small-$d$ behavior might extend to all $d$, so that the best possible bound $\De_0\leq 2d$ is realized.  In other words, it might be the case that the anomalous dimension $\g_{\Phi^\dag\Phi}=\dim(\Phi^\dag\Phi)-2\dim(\Phi)$ is always non-positive.  This possibility was investigated recently for theories with a weakly-coupled gravity dual in~\cite{Fitzpatrick:2011hh}, with inconclusive results; effective field theories in $\AdS_5$ allow for both positive and negative contributions to $\g_{\Phi^\dag\Phi}$.  However, it's possible that additional constraints might be present in those theories which admit a consistent UV completion.

Another possibility is that the bound converges above the factorization line, with a shape similar to the $k=11$ curve in figure~\ref{fig:SUSYU1dim}.  In that case, one might wonder about the significance of the cusp near $d=1.4$, which appears to be a common feature of each curve with $k\geq 4$.  A previous example of a dimension bound with a cusp is the 2D real scalar dimension bound, presented in~\cite{Rychkov:2009ij} (building on the first 2D results of~\cite{Rattazzi:2008pe}).   There, an actual theory, the 2D Ising model, exists very near the cusp, so that the bound is close to the best possible at that value of $d$.  By analogy, one might speculate that an $\cN=1$ SUSY `minimal model' exists in the cusp in figure~\ref{fig:SUSYU1dim}.

\subsubsection{Phenomenological Applications}

Our bound on $\dim(\Phi^\dag \Phi)$ has implications for several models that use strong superconformal dynamics to tailor soft parameters in the MSSM.  One example is the solution to the $\mu/B\mu$ problem in gauge mediation proposed in \cite{Roy:2007nz,Murayama:2007ge} and further developed in~\cite{Perez:2008ng,Kim:2009sy,Craig:2009rk,Hanaki:2010xf}.  In this scenario, SUSY breaking is communicated to the visible sector via a chiral field $X$ which develops a SUSY-breaking VEV $\<X\>=F\th^2$ at some scale $\L_\mathrm{IR}$.  In matching to the MSSM at $\L_\mathrm{IR}$, the effective operators
\be
\cO_{X} = c_{X}\int d^4 \th \frac{X^\dag H_u H_d}{M_*}+\mathrm{h.c.}
\qquad
\textrm{and}
\qquad
\cO_{X^\dag X}=c_{X^\dag X}\int d^4\th \frac{X^\dag X H_u H_d}{M_*^2},
\ee
contribute to $\mu$ and $B\mu$, respectively.  Here, $M_*$ is the scale where these operators originate (typically the messenger scale). Many of the simplest gauge-mediated models generate both $\cO_{X}$ and $\cO_{X^\dag X}$ at one-loop at the messenger scale, so that na\"ively $c_X\sim c_{X^\dag X}\sim \frac {\l^2}{16\pi^2}$, with $\l$ an $O(1)$ coupling constant.  However, this then leads to the problematic relation $B\mu/\mu^2 \sim 16\pi^2$, which precludes viable electroweak symmetry breaking.

\begin{figure}[h!]
\begin{center}
\begin{psfrags}
\def\PFGstripminus-#1{#1}%
\def\PFGshift(#1,#2)#3{\raisebox{#2}[\height][\depth]{\hbox{%
  \ifdim#1<0pt\kern#1 #3\kern\PFGstripminus#1\else\kern#1 #3\kern-#1\fi}}}%
\providecommand{\PFGstyle}{}%
%
\psfrag{d}[cl][cl]{\PFGstyle $d$}%
\psfrag{dist}[bc][bc]{\PFGstyle $\L_\mathrm{UV}/\L_\mathrm{IR}$}%
\psfrag{muBmurunni}[bc][bc]{\PFGstyle $\text{Running distance needed to solve $\mu/B\mu$}$}%
\psfrag{x11}[tc][tc]{\PFGstyle $1$}%
\psfrag{x121}[tc][tc]{\PFGstyle $1.2$}%
\psfrag{x141}[tc][tc]{\PFGstyle $1.4$}%
\psfrag{x161}[tc][tc]{\PFGstyle $1.6$}%
\psfrag{x181}[tc][tc]{\PFGstyle $1.8$}%
\psfrag{y1000}[cr][cr]{\PFGstyle $10^3$}%
\psfrag{y1}[cr][cr]{\PFGstyle $1$}%
\psfrag{ySuperscriA}[cr][cr]{\PFGstyle $10^9$}%
\psfrag{ySuperscriB}[cr][cr]{\PFGstyle $10^{12}$}%
\psfrag{ySuperscriC}[cr][cr]{\PFGstyle $10^{15}$}%
\psfrag{ySuperscri}[cr][cr]{\PFGstyle $10^6$}%
\hspace{-10mm}\includegraphics[width=0.9\textwidth]{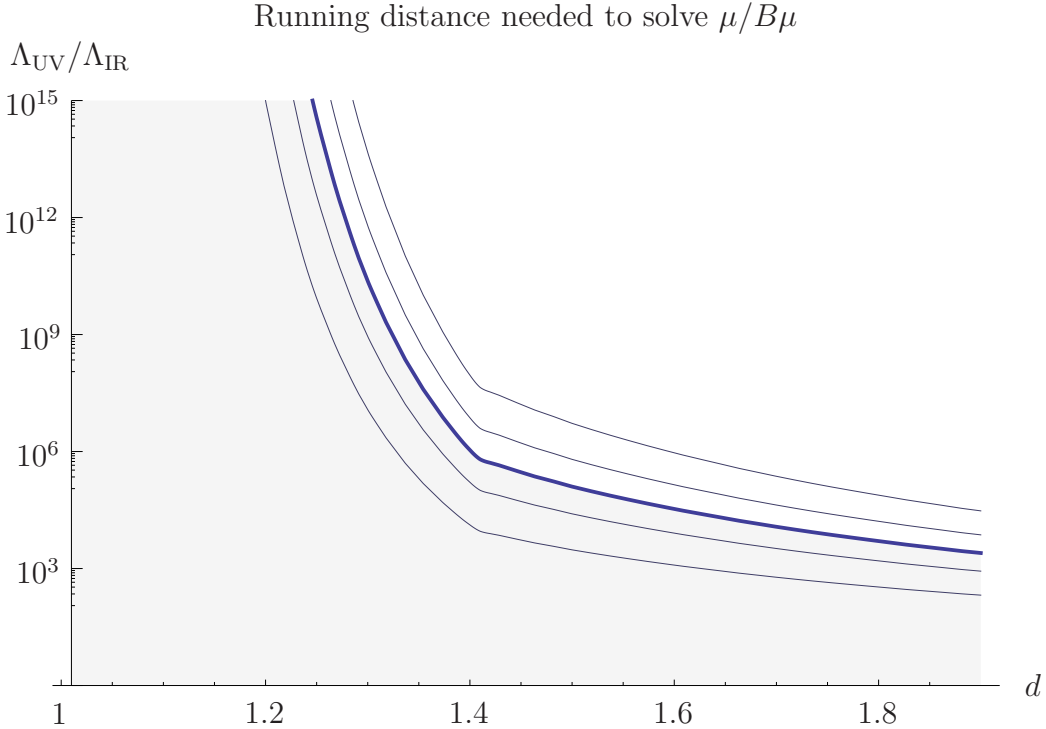}
\end{psfrags}
\end{center}
\caption{An approximate lower bound on the running distance required for solving the $\mu/B\mu$ problem with strong conformal dynamics, as a function of $d=\dim(X)$.  The middle curve corresponds to a loop factor suppression: $c_{X^\dag X}(\L_\mathrm{IR})=\frac 1{16\pi^2}c_X^2(\L_\mathrm{IR})$, while the outer curves correspond to suppressions within factors of $2$ and $5$ of a loop factor.}
\label{fig:muBmurunning}
\end{figure}

The solution proposed in~\cite{Roy:2007nz,Murayama:2007ge} is that $X$ should participate in strong conformal dynamics over some range of scales $\L_\mathrm{IR}<\mu<\L_\mathrm{UV}$, with $\L_\mathrm{UV}\leq M_*$.  If the anomalous dimension $\g_{X^\dag X}\equiv \dim(X^\dag X)- 2\dim(X)$ is positive, then the operator $\cO_{X^\dag X}$ will be suppressed relative to $\cO_X$, and $B\mu/\mu^2$ can be close to unity at the matching scale $\L_\mathrm{IR}$.  In particular, to restore proper electroweak symmetry breaking, we should approximately have
\be
\label{eq:viableEWSB}
\p{\frac{\L_\mathrm{IR}}{\L_\mathrm{UV}}}^{\g_{X^\dag X}} &\approx& \frac{1}{16\pi^2}.
\ee
Using this relation, our upper bound on $\dim(X^\dag X)$ in figure~\ref{fig:SUSYU1dim} translates into a lower bound on the running distance $\L_\mathrm{UV}/\L_\mathrm{IR}$, shown in figure~\ref{fig:muBmurunning}.  Note in particular that a small $\dim(X)$ requires a very large running distance, since our bound on $\g_{X^\dag X}$ approaches zero as $\dim(X)\to 1$.  Consequently, viable models should at least have $\dim(X)\gtrsim 1.3$.  Note that $\dim(X)$ can almost always be calculated using $a$-maximization in concrete examples, so a bound on the required running distance can be easily read from figure~\ref{fig:muBmurunning} for specific models.

Our bound can also apply to models of conformal sequestering \cite{Luty:2001jh,Luty:2001zv,Dine:2004dv,Sundrum:2004un,Ibe:2005pj,Ibe:2005qv,Schmaltz:2006qs,Murayama:2007ge,Kachru:2007xp} which contain chiral gauge singlets, where the idea is that a large $\dim(X^\dag X)$ can lead to suppression of flavor-dependent soft-mass operators,
\be
c_{ij}\int d^4\th \frac{1}{M_*^2} X^\dag X \f_i^\dag \f_j.
\ee
Let us for example assume a gravity mediated scenario, where the cutoff scale is $M_*\sim M_\mathrm{pl}$ and conformal running occurs between $M_\mathrm{pl}$ and an intermediate scale $\L_\mathrm{int} \sim 10^{11} \,\mathrm{GeV}$.  Viable flavor physics then roughly requires $\dim(X^\dag X)-2\gtrsim 1$~\cite{Schmaltz:2006qs}, and from figure~\ref{fig:SUSYU1dim} we see that such models should also have $\dim(X)\gtrsim 1.35$ or so.\footnote{However, it's possible that one could avoid these constraints by having `safe' flavor currents appear in the OPE (as discussed in~\cite{Schmaltz:2006qs}).}  Our bounds similarly constrain the possible suppression of these operators in superconformal flavor models~\cite{Nelson:2000sn, Kobayashi:2001kz, Nelson:2001mq,Kobayashi:2002iz,Poland:2009yb,Craig:2010ip,Kobayashi:2010ye, Aharony:2010ch,Dudas:2010yh}, where the visible sector fields participate in the strong conformal dynamics.  Once again, in all of these situations a comparison to our bounds can be checked in concrete examples using $a$-maximization.

\section{Bounds on OPE Coefficients}
\label{sec:opebounds}

In this section we will turn our attention away from bounding operator dimensions and instead explore some of the more basic bounds on OPE coefficients obtainable using these methods.  We'll begin by reproducing (and strengthening) the upper bounds on scalar OPE coefficients for general CFTs previously presented in~\cite{Caracciolo:2009bx}.  Then we'll focus on something qualitatively new --- the possibility of placing {\it lower} bounds on OPE coefficients in theories that have a gap in the spectrum of operator dimensions.  In fact, this happens naturally in supersymmetric theories for protected operators appearing in the $\Phi \times \Phi$ OPE, where a gap is forced by unitarity.  We will then demonstrate that there are extremely constraining upper and lower bounds on the OPE coefficients of these operators when $\dim(\Phi) < 3/2$.

\subsection{Scalar Operators in General Theories}

\begin{figure}[h!]
\begin{center}
\begin{psfrags}
\def\PFGstripminus-#1{#1}%
\def\PFGshift(#1,#2)#3{\raisebox{#2}[\height][\depth]{\hbox{%
  \ifdim#1<0pt\kern#1 #3\kern\PFGstripminus#1\else\kern#1 #3\kern-#1\fi}}}%
\providecommand{\PFGstyle}{}%
%
\psfrag{D0}[cl][cl]{\PFGstyle $\text{D0}$}%
\psfrag{maxlambda0}[bc][bc]{\PFGstyle $\l_{\cO_0}$}%
\psfrag{maxscalarO}[bc][bc]{\PFGstyle $\text{Upper bounds on scalar OPE coefficients, $d=1.01,\dots,1.66$}$}%
\psfrag{x11}[tc][tc]{\PFGstyle $1$}%
\psfrag{x151}[tc][tc]{\PFGstyle $1.5$}%
\psfrag{x21}[tc][tc]{\PFGstyle $2$}%
\psfrag{x251}[tc][tc]{\PFGstyle $2.5$}%
\psfrag{x31}[tc][tc]{\PFGstyle $3$}%
\psfrag{x351}[tc][tc]{\PFGstyle $3.5$}%
\psfrag{x41}[tc][tc]{\PFGstyle $4$}%
\psfrag{y0}[cr][cr]{\PFGstyle $0$}%
\psfrag{y11}[cr][cr]{\PFGstyle $1$}%
\psfrag{y21}[cr][cr]{\PFGstyle $2$}%
\psfrag{y31}[cr][cr]{\PFGstyle $3$}%
\psfrag{y41}[cr][cr]{\PFGstyle $4$}%
\psfrag{D0}[cr][cr]{\PFGstyle $\De_0$}%
\includegraphics[width=0.9\textwidth]{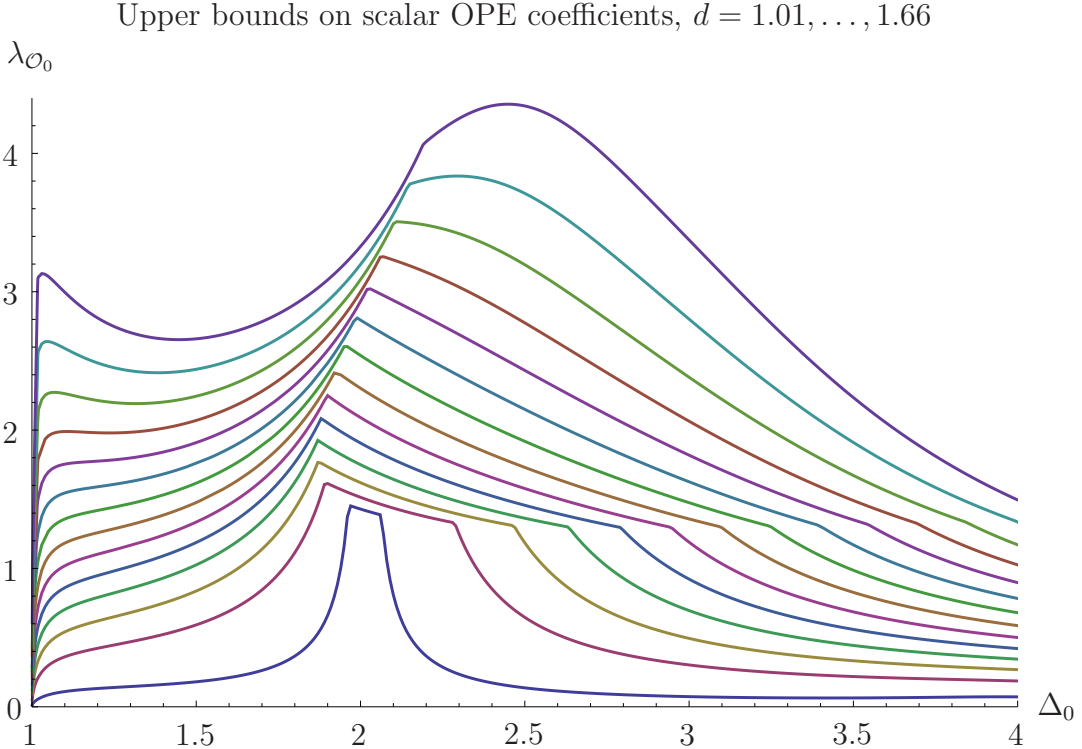}
\end{psfrags}
\end{center}
\caption{Upper bounds on the OPE coefficient of a scalar operator $\cO_0\in \f\x\f$ (not necessarily of lowest dimension).  Each curve is for a different value  $d=1.01,\dots,1.66$, with a spacing of $0.05$ and $d=1.01$ corresponding to the lowest curve.  Here we have taken $k=11$.}
\label{fig:realscalarOPE}
\end{figure}

Let us begin by producing bounds on OPE coefficients of scalar operators $\cO_0$ of dimension $\De_0$ appearing in the $\phi \times \phi$ OPE, where $\phi$ is a scalar operator of dimension $d$.  As we saw in Eq.~(\ref{eq:upperboundonlambda}), by applying a linear functional $\alpha$ to the CFT crossing relations we can obtain an upper bound $\lambda_{\cO_0}^2 \leq -\alpha(F_{0,0})$.  In figure~\ref{fig:realscalarOPE} we show the best upper bounds on $\lambda_{\cO_0}$ as a function of $\De_0$ that we have obtained so far, for $d=1.01,\ldots,1.66$ with a spacing of 0.05.  These bounds are obtained using $k=11$, corresponding to a $66$-dimensional search space.  This plot strengthens bounds previously presented in~\cite{Caracciolo:2009bx}.

Figure~\ref{fig:realscalarOPE} clearly contains a lot of interesting structure.  First, as $d \rightarrow 1$, the curve becomes more and more sharply peaked around $\De_0 \simeq 2$, with the height of the peak converging to the free value $\lambda_0 = \sqrt{2}$.\footnote{Note that the free OPE coefficient is $\sqrt{2}$ rather than $1$ because we have required the $\phi^2$ operator to have a canonically normalized two-point function, rather than the normalization inherited from Wick contractions.}  On the other hand, as $\De_0 \rightarrow 1$ all of the curves drop sharply to zero (first peaking at larger values of $d$), corresponding to the fact that a free operator cannot appear in the OPE.  All of the bounds also increase in strength as $\De_0$ becomes large, possibly asymptoting to zero.  Finally, as $d$ increases at fixed $\De_0$ the bounds monotonically decrease in strength.  Note that in the present study we have found the region $d > 1.66$ to be numerically more difficult (though very weak bounds appear to exist at least up to $d \sim 1.86$), and we postpone a full investigation of this region to future work.

Let us take a moment to understand a way in which our method fails to fully pick out the spectrum of free theories as $d \rightarrow 1$.  While our upper bound becomes nicely peaked around the free value in this limit, our algorithm cannot easily distinguish between a single $\De_0 \simeq 2$ operator with $\l_0 \simeq \sqrt{2}$, and a broader spectrum of operators, each having $\De_0$ somewhat close to $2$ and $\l_0 < \sqrt{2}$.  The issue is that both of these scenarios can lead to very similar conformal block contributions to the 4-point functions that we are studying.  On the other hand, if we knew that there was only a {\it single} operator appearing in the OPE up to a certain dimension, this ambiguity could not occur and we would be able to also place lower bounds on its OPE coefficient.  In the next subsection we will study this possibility in more detail, focusing on protected operators appearing in the $\Phi \times \Phi$ OPE in SCFTs.

\subsection{Protected Operators in Superconformal Theories}

As we reviewed in section~\ref{sec:crossrelSCFTs}, if $\Phi$ is a chiral superconformal primary of dimension $d$ in an $\cN=1$ SCFT, the $\Phi \times \Phi^{\dagger}$ OPE contains superconformal primaries of dimension $\De \geq \ell + 2$ and their descendants.  On the other hand, the $\Phi \times \Phi$ OPE can contain a chiral $\Phi^2$ operator of dimension $2d$, superconformal descendants $\bar{Q} \cO_{\ell}$ of protected operators having dimension $2d+\ell$, and superconformal descendants $\bar{Q}^2 \cO$ of unprotected operators with a dimension satisfying $\De \geq |2d-3|+3+\ell$.  

Notice that, as long as $d < 3/2$, there is necessarily a gap between the dimensions of the protected operators appearing in the $\Phi \times \Phi$ OPE and the dimensions of the unprotected operators.  This gap is a consequence of the unitarity constraints on operator dimensions in SCFTs.  Because of this gap, no other operators appearing in the OPE can give similar conformal block contributions to the four-point function $\< \Phi \Phi^{\dagger} \Phi \Phi^{\dagger}\>$, and we can attempt to derive {\it lower} bounds on the OPE coefficients $\l_{\Phi^2}$ and $\l_{\bar{Q} \cO_{\ell}}$, in addition to upper bounds.  

The logic used to obtain a lower bound requires only a slight modification to the procedure described in section~\ref{sec:review}.  Since one could in principle attempt to obtain a lower bound in any theory with a dimension gap, let us first describe the logic for the simplest case of the real scalar crossing relation in general CFTs.  To obtain a lower bound on an OPE coefficient $\l_{\cO_0}^2$, we can again consider applying a linear functional to the real scalar crossing relation, as in Eq.~(\ref{eq:linfunctionalapplied}).  However, instead of imposing the constraints~(\ref{eq:alphaconstraint1}) and~(\ref{eq:alphaconstraint2}), we can alternatively require
\be
\label{eq:loweralphaconstraint1}
\a(F_{\De_0,\ell_0}) &=& 1,\qquad\textrm{and}\\
\label{eq:loweralphaconstraint2}
\a(F_{\De,\ell}) &\leq& 0, \qquad\textrm{for all other operators in the spectrum,}
\ee
which leads to the lower bound
\be
\label{eq:lowerboundonlambda}
\l_{\cO_0}^2 =-\a(F_{0,0})-\sum_{\cO\neq \cO_0}\textrm{pos.}\x\textrm{neg.} \geq -\a(F_{0,0}).
\ee
Note that (\ref{eq:loweralphaconstraint1}) and (\ref{eq:loweralphaconstraint2}) are only compatible with each other if we know that there is a gap between $\De_0$ and the $\De$'s for all other operators in the spectrum.

Generalizing to the superconformal crossing relation of Eq.~(\ref{eq:SCFTcrossing}), if we isolate a protected operator $\cO_0$ of spin $\ell_0$ and require
\be
\label{eq:lowersusyalphaconstraint1}
\a \p{
\begin{array}{c}
0\\
 F_{2d+\ell_0,\ell_0}\\
- H_{2d+\ell_0,\ell_0}
\end{array}
} &=& 1,\qquad\textrm{}\\
\label{eq:lowersusyalphaconstraint2}
\a  \p{
\begin{array}{c}
0\\
 F_{\De,\ell}\\
- H_{\De,\ell}
\end{array}
} & \leq & 0,\qquad\textrm{for all other operators in $\Phi \times \Phi$, and}\\
\label{eq:lowersusyalphaconstraint3}
\a \p{
\begin{array}{c}
\cF_{\De,\ell}\\
\tl\cF_{\De,\ell} \\
\tl \cH_{\De,\ell}
\end{array}
}& \leq & 0,\qquad\textrm{for all (non-unit) operators in $\Phi \times \Phi^{\dagger}$,}
\ee
we obtain the lower bound
\be
\l_{\cO_0}^2 &\geq& -\a \p{
\begin{array}{c}
\cF_{0,0}\\
\tl\cF_{0,0} \\
\tl\cH_{0,0}
\end{array}
} .
\ee
Meanwhile, reversing the inequalities in~(\ref{eq:lowersusyalphaconstraint2}) and~(\ref{eq:lowersusyalphaconstraint3}) leads to an upper bound on $\l_{\cO_0}^2$, following our usual logic.  

\begin{figure}[h!]
\begin{center}
\begin{psfrags}
\def\PFGstripminus-#1{#1}%
\def\PFGshift(#1,#2)#3{\raisebox{#2}[\height][\depth]{\hbox{%
  \ifdim#1<0pt\kern#1 #3\kern\PFGstripminus#1\else\kern#1 #3\kern-#1\fi}}}%
\providecommand{\PFGstyle}{}%
%
\psfrag{d}[cl][cl]{\PFGstyle $d$}%
\psfrag{lam}[bc][bc]{\PFGstyle $\l_{\Phi^2}$}%
\psfrag{OPEcoeffic}[bc][bc]{\PFGstyle $\text{Upper and lower bounds on $\l_{\Phi^2}$}$}%
\psfrag{x111}[tc][tc]{\PFGstyle $1.1$}%
\psfrag{x11}[tc][tc]{\PFGstyle $1$}%
\psfrag{x121}[tc][tc]{\PFGstyle $1.2$}%
\psfrag{x131}[tc][tc]{\PFGstyle $1.3$}%
\psfrag{x141}[tc][tc]{\PFGstyle $1.4$}%
\psfrag{x151}[tc][tc]{\PFGstyle $1.5$}%
\psfrag{x161}[tc][tc]{\PFGstyle $1.6$}%
\psfrag{x181}[tc][tc]{\PFGstyle $1.8$}%
\psfrag{x21}[tc][tc]{\PFGstyle $2$}%
\psfrag{y0}[cr][cr]{\PFGstyle $0$}%
\psfrag{y11}[cr][cr]{\PFGstyle $1$}%
\psfrag{y151}[cr][cr]{\PFGstyle $1.5$}%
\psfrag{y251}[cr][cr]{\PFGstyle $2.5$}%
\psfrag{y21}[cr][cr]{\PFGstyle $2$}%
\psfrag{y31}[cr][cr]{\PFGstyle $3$}%
\psfrag{y5}[cr][cr]{\PFGstyle $0.5$}%
\includegraphics[width=0.9\textwidth]{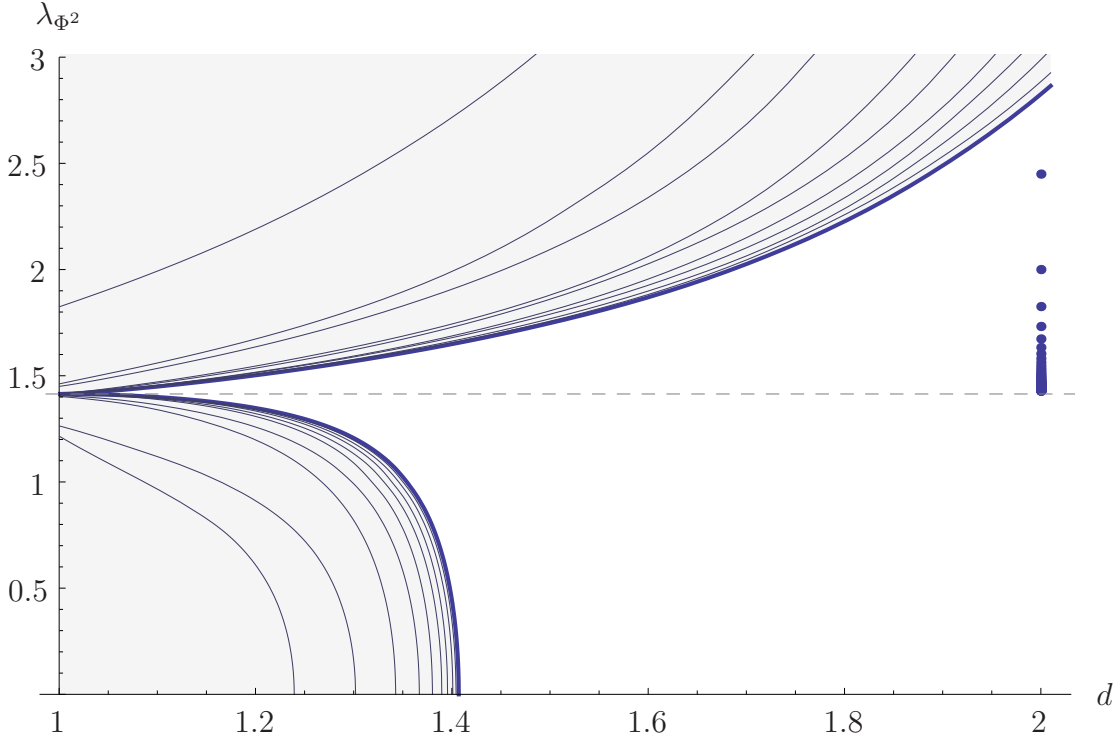}
\end{psfrags}
\end{center}
\caption{Upper and lower bounds on the OPE coefficient of $\Phi^2$ in $\Phi\x\Phi$, as a function of $d=\dim(\Phi)$.  The dashed line indicates the free value $\l_{\Phi^2}=\sqrt 2$.  The points shown at $d=2$ indicate the sequence of values $\lambda_{\Phi^2} = \sqrt{2 + \frac{4}{N}}$ realized for composite operators in free theories.  We give the bounds for $k=2,\ldots,11$.}
\label{fig:phi2OPE}
\end{figure}

In figure~\ref{fig:phi2OPE} we show the resulting upper and lower bounds on $\l_{\Phi^2}$, where we have taken $k=2,\ldots,11$ in the numerical optimization.  We can see that the strongest bounds are extremely constraining when $d=\dim(\Phi)$ is even somewhat close to 1, forcing $\l_{\Phi^2}$ to live very close to the free value $\l_{\Phi^2} = \sqrt{2}$. In particular, these results imply that it should not be possible to construct a weakly-coupled (Banks-Zaks) SCFT where both $d$ and $\l_{\Phi^2}$ are modified at the one-loop level.  Indeed, in all constructible examples $\l_{\Phi^2}$ receives its leading correction at second order in perturbation theory. On the other hand, we see that the lower bound disappears before $d = 3/2$, as expected, while the upper bound persists.

As $d \rightarrow 2$, we may also compare the upper bound to the OPE coefficients of composite operators in theories containing free chiral superfields.  In the simplest case, we can consider a single free field $Q$ and then identify $\Phi \equiv \frac{1}{\sqrt{2}} Q^2$.  In this case the operator $\Phi^2 \equiv \frac{1}{\sqrt{4!}} Q^4$ is canonically normalized, so the OPE is
\be
\left(\frac{1}{\sqrt{2}} Q^2 \right) \times \left(\frac{1}{\sqrt{2}} Q^2 \right) \sim \frac{\sqrt{4!}}{2} \left(\frac{1}{\sqrt{4!}} Q^4 \right) + \ldots,
\ee
and we have $\lambda_{\Phi^2} = \sqrt{6}$, which is consistent with the bound.  More generally, considering the dimension-$n$ operator $\Phi \equiv \frac{1}{\sqrt{n!}} Q^n$ leads to an OPE coefficient of $\lambda_{\Phi^2} = \frac{(2n)!^{1/2}}{n!}$, which the bound must respect at even higher integer values of $d$.

Another simple generalization is to consider meson operators $M \equiv \frac{1}{\sqrt{2 N}} Q^i Q_i$ built out of $N$ free quarks $Q^i$.  In this case Wick contractions give a two-point function $\<(M^2) (M^2)^\dagger\> \sim 2+\frac{4}{N}$, so the OPE in terms of canonically normalized operators is given by
\be
M \times M \sim \sqrt{2+\frac{4}{N}} \left(\frac{1}{\sqrt{2+\frac{4}{N}}} M^2 \right) + \ldots.
\ee
Thus, we can read off an OPE coefficient of $\lambda_{\Phi^2} = \sqrt{2+4/N}$, which is consistent with our bound for all values of $N$. It is interesting to see that while OPE coefficients of composite operators with $d \sim 2$ know about the underlying constituents of the operator, as $d \rightarrow 1$ the OPE coefficient necessarily loses memory of where the operator came from.  Indeed, free operators have no hair!

\begin{figure}[h!]
\begin{center}
\begin{psfrags}
\def\PFGstripminus-#1{#1}%
\def\PFGshift(#1,#2)#3{\raisebox{#2}[\height][\depth]{\hbox{%
  \ifdim#1<0pt\kern#1 #3\kern\PFGstripminus#1\else\kern#1 #3\kern-#1\fi}}}%
\providecommand{\PFGstyle}{}%
%
\psfrag{d}[cl][cl]{\PFGstyle $d$}%
\psfrag{lam}[bc][bc]{\PFGstyle $\l_{\bar Q\cO_\ell}$}%
\psfrag{OPEcoeffic}[bc][bc]{\PFGstyle $\text{Upper and lower bounds on $\l_{\bar Q\cO_\ell}$, $\ell=2,4,\dots,10$}$}%
\psfrag{x111}[tc][tc]{\PFGstyle $1.1$}%
\psfrag{x11}[tc][tc]{\PFGstyle $1$}%
\psfrag{x121}[tc][tc]{\PFGstyle $1.2$}%
\psfrag{x131}[tc][tc]{\PFGstyle $1.3$}%
\psfrag{x141}[tc][tc]{\PFGstyle $1.4$}%
\psfrag{x151}[tc][tc]{\PFGstyle $1.5$}%
\psfrag{x161}[tc][tc]{\PFGstyle $1.6$}%
\psfrag{x171}[tc][tc]{\PFGstyle $1.7$}%
\psfrag{x181}[tc][tc]{\PFGstyle $1.8$}%
\psfrag{x21}[tc][tc]{\PFGstyle $2$}%
\psfrag{yNumberFor}[cr][cr]{\PFGstyle $0.001$}%
\psfrag{yNumberForA}[cr][cr]{\PFGstyle $0.005$}%
\psfrag{yNumberForB}[cr][cr]{\PFGstyle $0.01$}%
\psfrag{yNumberForC}[cr][cr]{\PFGstyle $0.05$}%
\psfrag{yNumberForD}[cr][cr]{\PFGstyle $0.1$}%
\psfrag{yNumberForE}[cr][cr]{\PFGstyle $0.5$}%
\psfrag{yNumberForF}[cr][cr]{\PFGstyle $1$}%
\includegraphics[width=0.9\textwidth]{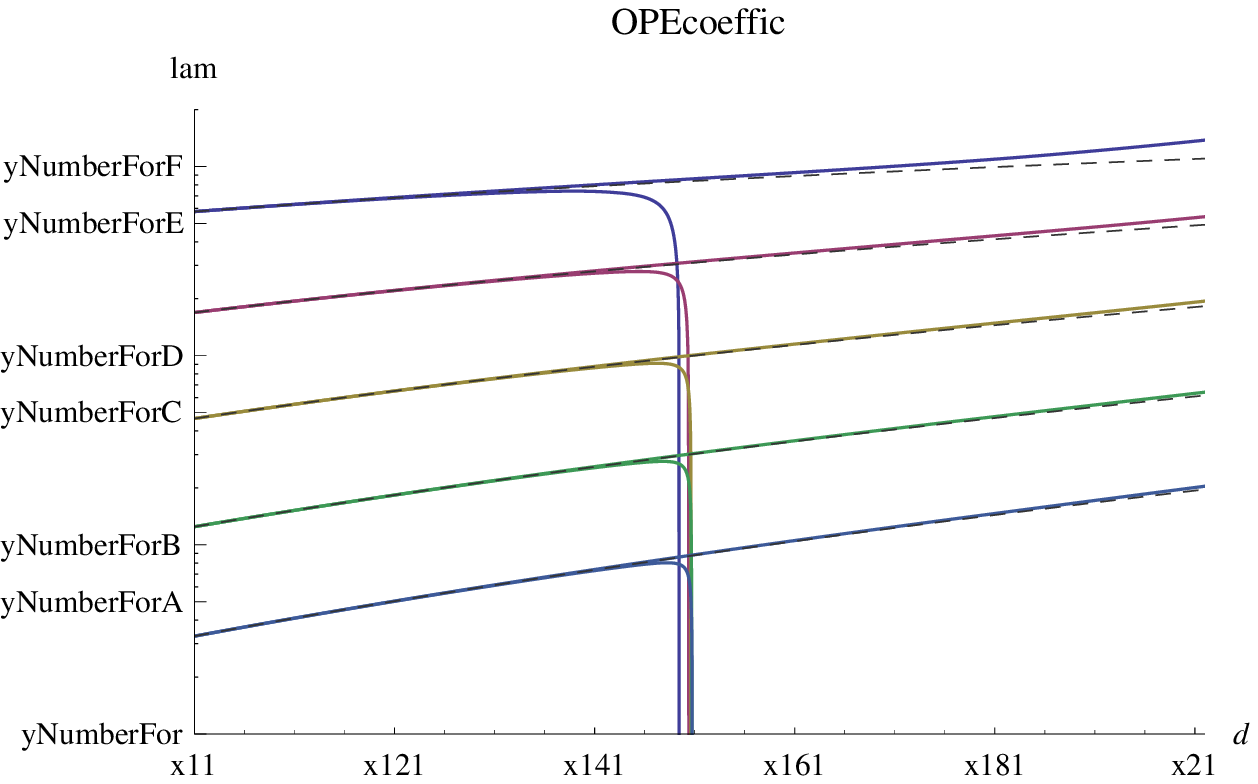}
\end{psfrags}
\end{center}
\caption{Upper and lower bounds on the OPE coefficients of protected operators $\bar Q\cO_\ell$ appearing in $\Phi\x\Phi$, along with their mean field theory values Eq.~(\ref{eq:MFTOPEcoefficients}) (dashed lines), for $\ell=2,4,\dots,10$.  Each curve goes continuously to the free value $\sqrt{2}\frac{\ell!}{(2\ell)!^{1/2}}$ at $d=1$.  All lower bounds vanish at $d=3/2$, since the gap in dimensions between $\bar Q\cO_\ell$ and non-protected operators disappears at that point.  Here we have taken $k=11$.}
\label{fig:SUSYprotectedOPE}
\end{figure}

In figure~\ref{fig:SUSYprotectedOPE} we extend these upper and lower bounds to OPE coefficients of the other protected operators $\bar{Q} \cO_{\ell}$ appearing in the $\Phi \times \Phi$ OPE.  Here we give the results for $\ell = 2,\ldots,10$ and have taken $k=11$ in the numerical optimization (though similar bounds also exist at larger values of $\ell$).  All of the bounds continuously interpolate to the free values as $d \rightarrow 1$, given by $\lambda_{\bar{Q} \cO_{\ell}} = \sqrt{2} \frac{\ell!}{(2\ell)!^{1/2}}$.  Notice that all lower bounds vanish before $d=3/2$, as they should.\footnote{Once they are computed, one can include information about these lower bounds in semidefinite programs for other quantities, like e.g. upper bounds on OPE coefficients of operators in the $\Phi\x\Phi^\dag$ OPE.  We found that this procedure does not significantly improve the results in practice.}

Taken together, the upper and lower bounds on $\l_{\bar Q\cO_\ell}$ are extremely strong, almost determining this coefficient when $d\lesssim 1.4$.   One can view this singling out of an essentially unique OPE coefficient as a remarkable success of the 4D conformal bootstrap program!  It is worth comparing the bounds to the known values of $\l_{\bar Q\cO_\ell}$ in supersymmetric mean field theories (MFTs), which occur in the planar limit of large-$N$ gauge theories.  There, the role of $\bar Q\cO_\ell$ is played by the `twist-$2d$' double-trace operators
\be
\cO^{(2)}_{\ell} &\equiv& \Phi\,\lrptl^{\mu_1}\cdots\lrptl^{\mu_\ell}\Phi-\textrm{traces},
\ee
with even spin $\ell$.  Their (squared) OPE coefficients in $\Phi\x\Phi$ are given by~\cite{Heemskerk:2009pn}
\be
\label{eq:MFTOPEcoefficients}
\l_{\cO^{(2)}_{\ell}}^2 &=& \frac{2\Gamma^2(d+\ell) \Gamma(2d+\ell-1)}{\Gamma^2(d) \Gamma(\ell+1)\Gamma(2d+2\ell-1)},
\ee
and these values of $\l_{\cO^{(2)}_{\ell}}$ are shown as dashed lines in figure~\ref{fig:SUSYprotectedOPE}, for $\ell=2,4,\dots,10$.  They are fully consistent with both our upper and lower bounds on $\l_{\bar Q\cO_\ell}$.  Note that the MFT value of $\l_{\cO^{(2)}_0}$ is equal to the free value $\sqrt2$, so it is consistent with our bounds in figure~\ref{fig:phi2OPE}.  

The striking agreement between our bounds and the mean field theory values of OPE coefficients at small $d$ has interesting implications for SCFTs with weakly-coupled AdS$_5$ duals.  In such theories, corrections to OPE coefficients away from their MFT values can be computed in perturbation theory using Witten diagrams.  Our bounds imply that corrections to $\l_{\cO^{(2)}_{\ell}}^2$ must vanish to very high order in $(d-1)$, particularly at large $\ell$.  If any corrections were nonzero at finite values of $(d-1)$, then we would obtain sharp bounds on bulk coupling constants.  We defer further exploration of these interesting constraints to future work.

\section{Bounds on Central Charges}
\label{sec:ccbounds}

In this section we explore bounds on the OPE coefficient appearing in front of the stress tensor $T^{\mu\nu}$, which is a conserved spin-$2$ operator of dimension $4$ that must be present in any CFT.  Since this OPE coefficient is fixed by a Ward identity in terms of the central charge $c$ of the theory (defined as the coefficient appearing in the two-point function $\<T^{\mu\nu} T^{\g\de} \> \propto c$), we will ultimately be deriving bounds on $c$.  Previously, lower bounds on the central charge in both general CFTs and SCFTs were explored in~\cite{Poland:2010wg,Rattazzi:2010gj,Vichi:2011ux}.  The main new results of this section will be to extend these analyses to situations with global symmetries, where we will show that there are bounds on the central charge that scale with the size of the global symmetry representation.  

\subsection{General Theories}

Let us begin by establishing some notation.  The stress tensor is typically normalized as
\be
\< T^{\mu\nu}(x)T^{\g\de}(0) \> &=& \frac{40 c}{\pi^4} \frac{I^{\mu\g}(x) I^{\nu\de}(x)}{x^8},
\ee
where $I^{\mu\g}(x) = \eta^{\mu\g} - 2 \frac{x^{\mu} x^{\g}}{x^2}$ and $c$ is the central charge appearing in the trace anomaly, $\<T^{\mu}_{\mu} \> = \frac{c}{16\pi^2} ( \textrm{Weyl} )^2 - \frac{a}{16\pi^2} ( \textrm{Euler} )$, when the theory is placed on a curved background.  In this normalization a free scalar has $c_{\textrm{free}}=\frac{1}{120}$ and a free Weyl fermion has $c_{\textrm{free fermion}}=\frac{1}{40}$.  

The stress tensor is the local current generating the dilatation charge, where in radial quantization $D = - \int d \Omega\, \hat{x}_{\mu} x_{\nu} T^{\mu\nu}$ (the integral is over a three-sphere surrounding the origin).  Requiring the action $D \phi(0) = d \phi(0)$ then fixes the OPE to have the form $T^{\mu\nu}(x) \phi(0) \sim - \frac{2d}{3\pi^2 x^6} \left(x^{\mu} x^{\nu} - \frac14 \eta^{\mu\nu} x^2 \right) \phi(0) + \ldots$, which leads to the stress tensor conformal block contribution
\be\label{eq:centralchargeconfblock}
x_{12}^{2d} x_{34}^{2d} \<\phi\phi\phi\phi\> \sim \frac{d^2}{360c} g_{4,2} \qquad \textrm{(general CFTs)} .
\ee
Generalizing to the situation where $\phi^i$ transforms under an $\SO(N)$ or $\SU(N)$ global symmetry, the stress tensor appears as an $S^+$ operator in the sum rules given in Eqs.~(\ref{eq:SONvectorialsumrule}) and~(\ref{eq:SUNsumrule}), again with OPE coefficient $\l_{T}^2 = \frac{d^2}{360c}$.  Note that a free real scalar transforming as an $\SO(N)$ fundamental or a complex scalar transforming as an $\SU(N/2)$ fundamental gives a contribution of $N c_{\textrm{free}}$ to the central charge.

\begin{figure}[h!]
\begin{center}
\begin{psfrags}
\def\PFGstripminus-#1{#1}%
\def\PFGshift(#1,#2)#3{\raisebox{#2}[\height][\depth]{\hbox{%
  \ifdim#1<0pt\kern#1 #3\kern\PFGstripminus#1\else\kern#1 #3\kern-#1\fi}}}%
\providecommand{\PFGstyle}{}%
%
\psfrag{Centralcha}[bc][bc]{\PFGstyle $\text{Lower bound on $c$ for a real scalar}$}%
\psfrag{d}[cl][cl]{\PFGstyle $\text{d}$}%
\psfrag{minc}[bc][bc]{\PFGstyle $c$}%
\psfrag{x11}[tc][tc]{\PFGstyle $1$}%
\psfrag{x121}[tc][tc]{\PFGstyle $1.2$}%
\psfrag{x141}[tc][tc]{\PFGstyle $1.4$}%
\psfrag{x161}[tc][tc]{\PFGstyle $1.6$}%
\psfrag{x181}[tc][tc]{\PFGstyle $1.8$}%
\psfrag{y0}[cr][cr]{\PFGstyle $0$}%
\psfrag{y12m1}[cr][cr]{\PFGstyle $0.012$}%
\psfrag{y14m1}[cr][cr]{\PFGstyle $0.014$}%
\psfrag{y16m1}[cr][cr]{\PFGstyle $0.016$}%
\psfrag{y1m1}[cr][cr]{\PFGstyle $0.01$}%
\psfrag{y2m2}[cr][cr]{\PFGstyle $0.002$}%
\psfrag{y4m2}[cr][cr]{\PFGstyle $0.004$}%
\psfrag{y6m2}[cr][cr]{\PFGstyle $0.006$}%
\psfrag{y8m2}[cr][cr]{\PFGstyle $0.008$}%
\includegraphics[width=1\textwidth]{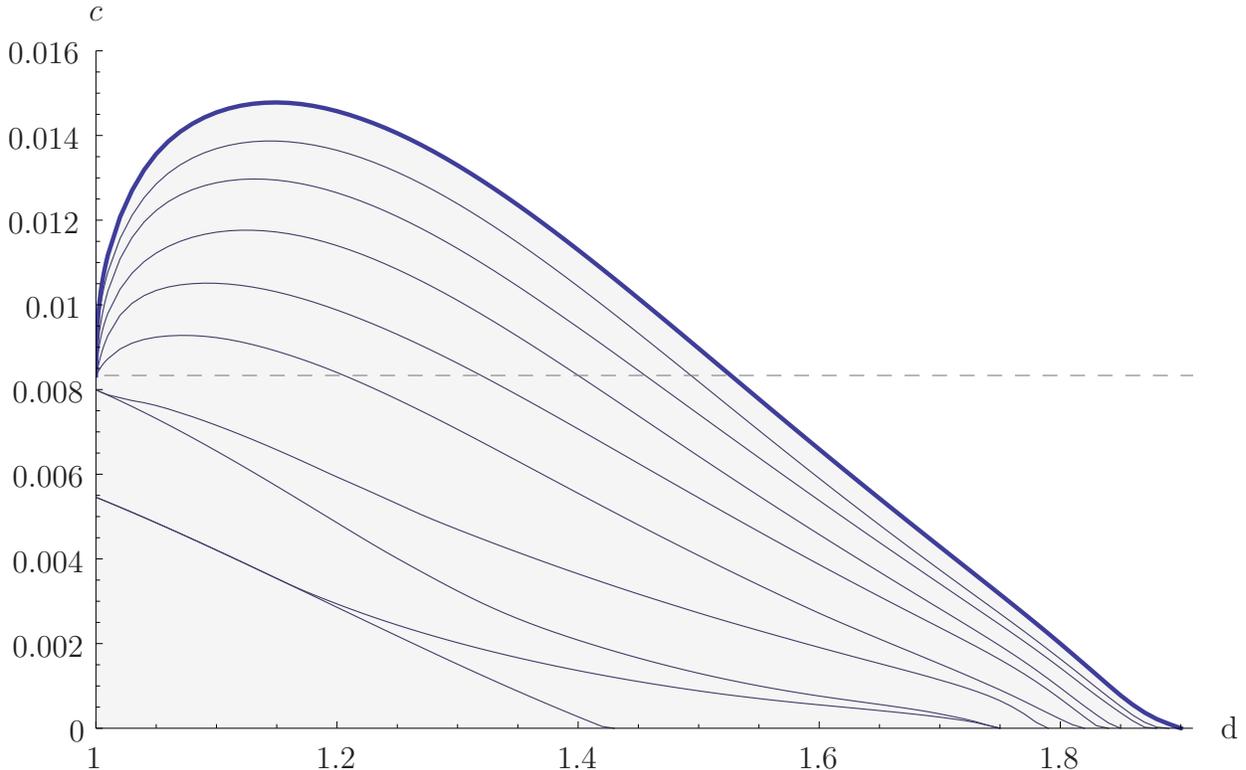}
\end{psfrags}
\end{center}
\caption{A lower bound on the central charge of a theory containing a scalar $\f$ of dimension $d$.  The dashed line indicates the value $c_{\text{free}}=1/120$, corresponding to the central charge of a free scalar. Here we show bounds for the values $k=2,\ldots,11$.}
\label{fig:RealscalarCCbound}
\end{figure}

To begin, in figure~\ref{fig:RealscalarCCbound} we show the bounds on $c$ obtained by applying our semidefinite programming algorithm to the case of a single real scalar $\phi$, where we show curves for $k=2,\ldots,11$ in the numerical optimization.  We see that for $k\geq 6$, the bounds smoothly approach the free value $c_{\textrm{free}}$ as $d \rightarrow 1$.  This is consistent with and improves upon the bounds on $c$ previously presented in~\cite{Poland:2010wg,Rattazzi:2010gj}.  Note that here we are only assuming that the dimensions of operators appearing in the $\phi \times \phi$ OPE satisfy the unitarity bound --- one could also assume that $\phi$ is the lowest dimension scalar in the theory to obtain somewhat stronger bounds at larger values of $d$ as was done in~\cite{Rattazzi:2010gj}.  However, here we make only the minimal assumption to allow for a more straightforward comparison to our other bounds.

\begin{figure}[h!]
\begin{center}
\begin{psfrags}
\def\PFGstripminus-#1{#1}%
\def\PFGshift(#1,#2)#3{\raisebox{#2}[\height][\depth]{\hbox{%
  \ifdim#1<0pt\kern#1 #3\kern\PFGstripminus#1\else\kern#1 #3\kern-#1\fi}}}%
\providecommand{\PFGstyle}{}%
%
\psfrag{centralcha}[bc][bc]{\PFGstyle $\text{Lower bounds on $c$ for $\SO(N)$ or $\SU(N/2)$, $N=2,\dots,14$}$}%
\psfrag{d}[cl][cl]{\PFGstyle $d$}%
\psfrag{minc}[bc][bc]{\PFGstyle $c$}%
\psfrag{x111}[tc][tc]{\PFGstyle $1.1$}%
\psfrag{x11}[tc][tc]{\PFGstyle $1$}%
\psfrag{x121}[tc][tc]{\PFGstyle $1.2$}%
\psfrag{x131}[tc][tc]{\PFGstyle $1.3$}%
\psfrag{x141}[tc][tc]{\PFGstyle $1.4$}%
\psfrag{x151}[tc][tc]{\PFGstyle $1.5$}%
\psfrag{x161}[tc][tc]{\PFGstyle $1.6$}%
\psfrag{x171}[tc][tc]{\PFGstyle $1.7$}%
\psfrag{y0}[cr][cr]{\PFGstyle $0$}%
\psfrag{y10cfree}[cr][cr]{\PFGstyle $10 c_{\text{free}}$}%
\psfrag{y12cfree}[cr][cr]{\PFGstyle $12 c_{\text{free}}$}%
\psfrag{y14cfree}[cr][cr]{\PFGstyle $14 c_{\text{free}}$}%
\psfrag{y16cfree}[cr][cr]{\PFGstyle $16 c_{\text{free}}$}%
\psfrag{y18cfree}[cr][cr]{\PFGstyle $18 c_{\text{free}}$}%
\psfrag{y20cfree}[cr][cr]{\PFGstyle $20 c_{\text{free}}$}%
\psfrag{y2cfree}[cr][cr]{\PFGstyle $2 c_{\text{free}}$}%
\psfrag{y4cfree}[cr][cr]{\PFGstyle $4 c_{\text{free}}$}%
\psfrag{y6cfree}[cr][cr]{\PFGstyle $6 c_{\text{free}}$}%
\psfrag{y8cfree}[cr][cr]{\PFGstyle $8 c_{\text{free}}$}%
\psfrag{yNull}[cr][cr]{\PFGstyle $\text{}$}%
\includegraphics[width=1\textwidth]{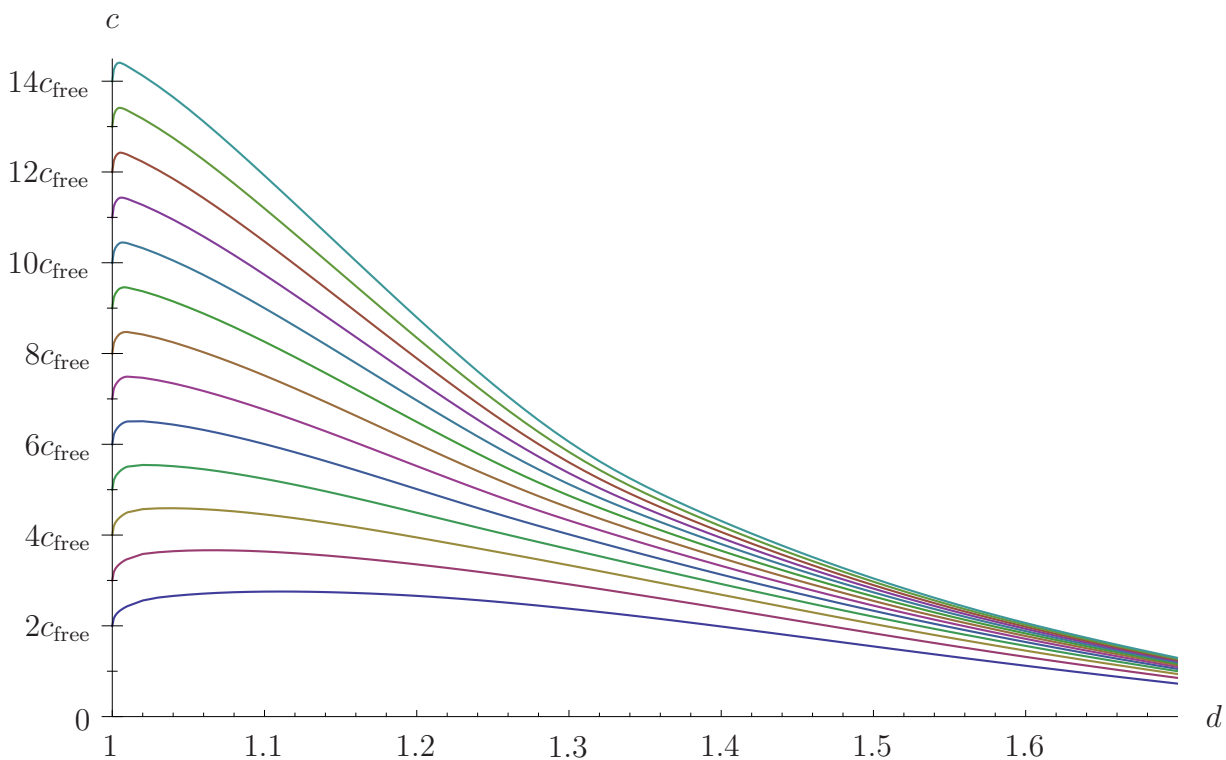}
\end{psfrags}
\end{center}
\caption{A lower bound on the central charge of a theory containing a scalar $\f_i$ of dimension $d$ transforming as a fundamental of an $\SO(N)$ or $\SU(N/2)$ global symmetry, for $N=2,\dots,14$.  In this plot $c_{\text{free}}=1/120$, corresponding to the central charge of a free scalar. Here we have taken $k=11$.}
\label{fig:SONcentralcharge}
\end{figure}

In figure~\ref{fig:SONcentralcharge} we show bounds on $c$ in the presence of $\SO(N)$ or $\SU(N/2)$ global symmetries for $N=2,\ldots,14$.  Here have taken $k=11$.  We see that the bounds also smoothly approach the free values as $d \rightarrow 1$, scaling linearly with $N$.  This greatly improves upon the bounds derived in~\cite{Poland:2010wg,Rattazzi:2010gj} (and given in figure~\ref{fig:RealscalarCCbound}) for theories with global symmetries.  The reason for the improvement is that here we have incorporated the constraints of crossing symmetry for {\it all} operators in the $\phi^i$ multiplet; without doing this the bounds of~\cite{Poland:2010wg,Rattazzi:2010gj} could not differentiate between the stress tensor and other spin 2 operators (e.g., the $\SO(N)$ symmetric tensor $\phi^{(i} \partial^{\mu} \partial^{\nu} \phi^{j)}$) that have an $O(1)$ OPE coefficient in the $d\rightarrow 1$ limit.

It is interesting to understand the implications of the bound of figure~\ref{fig:SONcentralcharge} for the AdS/CFT correspondence.  For theories with an AdS$_5$ dual description, the bulk Planck scale is proportional to $c$, the bulk gauge group is identified with the $\SO(N)$ or $\SU(N/2)$ global symmetry, and $d$ is related to the masses of bulk fields.  Our bound then says that theories with sufficiently light bulk excitations cannot have a gravitational scale that is arbitrarily small.  Moreover, if those fields transform as fundamentals under the bulk $\SO(N)$ or $\SU(N/2)$ gauge group (and correspond to operators with $d \sim 1$), then the Planck scale must scale at least linearly with $N$.  

It would be fascinating to identify CFTs that live close to these bounds, particularly in the large $N$ limit.  Unfortunately, in gauge theories believed to flow to conformal fixed points that also posses an $\SO(N)$ or $\SU(N/2)$ global symmetry, the central charge typically scales as $N^2$, at least near $d \sim 1$.  The reason is that conformality forces the size of the global symmetry to scale proportionally to the size of the gauge group, and gauge degrees of freedom live in adjoint representations of the gauge group which have $O(N^2)$ components.  We will see examples of this in the next subsection, where we extend the bounds to superconformal theories in which $c$ is explicitly calculable. 

\subsection{Superconformal Theories}
In $\cN=1$ SCFTs, the stress tensor is a superconformal descendant of the spin-$1$ $U(1)_R$ current, $T\sim (Q\bar Q J_R)_{\ell+1}$, as in Eq.~(\ref{eq:phiphidaggersuperOPE}).  Applying Eq.~(\ref{eq:superconformalrelations}) to~(\ref{eq:centralchargeconfblock}), we see that $J^{\mu}_R$ has an OPE coefficient of $\l_{R}^2 = \frac{d^2}{72 c}$, appearing as an $S^+$ operator in the superconformal sum rules of Eqs.~(\ref{eq:SCFTcrossing}) and~(\ref{eq:SUSYSUNsumrule}).  Since a free chiral superfield contains both a complex scalar and a Weyl fermion, it gives a contribution of $c_{\textrm{chiral}} = 2 \times \frac{1}{120} + \frac{1}{40} = \frac{1}{24}$.

\begin{figure}[h!]
\begin{center}
\begin{psfrags}
\def\PFGstripminus-#1{#1}%
\def\PFGshift(#1,#2)#3{\raisebox{#2}[\height][\depth]{\hbox{%
  \ifdim#1<0pt\kern#1 #3\kern\PFGstripminus#1\else\kern#1 #3\kern-#1\fi}}}%
\providecommand{\PFGstyle}{}%
%
\psfrag{d}[cl][cl]{\PFGstyle $d$}%
\psfrag{minc}[bc][bc]{\PFGstyle $c$}%
\psfrag{SUSYU1Cent}[bc][bc]{\PFGstyle $\text{Lower bound on $c$ for a chiral scalar}$}%
\psfrag{x11}[tc][tc]{\PFGstyle $1$}%
\psfrag{x121}[tc][tc]{\PFGstyle $1.2$}%
\psfrag{x141}[tc][tc]{\PFGstyle $1.4$}%
\psfrag{x161}[tc][tc]{\PFGstyle $1.6$}%
\psfrag{x181}[tc][tc]{\PFGstyle $1.8$}%
\psfrag{y0}[cr][cr]{\PFGstyle $0$}%
\psfrag{y2m1}[cr][cr]{\PFGstyle $0.02$}%
\psfrag{y4m1}[cr][cr]{\PFGstyle $0.04$}%
\psfrag{y6m1}[cr][cr]{\PFGstyle $0.06$}%
\psfrag{y8m1}[cr][cr]{\PFGstyle $0.08$}%
\includegraphics[width=0.9\textwidth]{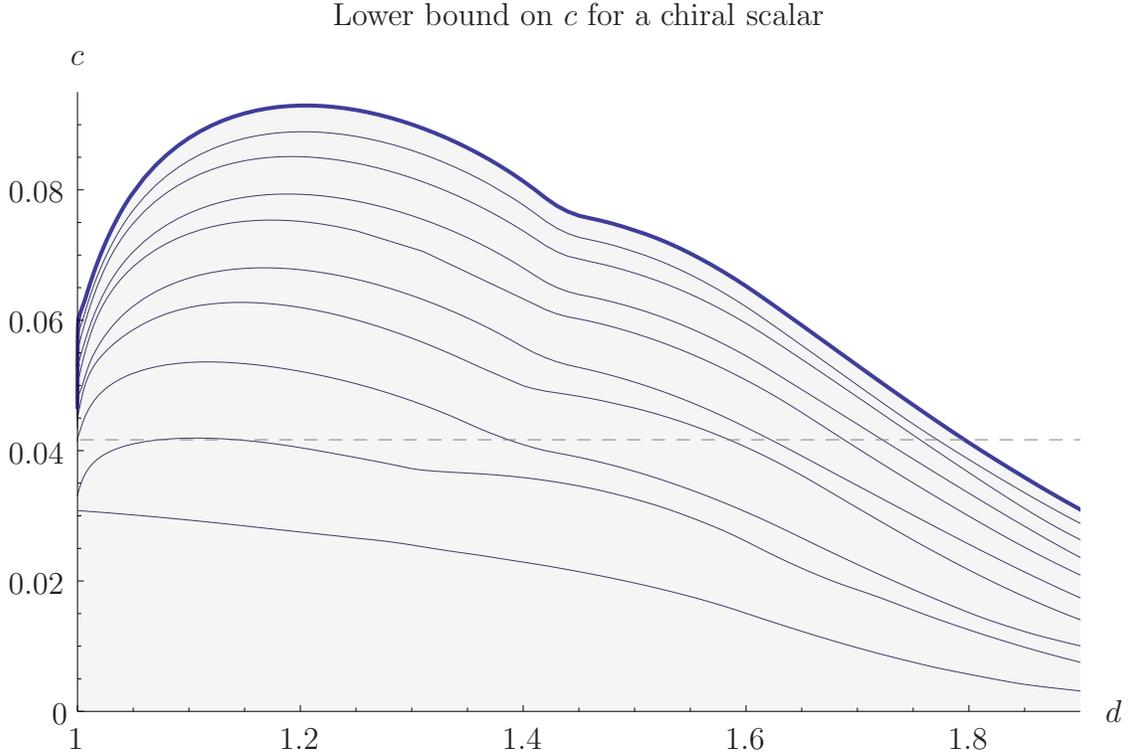}
\end{psfrags}
\end{center}
\caption{A lower bound on the central charge of any SCFT containing a chiral scalar $\Phi$ of dimension $d$.  The dashed line is at $c_{\textrm{chiral}}=1/24$, corresponding to the central charge of a free chiral superfield ($d=1$). Despite appearances at this zoom level, all the curves above drop sharply near $d=1$ and interpolate smoothly to the free value.  In this plot we have taken $k=2,\ldots,11$.}
\label{fig:SUSYU1centralcharge}
\end{figure}

\begin{figure}[h!]
\begin{center}
\begin{psfrags}
\def\PFGstripminus-#1{#1}%
\def\PFGshift(#1,#2)#3{\raisebox{#2}[\height][\depth]{\hbox{%
  \ifdim#1<0pt\kern#1 #3\kern\PFGstripminus#1\else\kern#1 #3\kern-#1\fi}}}%
\providecommand{\PFGstyle}{}%
%
\psfrag{dm1}[cl][cl]{\PFGstyle $d-1$}%
\psfrag{minc}[bc][bc]{\PFGstyle $c$}%
\psfrag{SUSYU1Cent}[bc][bc]{\PFGstyle $\text{Lower bound on $c$ for a chiral scalar}$}%
\psfrag{x1m1}[tc][tc]{\PFGstyle $0.01$}%
\psfrag{x1}[tc][tc]{\PFGstyle $1$}%
\psfrag{xSuperscriA}[tc][tc]{\PFGstyle $10^{-6}$}%
\psfrag{xSuperscriB}[tc][tc]{\PFGstyle $10^{-4}$}%
\psfrag{xSuperscri}[tc][tc]{\PFGstyle $10^{-8}$}%
\psfrag{y0}[cr][cr]{\PFGstyle $0$}%
\psfrag{y1}[cr][cr]{\PFGstyle $0.1$}%
\psfrag{y2m1}[cr][cr]{\PFGstyle $0.02$}%
\psfrag{y4m1}[cr][cr]{\PFGstyle $0.04$}%
\psfrag{y6m1}[cr][cr]{\PFGstyle $0.06$}%
\psfrag{y8m1}[cr][cr]{\PFGstyle $0.08$}%
\includegraphics[width=0.9\textwidth]{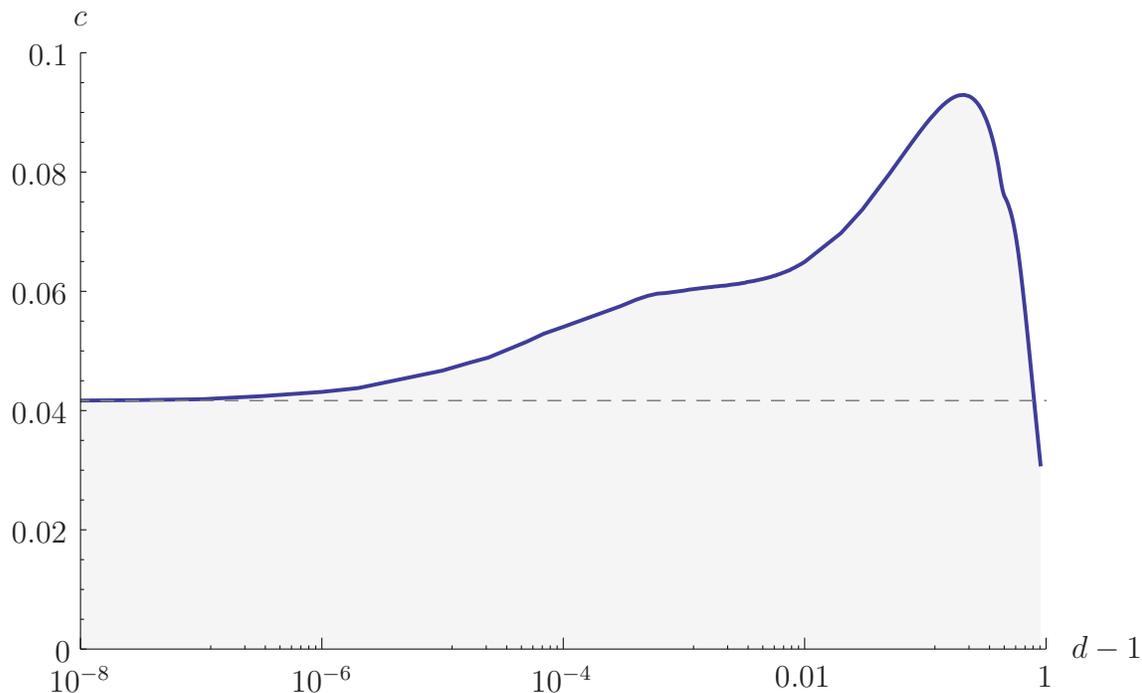}
\end{psfrags}
\end{center}
\caption{The $k=11$ curve of figure~\ref{fig:SUSYU1centralcharge}, where $(d-1)$ has been placed on a logarithmic scale. The bound smoothly approaches the free value $c_{\textrm{chiral}}=1/24$ very close to $d=1$.}
\label{fig:SUSYU1centralchargezoom}
\end{figure}

In figure~\ref{fig:SUSYU1centralcharge} we show the results of our semidefinite programming algorithm for obtaining bounds on the central charge of any theory containing a chiral scalar $\Phi$.  We give the results for $k=2,\ldots,11$, where all of the curves for $k>3$ drop sharply very close to $d\sim1$ and go just below the free value.  The $k=11$ curve significantly improves upon SCFT central charge bounds previously obtained in~\cite{Poland:2010wg,Vichi:2011ux}.  Note that the sharpness of the drop (reaching within $1\%$ of the free chiral value closer than $d\sim 1.0000002$) is strong evidence that the free theory is an isolated solution to the crossing relations.  This is intuitive from the perspective of constructing perturbations of the free theory --- all such perturbations leading to an interacting SCFT require additional matter, which increases the central charge.  In order to demonstrate that the bound does in fact approach the free value, in figure~\ref{fig:SUSYU1centralchargezoom} we also show the bound for $k=11$ where $(d-1)$ has been placed on a logarithmic scale. 

\begin{figure}[h!]
\begin{center}
\begin{psfrags}
\def\PFGstripminus-#1{#1}%
\def\PFGshift(#1,#2)#3{\raisebox{#2}[\height][\depth]{\hbox{%
  \ifdim#1<0pt\kern#1 #3\kern\PFGstripminus#1\else\kern#1 #3\kern-#1\fi}}}%
\providecommand{\PFGstyle}{}%
%
\psfrag{d}[cl][cl]{\PFGstyle $d$}%
\psfrag{minc}[bc][bc]{\PFGstyle $c$}%
\psfrag{SUSYSUNCen}[bc][bc]{\PFGstyle $\text{Lower bounds on $c$ for a SUSY $\SU(N)$ chiral scalar, $N=2,\dots,14$}$}%
\psfrag{x11}[tc][tc]{\PFGstyle $1$}%
\psfrag{x121}[tc][tc]{\PFGstyle $1.2$}%
\psfrag{x141}[tc][tc]{\PFGstyle $1.4$}%
\psfrag{x161}[tc][tc]{\PFGstyle $1.6$}%
\psfrag{x181}[tc][tc]{\PFGstyle $1.8$}%
\psfrag{y0}[cr][cr]{\PFGstyle $0$}%
\psfrag{y10cfree}[cr][cr]{\PFGstyle $10 c_{\text{chiral}}$}%
\psfrag{y12cfree}[cr][cr]{\PFGstyle $12 c_{\text{chiral}}$}%
\psfrag{y14cfree}[cr][cr]{\PFGstyle $14 c_{\text{chiral}}$}%
\psfrag{y16cfree}[cr][cr]{\PFGstyle $16 c_{\text{chiral}}$}%
\psfrag{y18cfree}[cr][cr]{\PFGstyle $18 c_{\text{chiral}}$}%
\psfrag{y20cfree}[cr][cr]{\PFGstyle $20 c_{\text{chiral}}$}%
\psfrag{y2cfree}[cr][cr]{\PFGstyle $2 c_{\text{chiral}}$}%
\psfrag{y4cfree}[cr][cr]{\PFGstyle $4 c_{\text{chiral}}$}%
\psfrag{y6cfree}[cr][cr]{\PFGstyle $6 c_{\text{chiral}}$}%
\psfrag{y8cfree}[cr][cr]{\PFGstyle $8 c_{\text{chiral}}$}%
\psfrag{yNull}[cr][cr]{\PFGstyle $\text{}$}%
\includegraphics[width=0.9\textwidth]{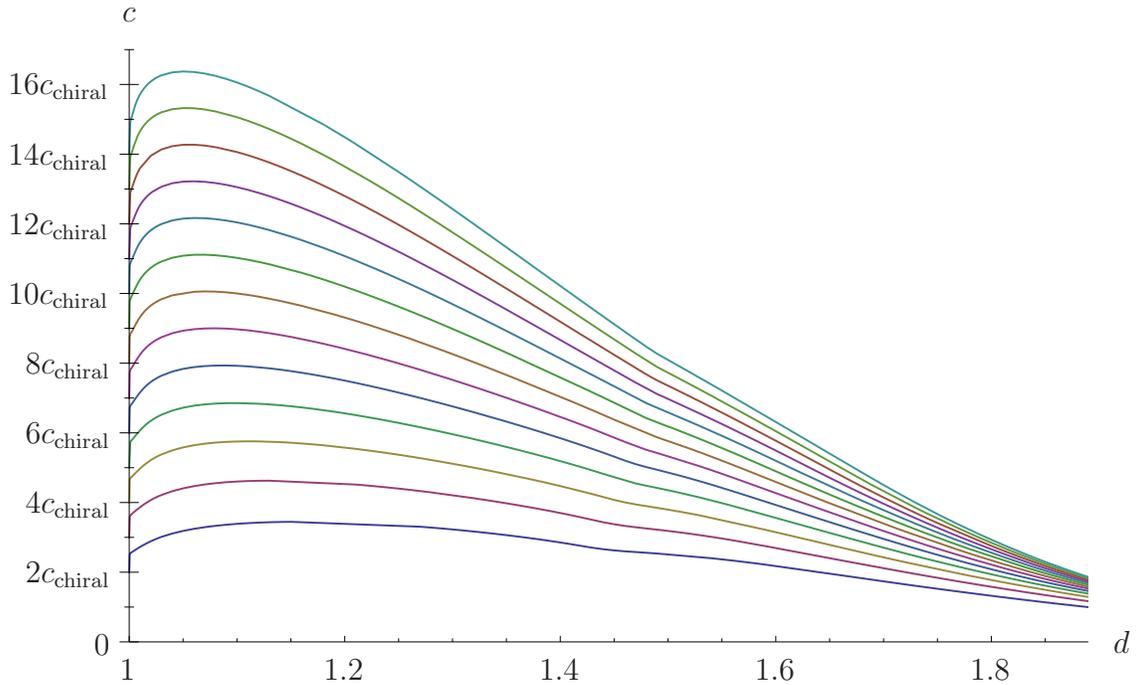}
\end{psfrags}
\end{center}
\caption{A lower bound on the central charge of any SCFT containing a chiral scalar $\Phi_i$ of dimension $d$ transforming as a fundamental of an $\SU(N)$ global symmetry, for $N=2,\dots,14$.  Here $c_{\text{chiral}}=1/24$ denotes the contribution to $c$ from a free chiral superfield.  Despite appearances at this zoom level, all the curves drop sharply very close to $d=1$ and interpolate continuously to the free values. In this plot we have taken $k=10$.}
\label{fig:SUSYSUNcentralcharge}
\end{figure}

We extend these bounds to the situation where $\Phi^i$ transforms as a fundamental under an $\SU(N)$ global symmetry in figure~\ref{fig:SUSYSUNcentralcharge}, where we have taken $k=10$ and show curves for $N=2,\ldots,14$.  All the curves interpolate to the free values $N c_{\text{chiral}}$ as $d \rightarrow 1$, in all cases with a very sharp drop in the bound close to $1$.  Again we see that the bounds scale linearly with $N$, and moreover the linear behavior extends out to larger values of $d$ compared to the non-supersymmetric bounds of figure~\ref{fig:SONcentralcharge}.

Let us now take a moment to compare these bounds to some concrete SCFTs.  The reason that such a comparison is possible is that both $d$ and $c$ are calculable in terms of the $U(1)_R$ symmetry --- $d$ is calculable because the dimensions of chiral superconformal primary operators are related to their $R$ charge as $d=\frac{3}{2} R$, and $c$ is calculable via 't Hooft anomaly matching using the relation $c =\frac{1}{32} ( 9 \Tr R^3 - 5 \Tr R )$ \cite{Anselmi:1997am,Anselmi:1997ys}.  The $U(1)_R$ symmetry can then often be determined using symmetry arguments, or more generally using $a$-maximization~\cite{Intriligator:2003jj}.

One of the simplest $\cN=1$ SCFTs is supersymmetric QCD with gauge group $\SU(N_c)$ and $N_f$ flavors of quarks $Q,\bar{Q}$ in the conformal window $\frac{3}{2} N_c \leq N_f \leq 3 N_c$~\cite{Seiberg:1994pq}.  In this case the gauge-invariant mesons $M=Q\tl{Q}$ have $d_M = 3(1-N_c/N_f)$, while the central charge is evaluated as $c = \frac{1}{16} (7 N_c^2 - 9 N_c^4/N_f^2 - 2)$.  The mesons are bi-fundamentals under the $\SU(N_f) \times \SU(N_f)$ symmetry group, so our bounds will apply by considering either of these groups.  

However, we immediately see that the central charge in SQCD grows like $O(N^2)$, so theories at large values of $N_f \sim N_c$ trivially satisfy the bounds.  On the other hand, all of the small $N$ theories still have a central charge larger than $1 = 24 c_{\text{chiral}}$, so the bound is also easily satisfied for these theories.  Part of the problem is that we have only included a subgroup of the full $\SU(N_f) \times \SU(N_f)$ global symmetry when deriving our bounds.  In a future publication~\cite{us:future} we hope to extend the bounds to bi-fundamentals transforming under an $\SU(N) \times \SU(N)$ symmetry group, in order to make closer contact with the values realized in SQCD and similar theories.

\section{Bounds on Current Two-point Functions}
\label{sec:currentbounds}

\subsection{General Theories}

Now let us turn to placing bounds on another set of fundamental OPE coefficients, namely those appearing in front of spin-1 conserved global symmetry currents.  In the OPE between $\SO(N)$ or $\SU(N)$ fundamentals, we should be careful to distinguish between the $\SO(N)$ or $\SU(N)$ symmetry currents living in the adjoint representation and singlet currents associated to some other global symmetry that we are not considering explicitly.

\subsubsection{Adjoint Currents}

Let us begin by focusing on the case of adjoint currents. Consider a CFT with some global symmetry, containing a scalar field $\phi_i$ transforming in some representation of this symmetry. We will denote by $T^{A}_{ij}$ the generators in this representation. The associated conserved currents transform as global symmetry adjoints. 
Ward identities completely fix the three-point functions with one current insertion:
\be
\<\phi_i(x_1) \phi_j(x_2) J_\mu^A(x_3) \> = -\frac{i}{2\pi^2} T_{ij}^A \frac{x_{12}^{2-2d}}{x_{13}^2 x_{23}^2} Z_{\mu},
\quad
\textrm{where}\quad 
 Z_\mu \equiv \frac{x_{13\mu}}{x_{13}^2}-\frac{x_{12\mu}}{x_{12}^2}.
\ee
With the above normalizations, the two-point function $\<J^A J^B\>$ contains undetermined coefficients $\tau^{AB}$ that roughly measure the amount of stuff charged under the global symmetry:
\be
\label{eq:taudef}
\<J^{A\mu}(x_1)J^{B\nu}(x_2)\> =\frac{3 \tau^{AB}}{4\pi^4}\frac{I^{\mu\nu}(x_{12})}{x_{12}^6}\,.
\ee
Let us write $\tau^{AB} \equiv \ka \Tr(T^A T^B)$, where $\ka$ can be viewed as a symmetry current `central charge.' As we did for the energy momentum tensor, we can rescale $J^A$ to have a canonically normalized two-point function and absorb $\ka$ into the OPE coefficient $\lambda_J^2$ associated with the current. In the end, the contribution of an adjoint current to a four-point function of $\phi_i$'s can be written
\be
x_{12}^{2d} x_{34}^{2d} \< \phi_i \phi_j \phi_k \phi_l \> \sim \frac{1}{6\ka} \Tr(T^A T^B)^{-1} T^{A}_{ij}T^{B}_{kl}\, g_{3,1} .
\ee 
In order to proceed further we need to specify the global symmetry group. For instance, for $\SO(N)$ and $\f_i$ in the vector representation, one can show that
\be
\Tr(T^A T^B)^{-1}T^{A}_{ij}T^B_{kl}=\frac 1 2 (\de_{il}\de_{jk}-\de_{ik}\de_{jl}),
\ee
and consequently, comparing to Eq.~(\ref{eq:SONconformalblock}), we have $\l_J^2 = \frac{1}{12\ka}$.  Similarly, for $\SU(N)$ and $\f_i$ in the fundamental representation, we have
\be
\Tr(T^A T^B)^{-1} (T^A)^i_j (T^B)_l^k &=& \de^i_l\de_j^k-\frac 1 N \de^i_j \de^k_l,
\ee
so that $\l_J^2 = \frac 1 {6\ka}$.  These relations hold for currents appearing in OPEs in general CFTs; we will discuss the generalization to $\cN=1$ superconformal theories below.  However, first we will consider the situation of singlet currents appearing in the OPE, namely currents corresponding to a global symmetry that is different from the $\SO(N)$ or $\SU(N)$ that  we are studying.

\subsubsection{Singlet Currents}

As mentioned above, the $\SO(N)$ or $\SU(N)$ global symmetry current is not the only conserved spin-1 operator of dimension $3$ that can contribute to the four-point function; additional currents, possibly transforming in different representations, may also exist. Clearly the presence of an additional conserved current implies the existence of a global symmetry beyond the one exploited to write the crossing symmetry constraints. The OPE coefficient associated to this operator not only contains the two-point function normalization, but also parametrizes our ignorance about the nature of the additional global symmetry.  Indeed, when the global symmetry is not specified the three-point function coefficient could in principle be arbitrary.

In the case of fundamentals transforming under an $\SO(N)$ global symmetry, spin-1 operators appearing in the OPE can only transform in the adjoint (antisymmetric) representation, corresponding to the $\SO(N)$ current itself.  In the case of $\SU(N)$ fundamentals, along with the adjoint current we also have the possibility of $\SU(N)$ singlet currents.\footnote{In addition, the OPE $\phi_i\times\phi_j$ could contain conserved spin-$1$ operators transforming in the antisymmetric representation of $\SU(N)$.  However, such currents (along with their complex conjugates) would generate charges which enhance $\SU(N)$ to a larger group $\SU(N)\to \SO(2N)$.  Thus, such theories necessarily fall under the class of CFTs with a global $\SO(2N)$ symmetry, which we consider separately.}  For example, we can think about a CFT with a global symmetry $\SU(N) \times \mathcal G$. If we consider scalar operators transforming in some representation of $\mathcal G$ with generators $\mathcal T^A$, then the $\mathcal G$-current is a singlet with respect to $\SU(N)$, and its contribution to the four-point function will be
\be
\frac{1}{6 \ka_{\mathcal G}} \Tr(\mathcal T^A \mathcal  T^B)^{-1}\mathcal  T^{A}_{ij}\mathcal T^{B}_{lm} \,\, g_{3,1} = \l^2_J \delta_{ij}\delta_{lm} \, g_{3,1} ,
\ee 
where $\ka_{\mathcal G}$ is the two-point function of the $\mathcal{G}$-current. Until we additionally specify the $\mathcal G$ symmetry group and charges, this parameter is arbitrary. However, we can collectively define, by analogy with the adjoint current, an {\it effective} current two-point function normalization $\ka_{\mathrm{eff}} \equiv 1/6{\l_J^2}$.  We will place bounds on $\ka_{\mathrm{eff}}$ when we give our results below.

\subsubsection{Free Theory and Numerical Results}

To clarify the above discussion, let us analyze in detail the theory of $N$ free complex scalars, using only information about the $\SU(N)$ global symmetry, which is contained in the larger $\SO(2N)$ symmetry of the theory.  The OPE $\phi_i\times \phi^{\bar j \dagger}$ contains an adjoint current and a singlet current,  both conserved:
\be
\	J_{\text{Ad}}^A \sim \phi^\dagger T^A \lrptl \phi,\,\qquad J_{\text{S}}\sim \phi^\dagger \lrptl \phi .
\ee 
The conformal block decomposition of the scalar four-point function directly gives us the values of the singlet and adjoint OPE coefficients for spin-$\ell$ currents:
\be
\	&&\l^2_{\text{Ad}}=\frac{(\ell!)^2}{(2\ell)!}\,,\qquad  \ka=\frac13,\\
\	&&\l^2_{\text{S}}=\frac{1}{N}\frac{(\ell!)^2}{(2\ell)!} \,,\qquad  \ka_{\text{eff}}=\frac{N}{3} .
\ee
The first point that we notice is the different scaling of the two above quantities with the size of the symmetry group. While the adjoint current two-point function normalization is independent of $N$, the singlet one grows with the dimension of the representation. We therefore expect lower bounds on $\ka_{\text{eff}}$ to scale with $N$, similarly to the way that the central charge bounds did in the previous section.

Let us now discuss the same theory, using the whole $\SO(N)$ global symmetry. This time only the adjoint current contributes to the four-point function, and its OPE coefficient (along with the other spin-$\ell$ adjoint operators) can be determined from the conformal block decomposition (see for instance~\cite{Rattazzi:2010yc}) as
\be
\	&&\l^2_{\text{Ad}}=\frac{(\ell!)^2}{(2\ell)!}\,,\qquad  \ka=\frac16 .
\ee

\begin{figure}[h!]
\begin{center}
\begin{psfrags}
\def\PFGstripminus-#1{#1}%
\def\PFGshift(#1,#2)#3{\raisebox{#2}[\height][\depth]{\hbox{%
  \ifdim#1<0pt\kern#1 #3\kern\PFGstripminus#1\else\kern#1 #3\kern-#1\fi}}}%
\providecommand{\PFGstyle}{}%
%
\psfrag{d}[cl][cl]{\PFGstyle $d$}%
\psfrag{k}[bc][bc]{\PFGstyle $\kappa$}%
\psfrag{Twompointf}[bc][bc]{\PFGstyle $\text{Lower bounds on $\kappa$ for $\SO(N)$ adjoint currents, $N=2,\dots,14$}$}%
\psfrag{x11}[tc][tc]{\PFGstyle $1$}%
\psfrag{x121}[tc][tc]{\PFGstyle $1.2$}%
\psfrag{x141}[tc][tc]{\PFGstyle $1.4$}%
\psfrag{x161}[tc][tc]{\PFGstyle $1.6$}%
\psfrag{x181}[tc][tc]{\PFGstyle $1.8$}%
\psfrag{y0}[cr][cr]{\PFGstyle $0$}%
\psfrag{y15}[cr][cr]{\PFGstyle $0.15$}%
\psfrag{y1}[cr][cr]{\PFGstyle $0.1$}%
\psfrag{y25}[cr][cr]{\PFGstyle $0.25$}%
\psfrag{y2}[cr][cr]{\PFGstyle $0.2$}%
\psfrag{y5m1}[cr][cr]{\PFGstyle $0.05$}%
\includegraphics[width=0.9\textwidth]{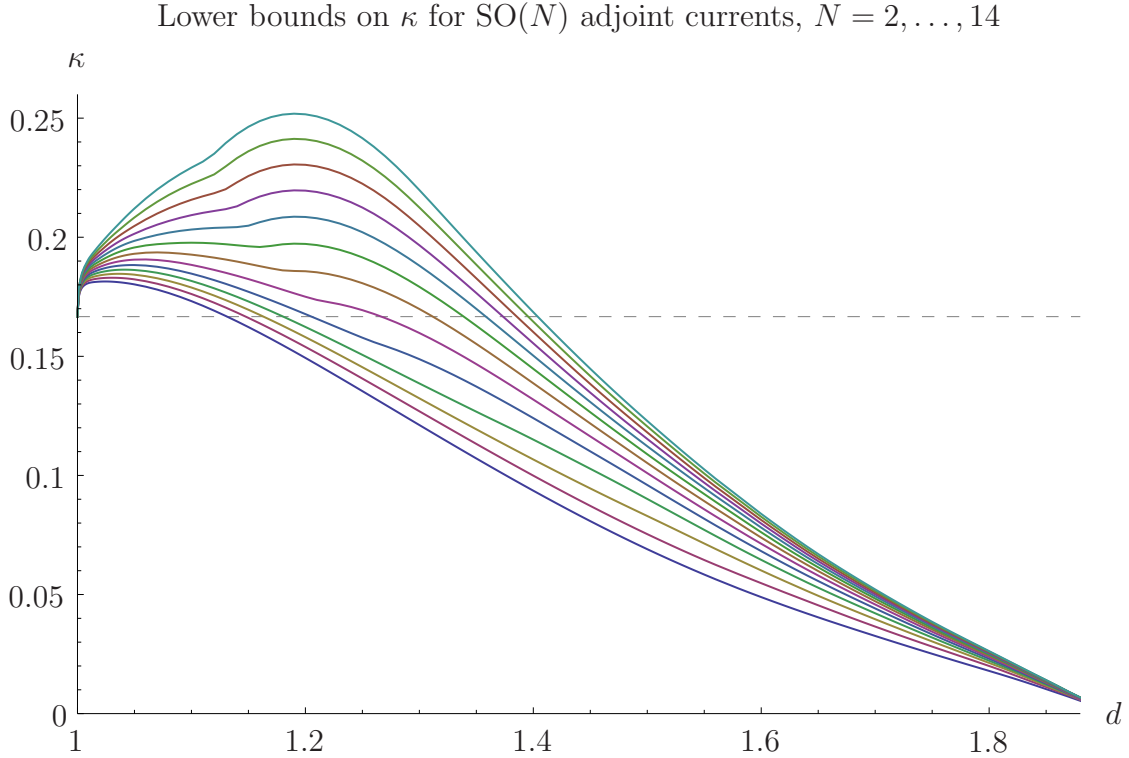}
\end{psfrags}
\end{center}
\caption{A lower bound on the two-point function coefficient $\<J^A_\mu J^B_\nu\>\propto \kappa \Tr(T^A T^B)$ of the $\SO(N)$ adjoint current appearing in $\f\x\f$, where $\f$ transforms in the fundamental of an $\SO(N)$ global symmetry group, for $N=2,\dots,14$. All curves smoothly approach the free $\SO(N)$ value $\kappa = 1/6$. Here we have taken $k=11$.}
\label{fig:SONkappa}
\end{figure}

Now that we have an intuition for the free values of $\ka$ and $\ka_\mathrm{eff}$, we are ready to present numerical bounds in several classes of theories.  In figure~\ref{fig:SONkappa}, we show a lower bound on the two-point function coefficient $\ka$ for a CFT with an $\SO(N)$ global symmetry for $N=2,\ldots,14$.  As expected, when $d \rightarrow 1$, all of the bounds drop sharply to the free $\SO(N)$ value $\ka = 1/6$.  The bounds get stronger as $N$ increases, while as $d$ varies away from $1$, they first become stronger and then weaken.

\begin{figure}[h!]
\begin{center}
\begin{psfrags}
\def\PFGstripminus-#1{#1}%
\def\PFGshift(#1,#2)#3{\raisebox{#2}[\height][\depth]{\hbox{%
  \ifdim#1<0pt\kern#1 #3\kern\PFGstripminus#1\else\kern#1 #3\kern-#1\fi}}}%
\providecommand{\PFGstyle}{}%
%
\psfrag{d}[cl][cl]{\PFGstyle $d$}%
\psfrag{k}[bc][bc]{\PFGstyle $\kappa_{\text{eff}}$}%
\psfrag{Twompointf}[bc][bc]{\PFGstyle $\text{Lower bounds on $\kappa_\mathrm{eff}$ for $\SU(N)$ singlet currents, $N=2,\dots,14$}$}%
\psfrag{x11}[tc][tc]{\PFGstyle $1$}%
\psfrag{x121}[tc][tc]{\PFGstyle $1.2$}%
\psfrag{x141}[tc][tc]{\PFGstyle $1.4$}%
\psfrag{x161}[tc][tc]{\PFGstyle $1.6$}%
\psfrag{x181}[tc][tc]{\PFGstyle $1.8$}%
\psfrag{y0}[cr][cr]{\PFGstyle $0$}%
\psfrag{y10kfree}[cr][cr]{\PFGstyle $10 \kappa_{\text{free}}$}%
\psfrag{y11kfree}[cr][cr]{}%
\psfrag{y12kfree}[cr][cr]{\PFGstyle $12 \kappa_{\text{free}}$}%
\psfrag{y13kfree}[cr][cr]{}%
\psfrag{y14kfree}[cr][cr]{\PFGstyle $14 \kappa_{\text{free}}$}%
\psfrag{y15kfree}[cr][cr]{}%
\psfrag{y16kfree}[cr][cr]{\PFGstyle $16 \kappa_{\text{free}}$}%
\psfrag{y17kfree}[cr][cr]{}%
\psfrag{y18kfree}[cr][cr]{\PFGstyle $18 \kappa_{\text{free}}$}%
\psfrag{y19kfree}[cr][cr]{}%
\psfrag{y20kfree}[cr][cr]{\PFGstyle $20 \kappa_{\text{free}}$}%
\psfrag{y2kfree}[cr][cr]{\PFGstyle $2 \kappa_{\text{free}}$}%
\psfrag{y3kfree}[cr][cr]{}%
\psfrag{y4kfree}[cr][cr]{\PFGstyle $4 \kappa_{\text{free}}$}%
\psfrag{y5kfree}[cr][cr]{}%
\psfrag{y6kfree}[cr][cr]{\PFGstyle $6 \kappa_{\text{free}}$}%
\psfrag{y7kfree}[cr][cr]{}%
\psfrag{y8kfree}[cr][cr]{\PFGstyle $8 \kappa_{\text{free}}$}%
\psfrag{y9kfree}[cr][cr]{}%
\psfrag{ykfree}[cr][cr]{}%
\includegraphics[width=0.9\textwidth]{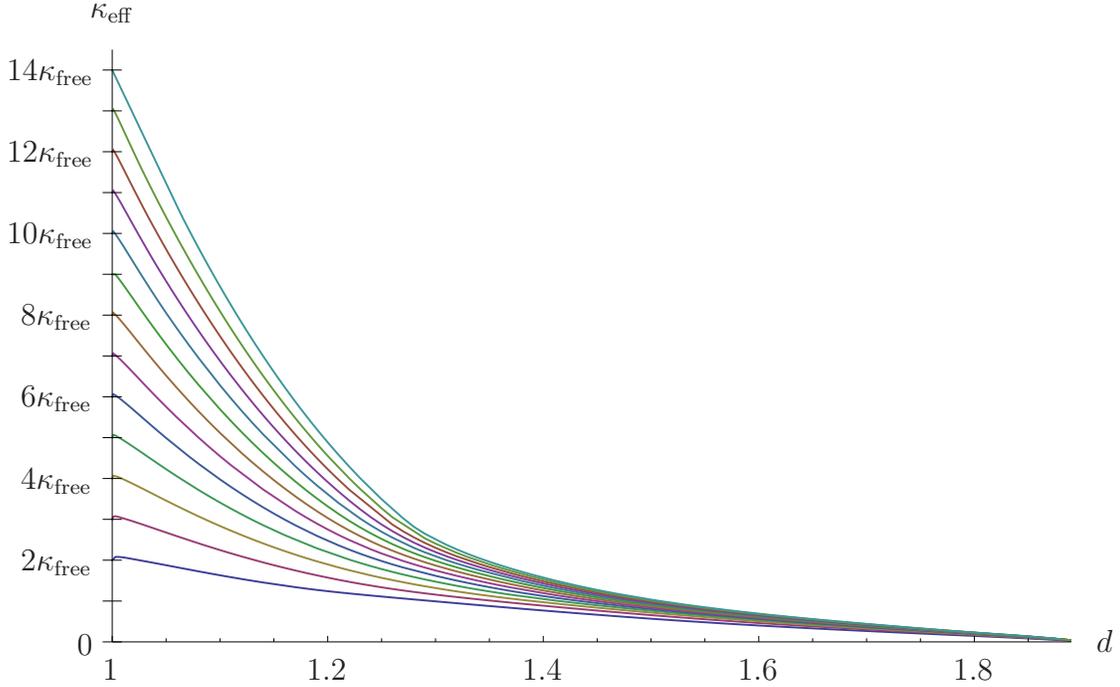}
\end{psfrags}
\end{center}
\caption{A lower bound on the effective two-point function coefficient $\ka_\mathrm{eff}=1/6\l_J^2$ of $\SU(N)$ singlet currents appearing in $\f_i\x\f^{\bar{\jmath}\dag}$, where $\f_i$ transforms in the fundamental of an $\SU(N)$ global symmetry group, for $N=2,\dots,14$. All curves interpolate continuously to the free values $N \kappa_{\text{free}}$ where $\kappa_{\text{free}} = 1/3$, and in this plot we have taken $k=11$.}
\label{fig:SUNkappaeff}
\end{figure}

As a second example, in figure~\ref{fig:SUNkappaeff} we consider the case of an $\SU(N)$ global symmetry and present lower bounds on $\ka_{\mathrm{eff}}$ for a singlet current. Our expectation that the constraints scale almost linearly with $N$ (when $d$ is close to 1) is confirmed.  Thus, this quantity serves as a rough measure of the number of degrees of freedom in the theory transforming under the symmetry, at least near $d = 1$.  One the other hand, the linear scaling disappears as $d$ increases.  

\subsection{Superconformal Theories}

Let us generalize the above bounds to theories with $\cN=1$ supersymmetry, where currents are descendants of scalar superconformal primaries of dimension $2$.  Consider four-point functions $\< \Phi_i \Phi^{\bar{\jmath}\dagger} \Phi_k \Phi^{\bar{l}\dagger}\>$ of chiral and anti-chiral operators transforming under an $\SU(N)$ global symmetry.  $\SU(N)$ adjoint currents give a superconformal block contribution
\be
x_{12}^{2d} x_{34}^{2d} \< \Phi_i \Phi^{\bar{\jmath}\dagger} \Phi_k \Phi^{\bar{l}\dagger}\> \sim \frac{1}{\ka} \Tr(T^A T^B)^{-1} (T^{A})_{i}^{\bar{\jmath}} (T^{B})_{k}^{\bar{l}} \, \mathcal{G}_{2,0} ,
\ee
while $\SU(N)$ singlet currents give an effective superconformal block contribution 
\be 
x_{12}^{2d} x_{34}^{2d} \< \Phi_i \Phi^{\bar{\jmath}\dagger} \Phi_k \Phi^{\bar{l}\dagger}\> \sim \frac{1}{\kappa_{\mathrm{eff}}} \delta_i^{\bar{\jmath}} \delta_k^{\bar{l}} \mathcal{G}_{2,0} .
\ee

\begin{figure}[h!]
\begin{center}
\begin{psfrags}
\def\PFGstripminus-#1{#1}%
\def\PFGshift(#1,#2)#3{\raisebox{#2}[\height][\depth]{\hbox{%
  \ifdim#1<0pt\kern#1 #3\kern\PFGstripminus#1\else\kern#1 #3\kern-#1\fi}}}%
\providecommand{\PFGstyle}{}%
%
\psfrag{d}[cl][cl]{\PFGstyle $d$}%
\psfrag{k}[bc][bc]{\PFGstyle $\kappa$}%
\psfrag{Twompointf}[bc][bc]{\PFGstyle $\text{Lower bounds on $\kappa$ for SUSY $\SU(N)$ adjoint currents, $N=2,\dots,14$}$}%
\psfrag{x11}[tc][tc]{\PFGstyle $1$}%
\psfrag{x121}[tc][tc]{\PFGstyle $1.2$}%
\psfrag{x141}[tc][tc]{\PFGstyle $1.4$}%
\psfrag{x161}[tc][tc]{\PFGstyle $1.6$}%
\psfrag{x181}[tc][tc]{\PFGstyle $1.8$}%
\psfrag{y0}[cr][cr]{\PFGstyle $0$}%
\psfrag{y1}[cr][cr]{\PFGstyle $0.1$}%
\psfrag{y2}[cr][cr]{\PFGstyle $0.2$}%
\psfrag{y3}[cr][cr]{\PFGstyle $0.3$}%
\psfrag{y4}[cr][cr]{\PFGstyle $0.4$}%
\psfrag{y5}[cr][cr]{\PFGstyle $0.5$}%
\psfrag{y6}[cr][cr]{\PFGstyle $0.6$}%
\psfrag{y7}[cr][cr]{\PFGstyle $0.7$}%
\psfrag{y11}[cr][cr]{\PFGstyle $1.0$}%
\psfrag{y151}[cr][cr]{\PFGstyle $1.5$}%
\psfrag{y21}[cr][cr]{\PFGstyle $2.0$}%
\includegraphics[width=0.9\textwidth]{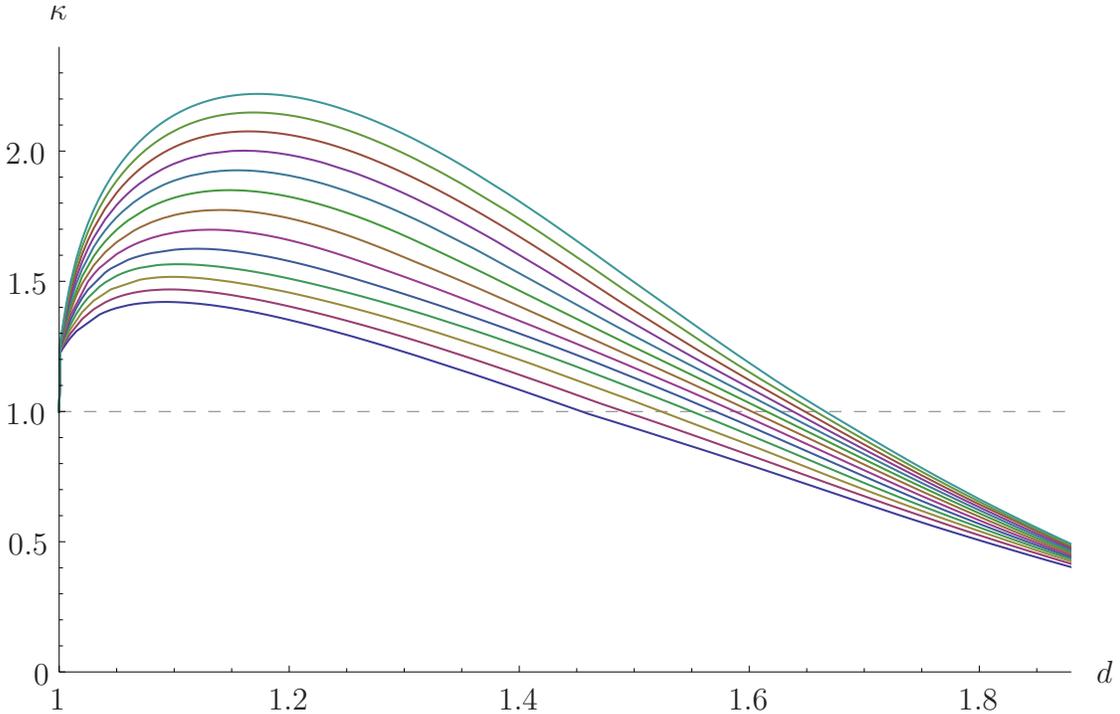}
\end{psfrags}
\end{center}
\caption{A lower bound on the two-point function coefficient $\<J^A J^B\>\propto \kappa \Tr(T^A T^B)$ of an $\SU(N)$ adjoint current appearing in $\Phi_i\x\Phi^{\bar{\jmath} \dag}$, where $\Phi_i$ is a chiral scalar transforming in the fundamental of an $\SU(N)$ global symmetry group in an SCFT, for $N=2,\dots,14$.  Despite appearances at this zoom level, all the curves above drop sharply near $d=1$ and interpolate continuously to the free value $\kappa_{\text{chiral}}=1$.  Here we have taken $k=10$.}
\label{fig:SUSYadjointcurrentkappa}
\end{figure}

In figure~\ref{fig:SUSYadjointcurrentkappa}, we show bounds on $\ka$ for adjoint currents appearing in $\Phi_i \x \Phi^{\jmath\dag}$, for SCFTs with an $\SU(N)$ global symmetry and $N=2,\dots,14$.  These bounds again increase strongly with $N$, growing as a roughly affine function.  For $d\lesssim 1.5$, $\ka$ must be substantially higher than its free value, with the bound dropping sharply to the contribution of a free chiral superfield $\ka_\mathrm{chiral} = 1$ near $d=1$.  Consequently, the free theory appears to be isolated in the space of SCFTs with an $\SU(N)$ flavor symmetry.  This accords with our intuition from theories with a Lagrangian description.  To couple a free $\SU(N)$ fundamental to a nontrivial interacting sector (and thus raise its dimension away from $d=1$), we need additional matter which must itself transform under $\SU(N)$.

\begin{figure}[h!]
\begin{center}
\begin{psfrags}
\def\PFGstripminus-#1{#1}%
\def\PFGshift(#1,#2)#3{\raisebox{#2}[\height][\depth]{\hbox{%
  \ifdim#1<0pt\kern#1 #3\kern\PFGstripminus#1\else\kern#1 #3\kern-#1\fi}}}%
\providecommand{\PFGstyle}{}%
%
\psfrag{d}[cl][cl]{\PFGstyle $d$}%
\psfrag{k}[bc][bc]{\PFGstyle $\kappa_{\text{eff}}$}%
\psfrag{Twompointf}[bc][bc]{\PFGstyle $\text{Lower bounds on $\kappa_\mathrm{eff}$ for SUSY $\SU(N)$ singlet currents, $N=2,\dots,14$}$}%
\psfrag{x11}[tc][tc]{\PFGstyle $1$}%
\psfrag{x121}[tc][tc]{\PFGstyle $1.2$}%
\psfrag{x141}[tc][tc]{\PFGstyle $1.4$}%
\psfrag{x161}[tc][tc]{\PFGstyle $1.6$}%
\psfrag{x181}[tc][tc]{\PFGstyle $1.8$}%
\psfrag{y0}[cr][cr]{\PFGstyle $0$}%
\psfrag{y10kfree}[cr][cr]{\PFGstyle $10 \kappa_{\text{chiral}}$}%
\psfrag{y11kfree}[cr][cr]{\PFGstyle $11 \kappa_{\text{chiral}}$}%
\psfrag{y12kfree}[cr][cr]{\PFGstyle $12 \kappa_{\text{chiral}}$}%
\psfrag{y13kfree}[cr][cr]{\PFGstyle $13 \kappa_{\text{chiral}}$}%
\psfrag{y14kfree}[cr][cr]{\PFGstyle $14 \kappa_{\text{chiral}}$}%
\psfrag{y15kfree}[cr][cr]{\PFGstyle $15 \kappa_{\text{chiral}}$}%
\psfrag{y16kfree}[cr][cr]{\PFGstyle $16 \kappa_{\text{chiral}}$}%
\psfrag{y17kfree}[cr][cr]{\PFGstyle $17 \kappa_{\text{chiral}}$}%
\psfrag{y18kfree}[cr][cr]{\PFGstyle $18 \kappa_{\text{chiral}}$}%
\psfrag{y19kfree}[cr][cr]{\PFGstyle $19 \kappa_{\text{chiral}}$}%
\psfrag{y20kfree}[cr][cr]{\PFGstyle $20 \kappa_{\text{chiral}}$}%
\psfrag{y2kfree}[cr][cr]{\PFGstyle $2 \kappa_{\text{chiral}}$}%
\psfrag{y3kfree}[cr][cr]{\PFGstyle $3 \kappa_{\text{chiral}}$}%
\psfrag{y4kfree}[cr][cr]{\PFGstyle $4 \kappa_{\text{chiral}}$}%
\psfrag{y5kfree}[cr][cr]{\PFGstyle $5 \kappa_{\text{chiral}}$}%
\psfrag{y6kfree}[cr][cr]{\PFGstyle $6 \kappa_{\text{chiral}}$}%
\psfrag{y7kfree}[cr][cr]{\PFGstyle $7 \kappa_{\text{chiral}}$}%
\psfrag{y8kfree}[cr][cr]{\PFGstyle $8 \kappa_{\text{chiral}}$}%
\psfrag{y9kfree}[cr][cr]{\PFGstyle $9 \kappa_{\text{chiral}}$}%
\psfrag{ykfree}[cr][cr]{\PFGstyle $\kappa_{\text{chiral}}$}%
\psfrag{yNull}[cr][cr]{\PFGstyle $\text{}$}%
\includegraphics[width=0.9\textwidth]{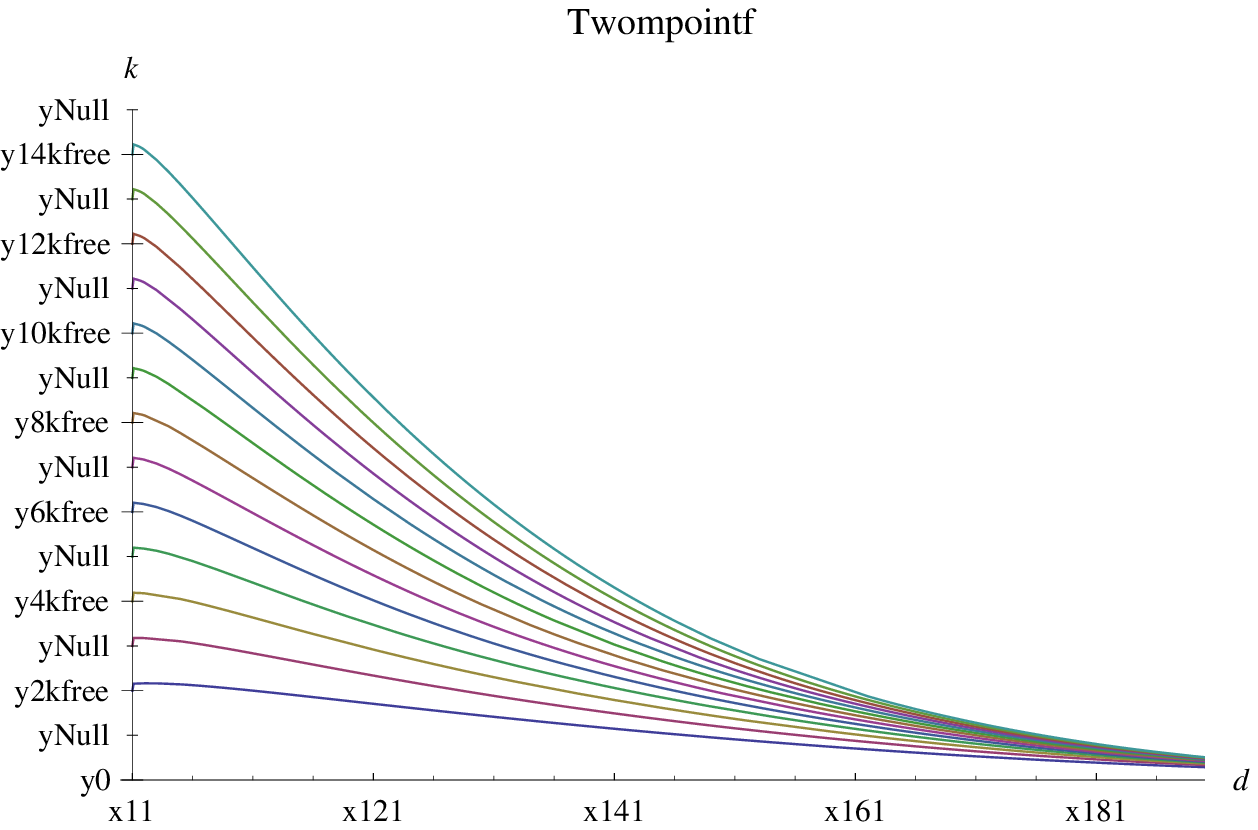}
\end{psfrags}
\end{center}
\caption{A lower bound on the effective two-point function coefficient $\ka_\mathrm{eff}=1/\l_J^2$ of $\SU(N)$ singlet currents appearing in $\Phi_i\x\Phi^{\bar{\jmath} \dag}$, where $\Phi_i$ is a chiral scalar transforming in the fundamental of an $\SU(N)$ global symmetry group, for $N=2,\dots,14$.  Despite appearances at this zoom level, all the curves above drop sharply near $d=1$ and interpolate continuously to the free values $N \kappa_{\text{chiral}}$ where $\kappa_{\text{chiral}} = 1$.  Here we have taken $k=10$.}
\label{fig:SUSYkappaeff}
\end{figure}

In figure~\ref{fig:SUSYkappaeff}, we also show a lower bound on $\ka_\eff$ for singlet currents appearing in $\Phi_i \times \Phi^{\bar{\jmath} \dagger}$.  Once again, we see that these bounds increase with $N$, scaling roughly linearly for small $d$.  As in the adjoint case above, the bounds drop very sharply to their free values $N \kappa_{\text{chiral}}$ near $d=1$, while the $N$ scaling disappears as $d$ increases.

\subsubsection{Comparison to SQCD}

As with central charges, our bounds on current two-point functions can be checked explicitly in a given superconformal theory.  For example, in SUSY QCD, $\SU(N_f)_L$ and $\SU(N_f)_R$ flavor currents appear in the OPE of a chiral meson and its conjugate 
\be
M^{\dag i}_{\tl \imath} \x M_j{}^{\tl \jmath} &\sim&\de^i_j (T^A)^{\tl\jmath}_{\tl\imath}J_R^A  + \de_{\tl\imath}^{\tl\jmath}  (T^A)^i_jJ_L^A+\dots.
\ee
Here, $i,j$ are indices for $\SU(N_f)_L$ and $\tl\imath,\tl\jmath$ are indices for $\SU(N_f)_R$.  We have not yet generated bounds that exploit the full $\SU(N_f)_L\x\SU(N_f)_R$ symmetry group of SQCD.  However, we can compare to our $\SU(N)$ bounds by `forgetting' one of the flavor groups, say $\SU(N_f)_R$, and examining the theory from the point of view of $\SU(N_f)_L$ alone.  Specifically, we shall set $\tl\imath=\tl \jmath=1$, so that the right-flavor currents $J_R^A$ are then singlet scalars in $M^{\dag i}_{1} \x M_j{}^{1}$, while the left-flavor currents $J_L^A$ are adjoints.  

The current two-point functions for $J_R^A$ and $J_L^A$ in SQCD both scale like $N_f$ (or $N_c$).  However, only our $\SU(N)$-singlet bounds scale with $N$, and thus have a chance of approaching the values for SQCD.  Consequently, we will focus on the contribution of $J_R^A$ to the conformal block expansion of meson four-point functions.  This reads
\be
x_{12}^{2d}x_{34}^{2d}\<M^{\dag i}_{1} M_j{}^{1} M^{\dag k}_{1} M_l{}^{1}\> &=& \tau_{AB}(T^A)_{1}^{1}(T^B)_{1}^{1} \de^i_j \de^k_l \cG_{2,0}+\dots,
\ee
where $\cG_{2,0}$ is the superconformal block for a conserved current multiplet and $\tau_{AB}=(\tau^{AB})^{-1}$ is the inverse two-point function coefficient for $J_R^A$.  In superconformal theories, $\tau^{AB}$ can be computed simply in terms of 't Hooft anomalies using $\tau^{AB}= -3 \Tr(R T^A T^B)$.  For $J_R^A$, this becomes
\be
\<J^A_R J^B_R\> \ \ \propto\ \ \tau^{AB} &=& \frac{3 N_c^2}{2N_f} \de^{AB},
\ee
where the $\SU(N_f)$ generators are normalized according to $\Tr(T^A T^B)=\frac 1 2 \de^{AB}$. Thus, we have
\be
\frac 1 {\ka_\eff} &=& \frac{N_f}{3N_c^2}\p{\de_1^1\de_1^1-\frac 1 {N_f}\de_1^1\de_1^1}\ \ =\ \ \frac{N_f-1}{3N_c^2}.
\ee

\begin{figure}[h!]
\begin{center}
\begin{psfrags}
\def\PFGstripminus-#1{#1}%
\def\PFGshift(#1,#2)#3{\raisebox{#2}[\height][\depth]{\hbox{%
  \ifdim#1<0pt\kern#1 #3\kern\PFGstripminus#1\else\kern#1 #3\kern-#1\fi}}}%
\providecommand{\PFGstyle}{}%
%
\psfrag{d}[cl][cl]{\PFGstyle $d$}%
\psfrag{k}[bc][bc]{\PFGstyle $\kappa_\mathrm{eff}$}%
\psfrag{Twompointf}[bc][bc]{\PFGstyle $\text{Lower bounds on $\kappa_\mathrm{eff}$ for SUSY $\SU(N)$ singlet currents and comparison to SQCD}$}%
\psfrag{x11}[tc][tc]{\PFGstyle $1$}%
\psfrag{x121}[tc][tc]{\PFGstyle $1.2$}%
\psfrag{x141}[tc][tc]{\PFGstyle $1.4$}%
\psfrag{x161}[tc][tc]{\PFGstyle $1.6$}%
\psfrag{x181}[tc][tc]{\PFGstyle $1.8$}%
\psfrag{y0}[cr][cr]{\PFGstyle $0$}%
\psfrag{y10kfree}[cr][cr]{\PFGstyle $10 \kappa_{\text{chiral}}$}%
\psfrag{y12kfree}[cr][cr]{\PFGstyle $12 \kappa_{\text{chiral}}$}%
\psfrag{y14kfree}[cr][cr]{\PFGstyle $14 \kappa_{\text{chiral}}$}%
\psfrag{y16kfree}[cr][cr]{\PFGstyle $16 \kappa_{\text{chiral}}$}%
\psfrag{y18kfree}[cr][cr]{\PFGstyle $18 \kappa_{\text{chiral}}$}%
\psfrag{y20kfree}[cr][cr]{\PFGstyle $20 \kappa_{\text{chiral}}$}%
\psfrag{y2kfree}[cr][cr]{\PFGstyle $2 \kappa_{\text{chiral}}$}%
\psfrag{y4kfree}[cr][cr]{\PFGstyle $4 \kappa_{\text{chiral}}$}%
\psfrag{y6kfree}[cr][cr]{\PFGstyle $6 \kappa_{\text{chiral}}$}%
\psfrag{y8kfree}[cr][cr]{\PFGstyle $8 \kappa_{\text{chiral}}$}%
\psfrag{yNull}[cr][cr]{\PFGstyle $$}%
\includegraphics[width=0.9\textwidth]{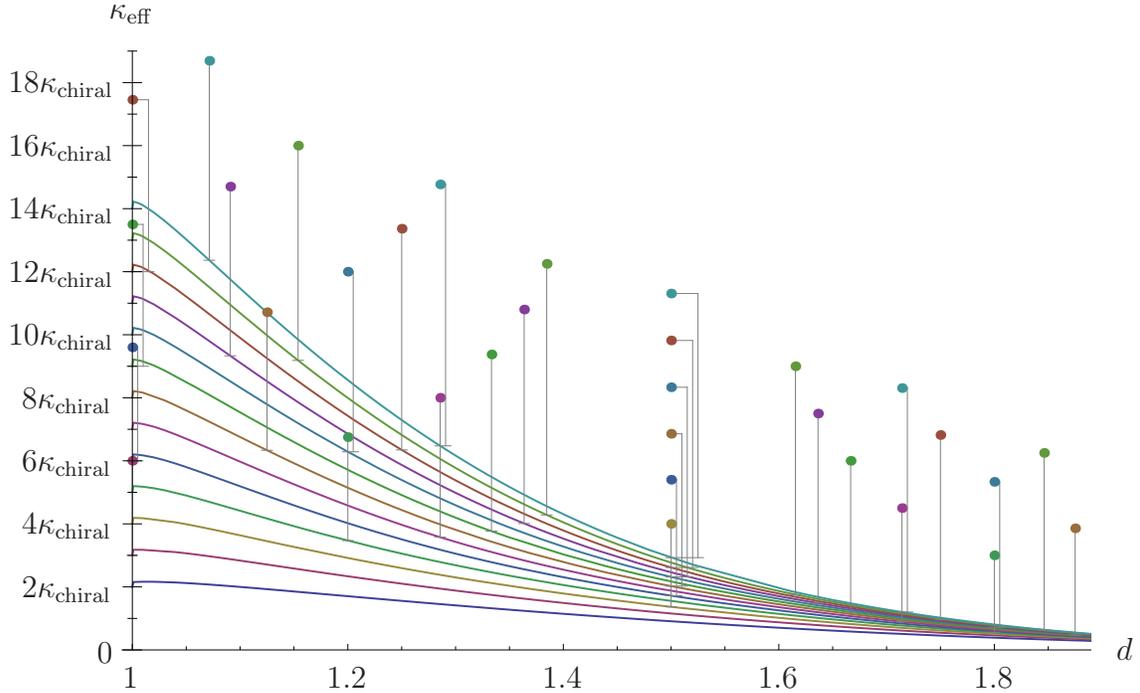}
\end{psfrags}
\end{center}
\caption{A lower bound on the effective two-point function coefficient $\ka_\mathrm{eff}= 1/\l_J^2$ of $\SU(N)$ singlet currents appearing in $\Phi_i\x\Phi^{\bar{\jmath} \dag}$, where $\Phi_i$ is a chiral scalar transforming in the fundamental of an $\SU(N)$ global symmetry group, for $N=2,\dots,14$.  Here we have taken $k=10$.  We have also plotted points corresponding to SQCD theories with various values of $N_f$ and $N_c$.  The lines below each point indicate the distance to the corresponding bound.  Many SQCD theories lie within an $O(1)$ factor from our bounds.}
\label{fig:SQCD2ptfncomparison}
\end{figure}

In figure~\ref{fig:SQCD2ptfncomparison} we compare this value of $\ka_\eff$ for several SQCD theories to our singlet current bounds from figure~\ref{fig:SUSYkappaeff}.  For many values of $N_f$ and $N_c$, our bound comes within an $O(1)$ factor of the SQCD value, with the smallest separation at small dimensions $d \sim 1$.  We expect our bound to become stronger with the added information of $\SU(N_f)_R$ symmetry, perhaps resulting in a hybrid of figures~\ref{fig:SUSYkappaeff} and \ref{fig:SUSYadjointcurrentkappa}.  It will be interesting to compare SQCD to these new bounds, and understand more about the structure of four-point functions in this important theory.

\section{Conclusions}
\label{sec:conclusions}

Let us briefly summarize our main results.   In this work we explored bounds on operator dimensions and OPE coefficients in 4D CFTs and $\cN=1$ SCFTs, building on the previous studies performed in~\cite{Rattazzi:2008pe,Rychkov:2009ij,Caracciolo:2009bx,Poland:2010wg,Rattazzi:2010gj,Rattazzi:2010yc,Vichi:2011ux}.   These bounds can be viewed as the initial stages of a concrete implementation of a 4D conformal bootstrap program.   Here we focused on bounds in the presence of $\SO(N)$ and $\SU(N/2)$ global symmetries, which had previously shown themselves to be more difficult (but not impossible~\cite{Vichi:2011ux}) to obtain using algorithms based on linear programming methods.  In order to push the program further, we presented a new algorithm based on semidefinite programming, which utilized the fact that derivatives of conformal blocks can be arbitrarily well approximated by positive functions times polynomials in the operator dimensions.  This new algorithm enabled us to show that there are completely general bounds on CFTs and SCFTs in the presence of global symmetries that are significantly stronger than were previously known to exist.

In particular, we greatly strengthened bounds on dimensions of singlet operators appearing in the OPE between fundamentals transforming under $\SO(N)$ or $\SU(N/2)$ global symmetries.  Bounds on dimensions of singlet operators in the presence of $\SO(4)$ or $\SU(2)$ global symmetries are relevant for models of conformal technicolor, and our bounds place severe constraints on these models, particularly when one does not assume any special flavor structure in four-fermion operators.  We refer readers to~\cite{rychkov:future} for further discussion of these constraints.  In fact, in the present work we saw that bounds on singlet operators were in general identical between $\SO(N)$ and $\SU(N/2)$ global symmetries.  We have so far not been able to construct a rigorous proof of this equivalence, so it would be good to gain a better understanding of it in future work.  

We also obtained similar bounds on operator dimensions in $\cN=1$ SCFTs, where we showed that there are bounds on the lowest-dimension scalar appearing in the $\Phi \times \Phi^\dagger$ OPE that appear to asymptote to the line $\De=2d$ near $d \sim 1$.  This result is particularly interesting in light of the discussion of~\cite{Fitzpatrick:2011hh} on positive anomalous dimensions of these operator in SCFTs --- our results demonstrate that this should not be possible when one is sufficiently close to the free limit.  

In this work we also initiated an exploration of both upper and lower bounds on OPE coefficients of protected operators appearing in the $\Phi \times \Phi$ OPE in SCFTs.  In this case, lower bounds are possible due to the fact that there is a gap in the dimensions of operators appearing in this OPE that is required by unitarity.  Because one can obtain bounds in both directions, we are able to see that the possible behavior is very tightly constrained even when one is only somewhat close to the free limit.  We expect that similar lower bounds should be possible in any situation (including non-supersymmetric theories) where one assumes that there is a dimension gap such that only a single operator can contribute to the conformal block decomposition up to a certain dimension.

We also explored bounds on central charges and current two-point function coefficients in the presence of operators transforming as fundamentals under $\SO(N)$ or $\SU(N/2)$ global symmetries, finding bounds that scale linearly with $N$ when the operator dimension is close to $1$.  An exception is the case when the current is the adjoint current corresponding to the $\SO(N)$ or $\SU(N/2)$ symmetry itself, in which case the bounds approach a value independent of $N$ in the free limit.  In superconformal theories, these bounds can be compared to concrete theories where the central charge $c$ and current two-point functions $\kappa$ are calculable using 't Hooft anomaly matching.  While the central charge bounds are still relatively far from their realized values, we showed concretely that our bounds on $\kappa$ are an $O(1)$ amount away from the values realized in supersymmetric QCD in the conformal window.

A clear future direction is to generalize these bounds on $\cN=1$ SCFTs to situations with bi-fundamentals transforming under $\SU(N) \times \SU(N)$ global symmetries (or adjoints transforming under $\SU(N)$ global symmetries).  Then one would hope to see bounds on the central charge that scale like $\sim N^2$, as well as significantly stronger bounds on current two-point functions.   It will be fascinating to see how these bounds compare to concrete $\cN=1$ theories such as supersymmetric QCD in the conformal window, particularly if one can find theories that nearly saturate the bounds.  One could also input all known information about these theories and attempt to find even stronger constraints on the dimensions of unprotected operators.\footnote{An alternate approach to learning about these dimensions  is to look for hidden structure such as integrability (e.g., see~\cite{Beisert:2010jr}) that makes the theory more solvable than one na\"ively expects.  We recently started exploring the possibility of such structure in $\cN=1$ SQCD in~\cite{Poland:2011kg}.} We plan to explore these bounds in a future publication~\cite{us:future}.

Another interesting direction would be to apply these methods to four-point functions of operators with spin, such as symmetry currents or the stress tensor.  To do this, one needs a tractable way of working with higher-spin conformal blocks.  Recently, some progress has been made in this direction~\cite{Costa:2011mg, Costa:future}, though more work may be needed in order to make a completely general analysis possible.  However, if this program could be carried out, one could for example start to study whether crossing symmetry of stress tensor four-point functions is connected to the bounds on $a/c$ obtained in~\cite{Hofman:2008ar}.  In fact, it may be more immediately tractable to begin such explorations for theories with $\cN=2$ supersymmetry, where the stress tensor is contained in a scalar multiplet.  Similarly, one can explore crossing symmetry of current four-point functions in $\cN=1$ theories, where progress at deriving the relevant superconformal blocks was made recently in~\cite{Fortin:2011nq}.  We believe that these directions may be worth pursuing in future work.

One would additionally like to generalize these bounds to 3D CFTs, where progress has been recently made at understanding the properties of 3D conformal blocks~\cite{Dolan:2011dv}.  While closed-form expressions are not yet known, it is likely that recursion relations similar to what we used in the present study could make a numerical exploration tractable.  It would be particularly interesting to see if one could learn more about the 3D Ising model using these methods, or whether bounds could be placed on the behavior of real-world condensed matter systems.  It is also interesting to explore these bounds in 2D (expanding on the preliminary studies of~\cite{Rattazzi:2008pe,Rychkov:2009ij}) or in 6D, where perhaps progress can be made at unraveling the structure of the mysterious 6D $(2,0)$ SCFTs. 

Of course, it would be nice to have a better analytical understanding of the structure of the optimal bounds.  While such an understanding has eluded us so far, it is possible that a new approach (such as studying the Mellin representation as in~\cite{Mack:2009mi,Penedones:2010ue,Fitzpatrick:2011ia}) could shed light on the origin of these bounds.  Less ambitiously, it would be good to study whether expansions of the crossing relation around other points in $(z,\bar{z})$ space may provide a more efficient way to find an optimal linear functional.  A related question is to understand whether any of the multiple crossing relations that we have used in cases of global symmetries are redundant or unnecessary for obtaining an optimal bound.  We leave such questions to future work.

Finally, we hope that progress can be made at understanding where these bounds fit in the context of the AdS/CFT correspondence~\cite{Maldacena:1997re,Gubser:1998bc,Witten:1998qj}.  Bounds on the central charge and current two-point function coefficients can be mapped to limitations on the strength of gravitational or gauge forces in the presence of light bulk excitations.  In the present work, we have obtained bounds that scale with the sizes of global symmetry representations, which in AdS corresponds to scaling with the size of the bulk gauge group.  While many of our bounds necessarily apply in a highly quantum regime, we have seen that there are at least some bounds (e.g., bounds on operator dimensions in SCFTs) that constrain deviations from the large-$N$ factorization limit, where an AdS description would be weakly coupled.  It would then be good to find alternate ways of arriving at these bounds in the context of AdS, particularly since these constraints are not obvious from the perspective of effective field theory~\cite{Fitzpatrick:2011hh}.  One hopes that thinking more along these lines will lead to a deeper understanding of which low-energy theories may admit consistent UV completions, particularly in the context of quantum gravity.

\section*{Acknowledgements}

We thank Nima Arkani-Hamed, Diego Hofman, Ken Intriligator, Juan Maldacena, Riccardo Rattazzi, Slava Rychkov, and Matt Strassler for helpful comments and conversations.  The computations in this paper were run on the Odyssey cluster supported by the FAS Science Division Research Computing Group at Harvard University.  We would like to thank John Brunelle in particular for technical support.  This work is supported in part by the Harvard Center for the Fundamental Laws of Nature, NSF grant PHY-0556111, the Swiss National Science Foundation under contract No. 200021-125237, and by the Director, Office of Science, Office of High Energy and Nuclear Physics, of the US Department of Energy under Contract DE-AC02-05CH11231.

\appendix

\section{Polynomial Approximation Details}
\label{app:polynomialapproximations}

In this appendix we give further details of our implementation of the optimization problem discussed in section~\ref{sec:semidefiniteprogramming} using semidefinite programming.  In all of the situations we consider, the problem is to find the optimal set of coefficients $a_{mnk}$, which minimizes the combination $-a_{mnk} V^{S^+,mnk}_0(0)$, subject to the constraints
\be
\label{eq:SDPSUNalphaconstraint1}
a_{mnk} V^{I_0,mnk}_{\ell_0}(\De_0) &=& 1,\qquad\textrm{}\\
\label{eq:SDPSUNalphaconstraint2}
a_{mnk} V^{I,mnk}_{\ell}(\De) & \geq & 0,\qquad\textrm{for all other (non-unit) operators in the spectrum.}
\ee
Here $V^{I,mnk}_{\ell}(\De) = \partial_{z}^m \partial_{\bar{z}}^n V_{\De,\ell}^{I,k}$ denotes derivatives of the $k$-th component of the appropriate vector $V^I_{\De,\ell}$, which may be any of the functions $\{F_{\De,\ell}, H_{\De,\ell}, \cF_{\De,\ell}, \cH_{\De,\ell},\tl\cF_{\De,\ell}, \tl\cH_{\De,\ell}\}$.  The index $I$ denotes possible global symmetry representations.

As discussed in section~\ref{sec:semidefiniteprogramming}, to apply semidefinite programming we must approximate $V_\ell^{I,mnk}(\De)$ as $\chi_\ell(\De)P^{I,mnk}_\ell(\De)$, where $\chi_\ell(\De)$ is a strictly positive function, and $P^{I,mnk}_\ell(\De)$ is a polynomial in $\De$.  Let us begin by discussing derivatives of $F_{\De,\ell}$ and $H_{\De,\ell}$.  It is convenient to first rescale each of these by a $(\De,\ell)$-independent function of $z$ and $\bar{z}$, so that they become sums of terms that factorize:
\be\label{eq:Fexpansion}
E_{\De,\ell,+}(z,\bar{z}) &\equiv& \left[\frac{(z-\bar{z})}{[(1-z)(1-\bar{z})]^d} - \frac{(z-\bar{z})}{(z \bar{z})^d} \right] F_{\De,\ell}(z,\bar{z}) \nn\\
&=& \left[ \frac{k_{\De+l}(z) k_{\De-l-2}(\bar{z})}{(z\bar{z})^{d-1}} + \frac{k_{\De+l}(1-z) k_{\De-l-2}(1-\bar{z})}{[(1-z)(1-\bar{z})]^{d-1}}\right] - (z \leftrightarrow \bar{z}), \\
\label{eq:Hexpansion}
E_{\De,\ell,-}(z,\bar{z}) &\equiv& \left[\frac{(z-\bar{z})}{[(1-z)(1-\bar{z})]^d} + \frac{(z-\bar{z})}{(z \bar{z})^d} \right] H_{\De,\ell}(z,\bar{z})\nn\\
&=& \left[ \frac{k_{\De+l}(z) k_{\De-l-2}(\bar{z})}{(z\bar{z})^{d-1}} - \frac{k_{\De+l}(1-z) k_{\De-l-2}(1-\bar{z})}{[(1-z)(1-\bar{z})]^{d-1}}\right] - (z \leftrightarrow \bar{z}),
\ee
where $k_{\b}(z) \equiv z^{\b/2} {}_2F_1(\b/2,\b/2,\b,z)$.  Derivatives of these quantities at $(1/2,1/2)$ can then be straightforwardly  evaluated using~\cite{Poland:2010wg}
\be\label{eq:Ceq}
C^{n}_{\beta,d} &\equiv& \ptl_z^n \left[ z^{1-d+\b/2} {}_2F_1(\b/2,\b/2,\b,z) \right]_{z=1/2} \nn\\
&=& 2^{n+(d-1)-\beta/2}\frac{\Gamma(\beta/2+2-d)}{\Gamma(\beta/2+2-d-n)} {}_3F_2(\beta/2+2-d,\beta/2,\beta/2;\beta/2+2-d-n,\beta;1/2) \nn\\
&=& 2(5-2d-n)C^{n-1}_{\b,d}+2\p{\b(\b-2)+2 n(n-3)-2 d^2+8d-2}C^{n-2}_{\b,d}\nn\\
&& +\,8(n-2)(n+d-4)^2 C^{n-3}_{\b,d} \nn\\
&=& P^n_{d}(\b) k_\b(1/2) + Q^n_{d}(\b) k'_\b(1/2) .
\ee
Here $P^n_{d}(\b)$ and $Q^n_{d}(\b)$ are polynomials in $\b$ that can be determined through the above recursion relation for $C^n_{\b,d}$.\footnote{This recursion relation follows from the hypergeometric differential equation for $k_\b(z)$, which itself is a consequence of the fact that conformal blocks are eigenfunctions of the quadratic Casimir of the conformal group.}
Note that taking $z \rightarrow 1-z$ simply introduces an overall factor of $(-1)^n$.  

In Eq.~(\ref{eq:Ceq}), we have written derivatives of $k_\b(z)$ at $z=1/2$ in terms of polynomials in $\b$, up to two non-polynomial quantities: $\b k_\b(1/2)$ and $k_\b'(1/2)$.  For the purposes of writing positivity constraints, we are free to divide by $k_\b'(1/2)/\b$, which is positive for all $\b$ that occur in unitary theories ($\b\geq -1$).   Now, the crucial fact for us is that the remaining non-polynomial quantity $K_\b\equiv\b k_{\b}(1/2)/k'_{\b}(1/2)$ is meromorphic in $\b$, and admits a simple approximation in terms of rational functions
\be\label{eq:rationalapprox}
K_\b\equiv \frac{\b k_{\b}(1/2)}{k'_\b(1/2)} \simeq \frac{1}{\sqrt{2}}  \prod_{j=0}^M\frac{ (\b - r_j) }{(\b - s_j)}  \equiv \frac{N_M(\b)}{D_M(\b)},
\ee
where $N_M(\b)$ and $D_M(\b)$ are polynomials in $\b$ of degree $M+1$.
Here, $r_j$ is the $j$'th zero of $\b k_{\b}(1/2)$ and $s_j$ is the $j$'th zero of $k'_{\b}(1/2)$, both of which are close to $-2j-1$.  Ordinarily we would need to account for both the zeros and poles of $\b k_\b(1/2)$ and $k'_\b(1/2)$ in the above product representation.  However, the poles of $\b k_\b(1/2)$ and $k'_\b(1/2)$ coincide at the negative odd integers, and so cancel between numerator and denominator.\footnote{The factor $1/\sqrt 2$ is $\lim_{\b\to\oo}K_\b$ (with an arbitrary phase for $\b$), as can be verified using the standard integral formula for ${}_2F_1$ hypergeometric functions.  Since this limit exists, $K_\b$ is meromorphic on the Riemann sphere, not just $\C$.}

The approximation Eq.~(\ref{eq:rationalapprox}) becomes arbitrarily good as more zeros are included, and moreover converges very quickly.  In fact, one can show that
\be
r_j,s_j &=& -1-2j+O(2^{-5.5j})\qquad j=0,1,2,\dots,
\ee
so that
\be
K_\b &=& \frac{N_M(\b)}{D_M(\b)}\times\p{1+O\p{\frac{2^{-5.5 M}}{\b+2M+1}}} \qquad(\b\geq -2M-1).
\ee
Consequently, it is sufficient to take $M\sim\textrm{a few}$ to achieve an accurate rational approximation for $K_\b$ that holds uniformly for all physical values $\b\geq -1$.  In practice, we found that $M=3$ or $4$ gives excellent results, which remain effectively unchanged when $M$ is increased.  Henceforth, we will assume that some appropriate $M$ has been chosen, and write simply $N(\b)$ and $D(\b)$ for brevity.

Combining Eqs.~(\ref{eq:Ceq}) and (\ref{eq:rationalapprox}), we can now write
\be\label{eq:Cpol}
C^{n}_{\beta,d} &=& \frac{k'_\b(1/2)}{\b D(\b)} u_d^n(\b),
\ee
where 
\be
u_d^n(\b) \equiv N(\b) P^n_{d}(\b)  +  \b D(\b) Q^n_{d}(\b)
\ee
is a polynomial in $\b$, and it can be verified that the pre-factor $k'_\b(1/2)/\b D(\b)$ is positive for all $\b\geq -1$.  Note that the degree of $u_d^n(\b)$ depends on the number of roots $M+1$ included in the approximation of Eq.~(\ref{eq:rationalapprox}).

Derivatives of $E_{\De,\ell,\pm}(z,\bar{z})$ at $(1/2,1/2)$ can now be written
\be
\ptl_z^m\ptl_{\bar z}^n E_{\De,\ell,\pm}(1/2,1/2) &=& \chi_\ell(\De) U^{mn}_{\ell,d,\pm}(\De),
\ee
where
\be
\chi_\ell(\De) &\equiv& \frac{k'_{\De+\ell}(1/2) k'_{\De-\ell-2}(1/2)}{(\De+\ell)(\De-\ell-2)D(\De+\ell) D(\De-\ell-2)}
\ee
is positive, and
\be
U^{mn}_{\ell,d,\pm}(\De) &\equiv&\left(1 \pm (-1)^{m+n} \right)\left[u_d^m(\De+\ell) u_d^n(\De-\ell-2) - (m \leftrightarrow n)\right]
\ee
is a polynomial in $\De$.  The inequalities $a_{mnk} V^{I,mnk}_{\ell}(\De) \geq 0$ given in~(\ref{eq:SDPSUNalphaconstraint2}) are then equivalent to a set of polynomial inequalities, which can be rewritten in terms of a semidefinite program as described in section~\ref{sec:semidefiniteprogramming}.

Next let us consider derivatives of the functions $\cF_{\De,\ell}(z,\bar{z})$ and $\cH_{\De,\ell}(z,\bar{z})$, appearing in superconformal crossing relations.  We can again take derivatives using Eq.~(\ref{eq:Cpol}) after rescaling by the same functions of $z$ and $\bar{z}$ appearing in Eqs.~(\ref{eq:Fexpansion}) and~(\ref{eq:Hexpansion}).  Applying $\partial_z^m \partial_{\bar{z}}^n$ at $(1/2,1/2)$ to the resulting functions gives
\begin{align}
\chi_\ell(\De)& \left[\phantom{\frac 1 2}U^{mn}_{\ell,d,\pm}(\De)\right.\nn\\
&+\frac{(\De+\ell)}{4(\De+\ell+1)} \frac{D(\De+\ell)}{D(\De+\ell+2)}  \cK_{\De+\ell}\ U^{mn}_{\ell+1,d,\pm}(\De+1)\nn\\
&+\frac{(\De-\ell-2)}{4(\De-\ell-1)}  \frac{D(\De-\ell-2)}{D(\De-\ell)} \cK_{\De-\ell-2}\ U^{mn}_{\ell-1,d,\pm}(\De+1)\nn\\
&+ \left.\frac{(\De+\ell)(\De-\ell-2)}{16(\De+\ell+1)(\De-\ell-1)} \frac{D(\De+\ell)D(\De-\ell-2)}{D(\De+\ell+2)D(\De-\ell)} \cK_{\De+\ell} \cK_{\De-\ell-2}\ U^{mn}_{\ell,d,\pm}(\De+2) \right],\nn\\
\end{align}
where
\be
\cK_\b \equiv \frac{\b}{\b+2} \frac{k'_{\b+2}(1/2)}{k'_{\b}(1/2)}.
\ee
We can then use the fact that $\cK_\b$ can be arbitrarily well approximated by a rational function
\be\label{eq:rationalapproxSUSY}
\cK_\b \simeq (12-8\sqrt{2}) \frac{(\b+1) \prod_i (\b+2-s_i)}{\prod_j (\b - s_j)}.
\ee
Again, the approximation improves as more roots are included, and converges after only a few terms.  Thus, by isolating the polynomial numerator and denominator of the quantity
\be
\frac{\b}{4(\b+1)} \frac{D(\b)}{D(\b+2)} \cK_\b \equiv \frac{\cN(\b)}{\cD(\b)},
\ee
we can write the derivatives as a positive function times a polynomial in $\De$:
\begin{align}
  \frac{\chi_\ell(\De)}{\cD(\De+\ell) \cD(\De-\ell-2)} 
\times &\left[\ \ \cD(\De+\ell) \cD(\De-\ell-2)\ U^{mn}_{\ell,d,\pm}(\De)\right.\nn\\
 &+ \cN(\De+\ell) \cD(\De-\ell-2)\ U^{mn}_{\ell+1,d,\pm}(\De+1)  \nn\\ 
 &+ \cD(\De+\ell) \cN(\De-\ell-2)\ U^{mn}_{\ell-1,d,\pm}(\De+1)  \nn\\
  &\left.+ \cN(\De+\ell) \cN(\De-\ell-2)\ U^{mn}_{\ell,d,\pm}(\De+2) \right].
\end{align}
Finally, let us note that the results for $\tl{\cF}_{\De,\ell}(z,\bar{z})$ and $\tl{\cH}_{\De,\ell}(z,\bar{z})$ are identical, but with odd-spin terms having the opposite sign.  Thus, we see that we can reformulate any of the sum rules appearing in SCFTs as a semidefinite program, following the logic described in section~\ref{sec:semidefiniteprogramming}.

\section{Implementation in {\tt SDPA-GMP}}
\label{app:SDP}

In this appendix we'll give further details of our implementation of the SDP.  As we described in section~\ref{sec:semidefiniteprogramming} and appendix~\ref{app:polynomialapproximations}, the general problem (phrased as a SDP) is to minimize $-a_{i} V^{S^+,i}_{0}(0)$, subject to the constraints
\begin{align}
a_{i} V^{I_0,i}_{\ell_0}(\De_0) &= 1, \textrm{}\nn\\
a_{i} P^{I,i}_{\ell}(\De_{\ell}(1+x)) &=  [x]^T_{d_\ell} A_{\ell}^I [x]_{d_\ell}+x([x]^T_{d'_\ell} B_\ell^I [x]_{d'_\ell}) &\textrm{for $0 \leq \ell \leq L$,}\nn\\
A^I_\ell, B^I_\ell &\succeq  0 &\textrm{for $0\leq \ell \leq L$}.
\end{align}
For brevity we here we use the index $i=1,\ldots, k(k+1)/2 \times \textrm{dim}(V^I_{\De,\ell})$ to run over all of the $z$ and $\bar{z}$ derivatives under consideration, as well as the components of the vector $V^I_{\De,\ell}$.  $I$ runs over possible global symmetry representations, and $A^I_{\ell}$ and $B^I_{\ell}$ are positive semidefinite matrices.  We recall that $[x]_d$ is the vector with entries $(1,x,\ldots,x^d)$, and if the polynomial $P^{I,i}_\ell$ has degree $2\gamma_{\ell}+1-\e_{\ell}$ (with $\e_{\ell}=0,1$), then $d_\ell = 2\g_{\ell}$ and $d'_\ell = 2\g_\ell-2\e_\ell$.  

The middle constraint is an equality between polynomials in $x$, so in practice we will implement it by matching each polynomial coefficient:
\be
0 &=& \coeffs_x \left[ -a_i P_{\ell}^{I,i}(\De_{\ell}(1+x)) + \Tr(X_{d_\ell}(x) A_\ell^I)+x\Tr(X_{d'_\ell}(x)B_\ell^I) \right].
\ee
In this expression we have also defined the matrix $X_d(x) \equiv  [x]_d [x]^T_d$.  Since many SDP solvers only allow positive variables, in practice it will additionally be convenient to introduce a `slack variable' $s$, where without loss of generality we can replace $a_{i} \rightarrow a_{i} - s$ in the above expressions and require $a_i, s \geq 0$.

We solve the above semidefinite program using {\tt SDPA-GMP 7.1.2}~\cite{sdpa}, which utilizes the GNU Multiple Precision Arithmetic Library (GMP).  We use {\tt Mathematica 7.0} to compute the vectors $V^{I,i}_{\ell}$ and polynomials $P^{I,i}_{\ell}$, performing all computations using $100$ digits of precision.  When using the approximations of Eqs.~(\ref{eq:rationalapprox}) and~(\ref{eq:rationalapproxSUSY}) we keep four roots, leading to approximations that differ from the exact functions by $\sim 10^{-8} - 10^{-10}$, depending on the value of $\b$.  In our computations we have found it sufficient to take $L=20$; in addition we add constraints for $\ell=1000,1001$ in order to effectively include the asymptotic constraints at large $\ell$.  After setting up the problem in {\tt Mathematica}, we write the SDP to a file using the SDPA sparse data format.

When running {\tt SDPA-GMP}, we use the parameters:
\begin{center}
\begin{tabular}{c | c}
{\tt SDPA-GMP} Parameter & Value \\
\hline
maxIteration & $1000$ \\
epsilonStar & $10^{-10}$ \\
lambdaStar & $10^{20}$ \\
omegaStar & $10^{20}$ \\
lowerBound & $-10^{20}$ \\
upperBound & $10^{20}$ \\
betaStar & $0.1$ \\
betaBar & $0.3$ \\
gammaStar & $0.9$ \\
epsilonDash & $10^{-10}$ \\
precision & $200$
\label{tab:SDPAparams}
\end{tabular}
\end{center}
To make our plots, we run data points in parallel using the Odyssey computing cluster at Harvard University.  In the majority of our plots we use a horizontal spacing of $\delta d = 10^{-2}$, supplemented by a higher resolution scan with $\delta d = 10^{-3}$ for $d < 1.01$ ($\delta d \rightarrow \delta \De_0$ in figure~\ref{fig:realscalarOPE}).  To compute dimension bounds, we vary $\De_0$ using a binary search, terminating at a vertical resolution of $10^{-3}$.  In all cases that we have checked, increasing $L$ or including more roots in the polynomial approximation leads to a completely negligible ($\lesssim 10^{-4}$) change in the computed bound.



\end{document}